\documentclass[12pt,a4paper]{article}

\usepackage{jheppub,amsmath,amssymb,color,graphicx}

\usepackage[utf8]{inputenc}
\usepackage{tikz}
\usepackage{tikz-feynman}

\newcommand{\ETmiss}{{E\!\!\!/}_{\rm T}}

\title{Dimension-8 operators in $W^+W^-$ production via gluon fusion}
\date{\today}

\author[a]{Daniel Gillies,}
\author[a]{Andrea Banfi,}
\author[b]{Adam Martin,}
\author[a]{and Matthew A. Lim}

\affiliation[a]{Department of Physics and Astronomy, University of Sussex, Brighton BN1 9QH, U.K.}
\affiliation[b]{Department of Physics, University of Notre Dame, Notre Dame, IN 46556, USA}

 \abstract{We investigate the impact of dimension-8 operators on $W^+W^-$
   production at the LHC for the incoming gluon-gluon channel. To this end, we have identified all dimension-8 CP-even operators
   contributing to the process in question, and computed the
   corresponding tree-level helicity amplitudes for fully-leptonic
   decays of the $W$ bosons. These are implemented in the program
   MCFM-RE, which automatically incorporates the effect of a jet-veto to
   reduce the otherwise overwhelming $t\bar t$ background. We find that, unless we break the hierarchy of the effective field theory (EFT), the interference of the dimension-8 operators with the Standard Model is negligible across the considered
   distributions. This justifies including the square of dimension-6 operators
   when performing EFT fits with this channel. We then present new constraints on CP-even and CP-odd dimension-6 operators within the EFT regime. Lastly, we postulate a scenario in which the hierarchy of the EFT is broken, justified by the strong constraints on dimension-6 operators from existing on-shell Higgs data. In this scenario, we discuss the constraints that can be reasonably set on CP-even dimension-8 operators with current and future data. We remark that the effect of the jet-veto on the ability to constrain new physics in the $W^+W^-$ channel is quite dramatic and must be properly taken into account.
 } 
\begin{document}

\maketitle

\section{Introduction}
\label{sec:intro}

During the first two runs of the Large Hadron Collider (LHC) the
Standard Model (SM) has performed extremely well in predicting
cross-sections and other observables. Whilst there have been some
tensions between theory and data, there have been no $5\sigma$
deviations which would lead to the SM being rejected in favour of new
physics~\cite{ATLAS:2023fod, CMS:2024gzs, ATLAS:2019nkf,
  cms-xscts}. Future High Luminosity LHC (HL-LHC) runs will measure SM
parameters to even better precision and collect data to a luminosity
of up to $3\,\mathrm{ab}^{-1}$. This abundance of high-precision data,
alongside expected improvements in the control of both theoretical and
systematic uncertainties will allow us to push the SM to its
limits. With there being an extremely large space of UV-complete SM
extensions, and with the Standard Model having proved itself to be an
extremely good description of collider physics (at the energies probed
so far), Effective Field Theories (EFTs) have been used to categorise
possible deviations from the SM due to BSM physics at higher mass
scales than can currently be reached by colliders. The most popular of
these, the SM Effective Field Theory (SMEFT) relies on the assumptions
that SM gauge symmetries continue to apply at high energies, that
there is a gap between the electroweak (EW) scale and the scale
$\Lambda$ of physics beyond the SM (BSM), and that EW symmetry
breaking is linearly realised~\cite{Brivio:2017vri, BUCHMULLER1986621, Grzadkowski:2010es}. This gap allows for
a decoupling of the two scales generating an expansion of deformations
to the Standard Model which is finite at each order in the expansion
parameter $1/\Lambda$. These deformations manifest in operators which
modify the SM Lagrangian. These operators have mass dimension $D$
greater than four and have a coupling parameter inversely proportional
to $1/\Lambda^{(D-4)}$. At dimension-5 there is one independent
operator which can modify the SM Lagrangian and at dimension-6 there
are $59$ dimensions (assuming baryon and lepton number conservation
and flavour universality)~\cite{Grzadkowski:2010es}.

The power of these EFT methods is that as few assumptions as possible
are made about the UV-Complete theory. However, from these few
assumptions we can say that the first order effects of the infinite
number of possible UV-completions manifest in a $59$-dimensional space
of deformations to the SM. Each of these deformations will lead to
deviations in the SM in multiple observables and so we use many
collider channels to put constraints on this high-dimensional
space. The space of all possible operators can also be restricted
using theoretical arguments such as unitarity -- both theory and data
are thus able to constrain possible UV theories~\cite{CarrilloGonzalez:2023cbf, Adams:2006sv, Ellis:2018gqa}.

One class of observables which will benefit greatly from the increased
data available from the HL-LHC is diboson observables. In particular,
the high invariant-mass tails of diboson distributions can receive
contributions from EFT operators which grow with energy, in both $gg$
and $q\bar q$ channels. At dimension-6 $W^+W^-$ and other diboson processes have already been studied extensively~\cite{Butter:2016cvz, Green:2016trm, Zhang:2016zsp, Baglio:2017bfe, Ellis:2018gqa, Baglio:2018bkm, Baglio:2019uty, Bellan:2021dcy, Almeida:2021asy, Degrande:2012wf, Falkowski:2016cxu, Aoude:2019cmc}. Observables in $ZZ$, $Z\gamma$ and $W\gamma$ production
have already been studied at dimension-8~\cite{Ellis:2021dfa, Martin:2023tvi}.  On the
contrary, $W^+W^-$ production present more of a difficulty since it
is usually analysed in the context of a jet-veto. This jet-veto is
required to reduce the background from top-pair production but can
result in the introduction of a second jet-veto scale which breaks the
perturbative hierarchy of diagrams in $\alpha_s$, hence requiring a
jet-veto resummation. The interference of dimension-8 operators with the Standard Model has been studied for $q\bar q \to W^+W^-$~\cite{Degrande:2023iob} and several have been found to give contributions grow with energy. Only one analysis has been performed so far for
$gg \to W^+W^-$ production with higher order EFT
effects~\cite{Bellm:2016cks}. However, that study is incomplete as
only one dimensions-8 operator was considered.

In this paper we explore effective field theory (EFT) operators which
affect $gg \to W^+W^-$ production ($WW$ from now on) within the
context of the SM effective field theory. We focus on the case
in which the $W$ bosons decay into an electron and a muon and use
information from the tails of the distribution in the invariant mass
of the electron-muon system $M_{e\mu}$ to identify where the energy
dependence of the new physics operators becomes important.  The energy
dependence of the dimension-6 operators which enter into this process
has been studied in~\cite{Rossia:2023hen}. They identify six operators
which enter into this process at dimension-6. Of these operators, they
find that only two grow with energy, denoted by $\mathcal{O}_{tG}$ and
$\mathcal{O}_{GH}$ ($\mathcal{O}_{\varphi G}$
in~\cite{Rossia:2023hen}). They modify the $ggt$ coupling and
introduce a $ggh$ coupling respectively. The former enters into
diagrams at loop level with respect to the latter and so picks up a
large suppression. For this reason, we neglect it for this study. The
$\mathcal{O}_{GH}$ operator proceeds via an intermediate Higgs e.g.\
$gg \to h$ followed by $h \to WW$. Operators $\mathcal{O}_{GH}$ and
the anomalous $t\bar th$ coupling, generated by $\mathcal{O}_{tH}$
($\mathcal{O}_{t\varphi}$ in~\cite{Rossia:2023hen})
$\mathcal{O}_{H\Box}$, have already been constrained by both on-shell
and off-shell Higgs studies. However, at low energies, the $ggh$ and
$tth$ couplings become difficult to distinguish from each other and so
constraints are placed on both
together~\cite{Azatov:2013xha,Grojean:2013nya,Banfi:2013yoa,Azatov:2014jga,Azatov:2016xik}. For
this reason, we include both of these operators to see how the
constraints from the tails of distributions from $WW$ compare with
constraints from on-shell Higgs production.

At leading order in the SMEFT, the dimension-6 EFT operators first
enter into $ggWW$ at order $1/\Lambda^2$ by interfering with the
loop-induced SM contribution. However, many global SMEFT fits also use
the dimension-6 squared piece which formally enters as 
$1/\Lambda^4$~\cite{Celada:2024mcf, Ethier:2021bye}. This is the same order of the interference between the
SM and dimension-8 operators, which in this process generate $ggWW$
contact interactions. The effects of these operators grow with energy
and so should be accounted for in any analysis which aims to constrain the $gg$ induced
dimension-6 operators using their squared amplitudes. Moreover, dimension-6 CP-odd operators, which
enter $WW$ production only as squared contributions, should also be
included for a complete analysis.

Since the operators $\mathcal{O}_{GH}$ and $\mathcal{O}_{tH}$ also contribute to single Higgs production, they are highly constrained by current data. Taken in combination with
the fact that the dimension eight operators grow with energy as
$\hat s^2/\Lambda^4$ (where $\hat s$ is the partonic centre-of-mass
energy -- probed by some proxy for it), one may expect there is a kinematic
regime where dimension-8 effects are important, if not
dominant. However, when studying EFT effects, and especially those which grow
with energy in the tails of distributions, one must be vigilant about
the validity of the EFT expansion.

Last, as EFT effects manifest in small deviations from the SM, we need
the best possible SM predictions to have an accurate model of the background. Furthermore, any factorisable
effects that would modify BSM contributions should also be
included to the best of our abilities. To this end, we include both higher order EW
and jet-veto effects in our SM predictions and ensure that the latter
also applies to the colour initiated BSM signal in the presence of a
jet-veto. Electroweak effects have already been shown to be important in this channel, particularly in the high energy tails~\cite{Grazzini:2019jkl, Kallweit:2017khh, Banerjee:2024eyo}. With such a set-up we are able to study how these higher
order corrections and $WW$ specific analysis cuts affect the extracted
bounds.

In the following sections we will analyse the dimension-8 operators
which contribute to this process via gluon fusion. In
section~\ref{sec:EFT}, we provide expressions for the helicity
amplitudes for these operators and discuss the validity of including
these operators in the high invariant mass tail of the $M_{e\mu}$
distribution, and discuss the validity of the EFT regime. In
section~\ref{sec:Numerics}, we provide numerical predictions for the
state-of-the-art SM predictions and the dimension-6 and 8
contributions to this channel. We then perform fits with current data
and provide sensitivity studies at the HL-LHC
(section~\ref{sec:sensitivity}). We also discuss how systematic errors
and the jet-veto affects the ability to constrain these operators at
HL-LHC. Finally, in section~\ref{sec:constd8}, we consider a motivated
scenario where the constraints from Higgs on the dimension-6 operators
allow for the independent constraint of dimension-8 operators below
the mass scale already constrained for dimension-6.

\section{EFT Analysis of dimension-8 operators}
\label{sec:EFT}

In this section, we present dimension-8 operators contributing to $WW$
production via gluon fusion. We limit ourselves in this only to tree level processes which do not pick up a loop suppression.
The full set of
dimension-8 operators for the SM effective theory has been
determined in ref.~\cite{Lehman:2015coa}. From these we take those
which involve only the field strengths for the $W^I_{\mu}$ field and
the gluon field $A_\mu^a$, given by
\begin{subequations}
  \label{eq:WG}
  \begin{align}
  W^I_{\mu\nu}   & = \partial_\mu W^I_\nu-\partial_\nu W^I_\mu+ g_W \epsilon^{IJK} W_{\mu}^J W_\nu^K\,,\\
  G_{\mu\nu}^a & =  \partial_\mu A^a_\nu-\partial_\nu A^a_\mu+ g_s f^{abc} A_{\mu}^b A_\nu^c\,,
  \end{align}
  \end{subequations}
  as well as the Higgs field $H$. The corresponding dimension-8
  Lagrangian will contain terms:
\begin{equation}
  \label{eq:dim8-gg}
  \mathcal{L}\supset \sum_i \frac{c^{(GW)}_i}{\Lambda^4} \mathcal{O}_i +\sum_i \frac{\tilde c^{(GW)}_i}{\Lambda^4} \mathcal{\tilde O}_i\,,
\end{equation}
where $O_i$ are the CP-even operators, whereas $\tilde O_i$ are the
CP-odd ones. Note that only the CP-even contributions can interfere
with the SM when considering CP-even observables. Therefore, we will not consider the
contribution of CP-odd dimension-8 operators for the moment, leaving a discussion of their importance to section~\ref{sec:constd8}.

In section~\ref{sec:dim8ops} we introduce the CP-even dimension-8
operators we consider. Then, in section~\ref{sec:EFTValidity}, we
embed them in an effective Lagrangian, and investigate the validity of
the proposed EFT setup.
  
\subsection{Dimension-8 operators and their amplitudes}
\label{sec:dim8ops}

There are six CP-even dimension-8 operators contributing to $WW$
production via gluon fusion:
\begin{equation}
  \label{eq:gg-CP-even}
  \begin{split}
  \mathcal{O}_1  & = G_{\mu\nu}^a G^{a,\rho\sigma} W^{I,\mu\nu} W^{I}_{\rho\sigma}\,,\qquad\qquad
  \mathcal{O}_2   = G_{\mu\nu}^a G^{a,\mu\nu} W^{I,\rho\sigma} W^{I}_{\rho\sigma}\,,\\
  \mathcal{O}_3  & = G_{\mu\nu}^a \tilde G^{a,\mu\nu} W^{I,\rho\sigma} \tilde W^{I}_{\rho\sigma}\,,\qquad\qquad
  \mathcal{O}_4   = G_{\mu\nu}^a G^{a,\rho\sigma} \tilde W^{I,\mu\nu} \tilde W^{I}_{\rho\sigma}\,,\\
  \mathcal{O}_5  & = G_{\mu\rho}^a G^{a,\rho\nu} (D^{\mu} H)^{\dagger} (D_{\nu} H)\,,\qquad
  \mathcal{O}_6   = G_{\mu\nu}^a G^{a,\mu\nu}(D^{\rho} H)^{\dagger} (D_{\rho} H) \,,
  \end{split}
\end{equation}
where $\tilde T_{\mu\nu}=\frac 12 \epsilon_{\mu\nu\alpha\beta}T^{\alpha\beta}$ is
the dual of tensor $T_{\mu\nu}$.  In the unitary gauge, we set
\begin{equation}
  \label{eq:H}
  H(x) = \frac{1}{\sqrt 2} \left(
    \begin{array}{c}
      0 \\
      v+h(x)
    \end{array}
  \right)\,.
\end{equation}
Keeping only the terms that contribute to $WW$ production, we can rewrite the operators in eq.~\eqref{eq:gg-CP-even} in the form
\begin{equation}
  \label{eq:gg-CP-even-expanded}
  \begin{split}
  \mathcal{O}_1  & = 2 G_{\mu\nu}^a G^{a,\rho\sigma} W^{+,\mu\nu} W^-_{\rho\sigma} +\dots \,,\qquad
  \mathcal{O}_2   = 2 G_{\mu\nu}^a G^{a,\mu\nu} W^{+,\rho\sigma} W^-_{\rho\sigma} +\dots \,,\\
  \mathcal{O}_3  & = 2 G_{\mu\nu}^a \tilde G^{a,\mu\nu} W^{+,\rho\sigma} \tilde W^-_{\rho\sigma}+\dots \,,\qquad
  \mathcal{O}_4   = 2 G_{\mu\nu}^a G^{a,\rho\sigma}\tilde W^{+,\mu\nu} \tilde W^-_{\rho\sigma} +\dots \,,\\
  \mathcal{O}_5  & =  M_W^2 G_{\mu\rho}^a G^{a,\rho\nu} W^{+,\mu}  W^-_{\nu} +\dots \,,\quad\>
  \mathcal{O}_6  = M_W^2 G_{\mu\nu}^a G^{a,\mu\nu} W^{+,\rho}  W^-_{\rho} +\dots \,,
  \end{split}
\end{equation}
where we have introduced the short-hand notation
\begin{equation}
  \label{eq:Wmunu}
  W^\pm_{\mu\nu} = \partial_\mu W^\pm_\nu-\partial_\nu W^\pm_\mu +\dots \,,
\end{equation}
and used the SM relation $M_W = g_W v/2$, where $M_W$ is the mass of the
$W$ boson. This relation receives SMEFT corrections, but these enter at a higher order than we consider here. Also, the omitted terms in eqs.~\eqref{eq:gg-CP-even-expanded} and \eqref{eq:Wmunu} do not contribute to the process at
hand.

\begin{figure}[htbp]
  \begin{center}
    \begin{tikzpicture}[scale=2, transform shape]
      \begin{feynman}
        \vertex (a); 
        \vertex [right = 1cm  of a, dot] (v1) {};
        \vertex [right = 1cm  of v1] (c);
        \vertex [right = 0.5cm  of v1] (aux);
        \vertex [node font=\tiny, above = 0.55cm  of aux] (label1){\(W^+\)};
        \vertex [node font=\tiny, below = 0.55cm  of aux] (label2){\(W^-\)};
        \vertex [node font=\tiny, above = 0.9cm of a] (a1) {\(g\)};
        \vertex [node font=\tiny, below = 0.9cm of a] (a2) {\(g\)};
        \vertex [above = 0.9cm of c] (v2);
        \vertex [below = 0.9cm of c] (v3);
        \vertex [right = 0.8cm  of v2] (d);
       	\vertex [right = 0.8cm  of v3] (e);
       	\vertex [node font=\tiny, above = 0.6cm of d] (l1) {};
       	\vertex [node font=\tiny, right = 0.2cm of l1] (l12) {};
       	\vertex [node font=\tiny, below = 0.1cm of l12] (l13){\(\nu_{e}\)};
        \vertex [node font=\tiny, below = 0.6cm of d] (l2);
        \vertex [node font=\tiny, right = 0.2cm of l2] (l22) {};
        \vertex [node font=\tiny, above = 0.1cm of l22] (l23){\(e^+\)};
        \vertex [node font=\tiny, above = 0.6cm of e] (l3);
        \vertex [node font=\tiny, right = 0.2cm of l3] (l32) {};
        \vertex [node font=\tiny, above = 0cm of l32] (l33){\(\mu^{-}\)};
        \vertex [node font=\tiny, below = 0.6cm of e] (l4);
        \vertex [node font=\tiny, right = 0.2cm of l4] (l42) {};
        \vertex [node font=\tiny, below = 0cm of l42] (l43){\(\bar\nu_{\mu}\)};
        \diagram* {
          (a1) -- [gluon, momentum={\(p_1\)}] (v1), 
          (a2) -- [gluon, momentum'={\(p_2\)}] (v1), 
          (v1) -- [boson] (v2), 
          (v1) -- [boson] (v3), 
          (v2) -- [fermion, momentum={\(p_3\)}] (l1), 
          (v2) -- [anti fermion, momentum={\(p_4\)}] (l2), 
          (v3) -- [fermion, momentum={\(p_5\)}] (l3), 
          (v3) -- [anti fermion, momentum={\(p_6\)}] (l4),
        };
      \end{feynman}
       \end{tikzpicture}
  \end{center}
  \caption{Feynman diagram corresponding to the amplitude for the process $g(p_1)\>g(p_2)\! \to \! W^+(\to \nu(p_3)\>e^+(p_4))\>W^-(\to \mu^-(p_5)\> \bar \nu(p_6))$ occurring through the dimension-8 operators of eq.~(\ref{eq:gg-CP-even}).
  \label{fig:Feynman-dim8}
  }
\end{figure}
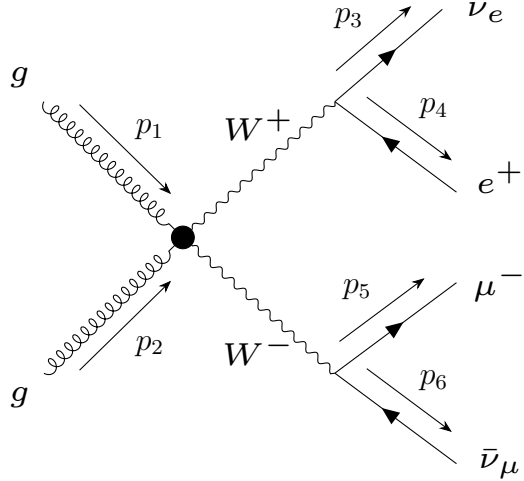

Each operator in eq.~(\ref{eq:gg-CP-even-expanded}) gives a contact
interaction between two incoming gluons of momenta $p_1,p_2$,
polarisation indices $\mu_1, \mu_2$, and colour indices $a_1,a_2$, and an
outgoing $W^+W^-$ pair. We consider the case in which $W^+$ decays into
two leptons of momenta $p_3$ and $p_4$, and $W^-$ into two leptons of
momenta $p_5$ and $p_6$. With this setup, the $W^+$ momentum is $p_{(34)}=p_3+p_4$ (and its
polarisation index $\mu_{(34)}$), and that of the $W^-$ is
$p_{(56)}=p_5+p_6$ (and its
polarisation index $\mu_{(56)}$). In terms of those momenta, the Feynman rules for
the different operators are:
\begin{equation}
  \label{eq:gg-FR}
  \begin{split}
    \mathcal{O}_1  : 8 i\,\frac{c^{(GW)}_1}{\Lambda^4} \delta_{a_1 a_2}\left[\right.&(p_1^{\mu_{(34)}} p_{(34)}^{\mu_1} -\eta^{\mu_1 \mu_{(34)}} (p_1 p_{(34)})) (p_2^{\mu_{(56)}} p_{(56)}^{\mu_2} -\eta^{\mu_2 \mu_{(56)}} (p_2 p_{(56)}))\\
    + &\left. (p_1^{\mu_{(56)}} p_{(56)}^{\mu_1} -\eta^{\mu_1 \mu_{(56)}} (p_1 p_{(56)})) (p_2^{\mu_{(34)}} p_{(34)}^{\mu_2} -\eta^{\mu_2 \mu_{(34)}} (p_2 p_{(34)}))\right]\\
    \mathcal{O}_2  : 16 i\,\frac{c^{(GW)}_2}{\Lambda^4} \delta_{a_1 a_2} & (p_1^{\mu_2} p_2^{\mu_1} -\eta^{\mu_1 \mu_2} (p_1 p_2)) (p_{(34)}^{\mu_{(56)}} p_{(56)}^{\mu_{(34)}} -\eta^{\mu_{(34)} \mu_{(56)}} (p_{(34)} p_{(56)})) \\
    \mathcal{O}_3  : 16 i \,\frac{c^{(GW)}_3}{\Lambda^4}\delta_{a_1 a_2} & \epsilon^{\mu_1\mu_2}_{\quad\>\>\alpha\beta} \, \epsilon^{\mu_{(34)}\mu_{(56)}}_{\quad\>\>\gamma\delta}\,p_1^\alpha p_2^\beta p_{(34)}^\gamma p_{(56)}^\delta\\
     \mathcal{O}_4  : 8 i \,\frac{c^{(GW)}_4}{\Lambda^4}\delta_{a_1 a_2}  \left[\right.&\left. \epsilon^{\mu_1\mu_{(34)}}_{\quad\>\alpha\beta} \, \epsilon^{\mu_2\mu_{(56)}}_{\quad\>\gamma\delta}\,p_1^\alpha p_{(34)}^\beta p_2^\gamma p_{(56)}^\delta +\epsilon^{\mu_1\mu_{(56)}}_{\quad\>\>\alpha\beta} \, \epsilon^{\mu_2\mu_{(34)}}_{\quad\>\>\gamma\delta}\,p_1^\alpha p_{(56)}^\beta p_2^\gamma p_{(34)}^\delta\right]\\
     \mathcal{O}_5  :  i \,\frac{c^{(GW)}_5 }{\Lambda^4}\delta_{a_1 a_2}  M_W^2 \left[\right.&
     \left((p_1 p_2) \eta^{\mu_1\mu_{(34)}}\eta^{\mu_2 \mu_{(56)}}+\eta^{\mu_1\mu_2}p_1^{\mu_{(34)}}p_2^{\mu_{(56)}}-\eta^{\mu_1\mu_{(34)}}p_1^{\mu_2} p_2^{\mu_{(56)}}-\eta^{\mu_2\mu_{(56)}}p_1^{\mu_{(34)}} p_2^{\mu_1}\right)\\ +
     &\left.\left((p_1 p_2) \eta^{\mu_1\mu_{(56)}}\eta^{\mu_2 \mu_{(34)}}+\eta^{\mu_1\mu_2}p_1^{\mu_{(56)}}p_2^{\mu_{(34)}}-\eta^{\mu_1\mu_{(56)}}p_1^{\mu_2} p_2^{\mu_{(34)}}-\eta^{\mu_2\mu_{(34)}}p_1^{\mu_{(56)}} p_2^{\mu_1}\right)\right]\,\\
     \mathcal{O}_6  : 4 i\,\frac{c^{(GW)}_6 }{\Lambda^4}\delta_{a_1 a_2}   M_W^2& (p_1^{\mu_2} p_2^{\mu_1}-(p_1p_2)\eta^{\mu_1 \mu_2}) \eta^{\mu_{(34)} \mu_{(56)}}
  \end{split}
\end{equation}
These Feynman rules can be used to construct the amplitude for
$gg\to WW$, with each $W$ boson decaying into a pair of leptons. This can be 
represented by the Feynman diagram in figure~\ref{fig:Feynman-dim8}.

Since decays of $W$ bosons give always left-handed fermions, we can
label the corresponding helicity amplitude $M_{\lambda_1,\lambda_2}$ for the process according to the
polarisation states of the incoming gluons $\lambda_1,\lambda_2=\pm$. Explicitly
\begin{equation}
  \label{eq:amplitude}
  \begin{split}    
M_{\lambda_1,\lambda_2}= \frac{g_W^2}{2}\, &\frac{  \delta_{a_1 a_2} }{\Lambda^4} \frac{i}{p_{(34)}^2-M_W^2 - i \Gamma_W M_W} \times \\ & \quad \times \frac{i}{p_{(56)}^2-M_W^2 - i \Gamma_W M_W}\, \sum_{i}  c^{(GW)}_i \mathcal{M}^{(i)}_{\lambda_1,\lambda_2}\,.
  \end{split}
\end{equation}
The subamplitudes $\mathcal{M}^{(i)}_{\lambda_1,\lambda_2}$ can be expressed in terms of the spinor products
\begin{equation}
  \label{eq:spinor-products}
  \langle ij\rangle \equiv \frac 12 \bar u(p_i)(1+\gamma^5) u(p_j)\,,\qquad [ij] \equiv
  \frac 12 \bar u(p_i)(1-\gamma^5) u(p_j)\,,
\end{equation}
and are given by
\begin{subequations}
  \label{eq:M1subamps}
  \begin{align}
    \mathcal{M}^{(1)}_{++} & = 4i\langle34\rangle\langle56\rangle\left(\left([14][26]\right)^2+\left([16][24]\right)^2\right) \,, \\
    \mathcal{M}^{(1)}_{--} & = 4i[34][56]\left(\left(\langle 13\rangle\langle 25\rangle \right)^2+\left(\langle 15\rangle \langle 23\rangle\right)^2\right)\,, \\
    \mathcal{M}^{(1)}_{+-} & = -4i\left(\langle34\rangle[56]\left(\langle25\rangle[14]\right)^2 + [34]\langle56\rangle\left(\langle23\rangle[16]\right)^2\right) \,, \\
    \mathcal{M}^{(1)}_{-+} & = -4i\left(\langle34\rangle[56]\left(\langle15\rangle[24]\right)^2 + [34]\langle56\rangle\left(\langle13\rangle[26]\right)^2\right)\,.
  \end{align}
\end{subequations}
\begin{subequations}
  \label{eq:M2subamps}
  \begin{align}
\mathcal{M}^{(2)}_{++} & = 8i[12]^2\left(\langle34\rangle\langle56\rangle[46]^2+[34][56]\langle35\rangle^2\right) \,, \\
    \mathcal{M}^{(2)}_{--} & = \frac{\langle 12\rangle^2}{[12]^2} \mathcal{M}^{(2)}_{++} = 8i\langle 12\rangle^2\left(\langle34\rangle\langle56\rangle[46]^2+[34][56]\langle35\rangle^2\right) \,, \\
    \mathcal{M}^{(2)}_{+-} & = \mathcal{M}^{(2)}_{-+} =  0 \,.
  \end{align}
\end{subequations}
\begin{subequations}
  \label{eq:M3subamps}
  \begin{align}
\mathcal{M}^{(3)}_{++} & = -8i[12]^2\left(\langle34\rangle\langle56\rangle[46]^2-[34][56]\langle35\rangle^2\right) \,, \\
     \mathcal{M}^{(3)}_{--} & = \frac{\langle 12\rangle^2}{[12]^2} \mathcal{M}^{(3)}_{++} = -8i\langle 12\rangle^2\left(\langle34\rangle\langle56\rangle[46]^2-[34][56]\langle35\rangle^2\right)\,, \\
    \mathcal{M}^{(3)}_{+-} & = \mathcal{M}^{(3)}_{-+} = 0\,.
  \end{align}                   
\end{subequations}
\begin{subequations}
  \begin{align}
    \mathcal{M}^{(4)}_{++} & = -\mathcal{M}^{(1)}_{++} = -4i\langle34\rangle\langle56\rangle\left(\left([14][26]\right)^2+\left([16][24]\right)^2\right) \,, \\
    \mathcal{M}^{(4)}_{--} & = -\mathcal{M}^{(1)}_{--} = -4i[34][56]\left(\left(\langle 13\rangle\langle 25\rangle \right)^2+\left(\langle 15\rangle \langle 23\rangle\right)^2\right)\,, \\
    \mathcal{M}^{(4)}_{+-} & = \mathcal{M}^{(1)}_{+-} = -4i\left(\langle34\rangle[56]\left(\langle25\rangle[14]\right)^2 + [34]\langle56\rangle\left(\langle23\rangle[16]\right)^2\right) \,, \\
    \mathcal{M}^{(4)}_{-+} & = \mathcal{M}^{(1)}_{-+} = -4i\left(\langle34\rangle[56]\left(\langle15\rangle[24]\right)^2 + [34]\langle56\rangle\left(\langle13\rangle[26]\right)^2\right)\,.
  \end{align}
\end{subequations}
\begin{subequations}
  \begin{align}
\mathcal{M}^{(5)}_{++} & = -M_W^2[12]^2\langle35\rangle\langle46\rangle \,,\\
    \mathcal{M}^{(5)}_{--} & = \frac{\langle 12\rangle^2}{[12]^2} \mathcal{M}^{(5)}_{++} = -M_W^2\langle 12\rangle^2\langle35\rangle\langle46\rangle \,, \\
    \mathcal{M}^{(5)}_{+-} & =  2M_W^2\langle23\rangle\langle25\rangle[14][16]\,, \\
    \mathcal{M}^{(5)}_{+-} & = 2M_W^2\langle13\rangle\langle15\rangle[24][26]\,.
  \end{align}
\end{subequations}
\begin{subequations}
  \begin{align}
    \mathcal{M}^{(6)}_{++} & = -4\mathcal{M}^{(5)}_{++} = 4M_W[12]^2\langle35\rangle[46]\,, \\
    \mathcal{M}^{(6)}_{--} & = -4\mathcal{M}^{(5)}_{--} = 4M_W\langle12\rangle^2\langle35\rangle[46] \,, \\
    \mathcal{M}^{(6)}_{+-} & = \mathcal{M}^{(6)}_{+-} = 0\,.
  \end{align}
\end{subequations}
Note that the subamplitudes corresponding to the (CP-even) dimension-6 operator
$\mathcal{O}_{GH}\equiv\left(H^\dagger H\right) G_{\mu\nu}^a G^{a,\mu\nu}$  have the same structure as
$\mathcal{M}_{\lambda_1\lambda_2}^{(6)}$. In fact, the latter
corresponds to an interaction mediated by the exchange of a very heavy
scalar boson coupling to a pair of gluons in a gauge-invariant
fashion. 

All helicity amplitudes have been implemented in a new version of MCFM-RE~\cite{mcfm_re_d8} and were cross checked with those
obtained automatically by feeding the appropriate
UFO~\cite{Darme:2023jdn} file to MadGraph~\cite{Alwall:2014hca} with
both the dimension-8 squared amplitude and with the interference with
the CP-even dimension-6 operator. Also, note that to simplify the
Levi-Civita symbols appearing in the helicity amplitudes for operators
$3$ and $4$, the relation~\eqref{eq:spinor_epsiilon_id} in
appendix~\ref{sec:AppendixO3O4} was used.

\subsection{Validity of the EFT formulation}
\label{sec:EFTValidity}

We study BSM effects induced by adding to the SM Lagrangian an effective
interaction Lagrangian that incorporates the effect of both dimension-6 and
dimension-8 operators:
\begin{equation}
  \label{eq:effective-Lkappa}
  \mathcal{L}\supset \frac{h}{v}\left[ -\delta\kappa_t m_t\bar t t+\kappa_g\frac{\alpha_s}{12\pi}G_{\mu\nu}^a G^{a,\mu\nu}+i\tilde \kappa_t m_t \bar t \gamma^5 t +\tilde \kappa_g\frac{\alpha_s}{8\pi}G_{\mu\nu}^a \tilde G^{a,\mu\nu}\right] +\sum_i \frac{c^{(GW)}_i}{\Lambda^4} \mathcal{O}_i \,,
\end{equation}
where the terms $\kappa_g$
and $\tilde\kappa_g$ encode the effects of the CP-even and CP-odd
dimension-6 operators which couple gluons to the Higgs. Introducing
the usual left-handed fermion doublet $T_L=(t_L,b_L)^T$ as well as $\tilde H = i\sigma_2 H^*$, the above
equation can be recast in terms of a SMEFT expansion as:
 \begin{multline}
  \label{eq:effective-L}
  \mathcal{L}\supset
  \frac{H^\dagger H}{\Lambda^2} \left[c_t \left(\bar T_L \tilde H t_R + \mathrm{h.c.}\right) +c^{(GH)}G_{\mu\nu}^a G^{a,\mu\nu} + \tilde c^{(GH)}G_{\mu\nu}^a \tilde G^{a,\mu\nu}\right] \\+\frac{c_H}{2\Lambda^2}\partial_\mu \left(H^\dagger H\right)\partial^\mu \left(H^\dagger H\right) +\sum_i \frac{c^{(GW)}_i}{\Lambda^4} \mathcal{O}_i \,,
\end{multline}
where we have introduced the scale of new physics $\Lambda$. By
comparing eqs.~\eqref{eq:effective-Lkappa} and~\eqref{eq:effective-L},
we can perform the identifications
 \begin{equation}
   \label{eq:kappaconversion}
   \begin{split}
   \delta \kappa_t = -\frac{v^2}{\Lambda^2}\left(\mathrm{Re}(c_t)+\frac{c_H}{2}\right)\,,\quad
   \tilde\kappa_t = \frac{v^2}{\Lambda^2}\mathrm{Im}(c_t)\,,\quad
   \kappa_g  = \frac{12\pi v^2 c^{(Gh)}_i}{\alpha_s\Lambda^2}\,,\quad \tilde\kappa_g = \frac{8\pi v^2 \tilde c^{(Gh)}_i}{\alpha_s\Lambda^2}\,.
   \end{split}
  \end{equation}
  For each of the dimension-6 operators
  in~\eqref{eq:effective-Lkappa}, a set of Feynman rules can be
  generated which create a $tth$ contact interaction for the
  $\kappa_t$ and $\tilde\kappa_t$ terms and a $ggh$ contact
  interaction for the $\kappa_g$ and $\tilde\kappa_g$ terms. Their
  contributions to physical amplitudes, denoted by
  $\mathcal{M}^{(gg)}_t$, $\mathcal{\tilde M}^{(gg)}_t$,
  $\mathcal{M}^{(gg)}_g$, have been extensively studied in the past
  ~\cite{ATLAS:2020ior, ATLAS:2021pkb, ATLAS:2023cbt,
    Arpino:2019fmo}. They are also implemented in
  the public code MCFM-RE~\cite{Arpino:2019fmo}.

\begin{figure}[htbp]
  \includegraphics[width=.8\textwidth]{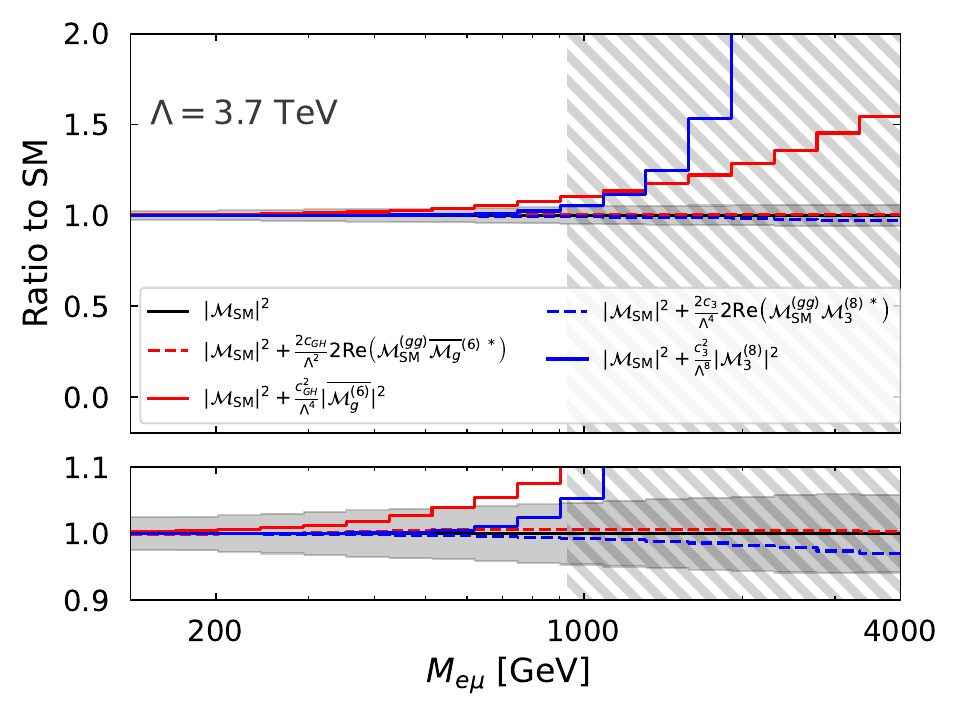}
  \centering
  \caption{Demonstration of breakdown of EFT regime. The contributions
    of the largest leading order (in the EFT expansion) operator's
    $(\mathcal{O}_{GH})$ squared contribution (red) and the largest
    next-to-leading operator's (dimension-8 operator 3) squared
    contribution (blue) are compared to the SM contribution
    (black). It can be seen at lower energies the EFT assumptions hold
    with the leading order term dominating. At energies of
    $\sim 1\,$TeV the next-to-leading order term is no longer
    negligible and at higher energies dominates over the leading order
    term. This signals the breakdown of the EFT regime. Using
    eq.~\eqref{eq:maxbin} this breakdown energy can be estimated and
    the dimension-6 operator can be constrained consistently. Note
    that we define $\overline{\mathcal{M}}^{(6)}_g$ such that
    $\frac{c_g}{\Lambda^2}\overline{\mathcal{M}}^{(6)}_g = \kappa_g
    \mathcal{M}^{(6)}_g$.}
  \label{fig:HLLHCdimsixeightcomp}
\end{figure}

  In this work, we want to assess to what extent it is possible to
  constrain dimension-8 operators from existing and future $WW$ data.
  Before doing this, it is important to explore how the ability to
  constrain the EFT amplitudes considered above is affected by the
  requirement of EFT validity. In order to establish the order of the
  effect of each operator within a systematic EFT expansion, we
  separate the various contributions to the amplitude
  $\mathcal{M}^{(gg)}$ for the $gg$ channel as follows:
\begin{equation}
  \label{eq:Mgg}
  \mathcal{M}^{(gg)}  = \mathcal{M}^{(gg)}_{\rm SM} + \delta \kappa_t \mathcal{M}^{(6,\,gg)}_t + \kappa_g  \mathcal{M}^{(6,\,gg)}_g + \tilde \kappa_t \tilde{\mathcal{M}}^{(6,\,gg)}_t + \tilde \kappa_g  \tilde{\mathcal{M}}^{(6,\,gg)}_g + \sum_i \frac{c^{(GW)}_i}{\Lambda^4} \mathcal{M}^{(8,\,gg)}_i\,.
\end{equation}
When we square the above amplitude, we obtain a second order
polynomial in all the BSM couplings:
\begin{equation}
  \label{eq:Mgg-squared}
  \begin{split}
    |\mathcal{M}^{(gg)}|^2  & = |\mathcal{M}^{(gg)}_{\rm SM}|^2\\ & +\underbrace{\delta \kappa_t\, 2\mathrm{Re}\left(\mathcal{M}^{(6,\,gg)}_t (\mathcal{M}^{(gg)}_{\rm SM})^*\right)+\kappa_g 2\mathrm{Re}\left(\mathcal{M}^{(6,\,gg)}_g (\mathcal{M}^{(gg)}_{\rm SM})^*\right)}_{\sim 1/\Lambda^2} \\
    & +\underbrace{ \left|\delta \kappa_t \mathcal{M}^{(6,\,gg)}_t +
        \kappa_g \mathcal{M}^{(6,\,gg)}_g\right|^2+\left|\tilde\kappa_t
        \tilde{\mathcal{M}}^{(6,\,gg)}_t + \tilde\kappa_g \tilde{
          \mathcal{M}}^{(6,\,gg)}_g\right|^2}_{\sim
      1/\Lambda^4} \\
     & +\underbrace{\sum_i \frac{c^{(GW)}_i}{\Lambda^4} 2\mathrm{Re}\left(
        \mathcal{M}^{(8,\,gg)}_i (\mathcal{M}^{(gg)}_{\rm
          SM})^*\right)}_{\sim
      1/\Lambda^4}\\
     & +\mathcal{O}\left(\frac{1}{\Lambda^6}\right)\,.
\end{split}
\end{equation}
What values of $\Lambda$ can be reasonably and consistently probed by
looking at physical distributions in $WW$ production?  We know that
the presence of higher-dimensional contributions to $WW$ production
results in deviations from SM expectations. These occur most
prominently in the distribution in $M_{WW}$, the invariant mass of the
$WW$ pair. However, this quantity cannot be measured when $W$ bosons
decay fully leptonically due to the presence of invisible
neutrinos. There are various observables that could act as proxies for
$M_{WW}$. One that is widely used is $M_{e\mu}$, the invariant mass of
the electron and muon. If we assume that $M_{e\mu}\simeq M_{WW}/2$,
and the EFT expansion parameter for amplitudes is
$c_i M^2_{WW}/\Lambda^2$. Imposing that this expansion parameter is
less than one gives us the possibility to probe
values of $\Lambda$ above:
  \begin{equation}
    \label{eq:maxbin}
    \Lambda_{\min}  = 2 \sqrt{c_i} M_{e\mu}\sim 2 M_{e\mu}.
   \end{equation}
   To demonstrate the need for this cut-off, we present predictions
   for the SM and the BSM predictions for both the dimension-6 squared
   and dimension-8 squared contributions to the $M_{e\mu}$
   distribution at $\Lambda=3.7\,$TeV in
   figure~\ref{fig:HLLHCdimsixeightcomp}. These predictions are
   obtained with the experimental cuts and parameter setup described
   in section~\ref{sec:SM} for $\sqrt s = 14\,$TeV, but the actual
   details of the calculation are not relevant for the moment. We also
   include in figure~\ref{fig:HLLHCdimsixeightcomp} the contribution
   from the dimension-6 and dimension-8 interference with the SM. We
   observe that they are both much smaller than the dimension-6
   squared contribution even though they are formally lower order and
   of the same order in the $1/\Lambda$ expansion respectively. This
   is due to the fact that the SM $gg$ contribution is
   loop-induced. As a consequence it decreases with increasing energy.
   The pure EFT terms are instead contact interactions and therefore
   do not suffer this suppression. This is discussed extensively in
   section~\ref{sec:GG}. This feature is process specific and cannot be naively extrapolated to other processes. 
   Also, the size of the interference of EFT contributions with the SM
   depends crucially on the overlap of the EFT amplitudes with the SM
   amplitude. Therefore, in order to probe the hierarchy of
   higher-dimensional operators, we find it more robust to use squared
   EFT amplitudes.

   The dimension-6 operator could be constrained very well from its
   squared amplitude using the high energy bins, since its
   contribution deviates significantly away from the SM
   prediction. However, the dimension-8 squared contribution is much
   larger in bins $M_{e\mu} > 1\,$TeV. This signals the breakdown of
   the EFT at around $M_{e\mu} \sim 1\,$TeV as expected from
   $\Lambda = 3.7\,$TeV. However by considering only the region where
   the dimension-8 term is negligible (the unshaded area in
   figure~\ref{fig:HLLHCdimsixeightcomp}), the dimension-6 term can
   still be safely excluded at this value of $\Lambda=3.7\,$TeV.

We can take advantage of these numerical predictions to test the
condition in eq.~\eqref{eq:maxbin} (taking $c_i = 1$), which relies on the assumption
that $M_{e\mu}\simeq M_{WW}/2$. To this end, we use an empirical
approach by finding the value of $\Lambda$ such that the largest
dimension-8 squared amplitude is no more than half of the dimension-6
squared amplitude. Comparing the dimension-6 squared amplitude with
the dimension-6-dimension-8 interference piece would give the same
result, but only in the case of perfect interference between
dimension-6 and dimension-8. For this reason, we use the higher order
dimension-8 squared piece. If we evaluate the contribution to each bin
of the largest dimension-6 operator ($\mathcal{O}_{GH}$) and of
dimension-8 operator 3 (which we have found to be the largest
dimension-8 operator), we can find a value of $\Lambda$ corresponding
to the above condition as:

 \begin{equation}
    \label{eq:empirical_lambdamin_derivation}
    \frac{(2\,\mathrm{TeV})^8\ \sigma^{(8)}_{3,\, \Lambda=2\,\mathrm{TeV}}}{\Lambda_{\min} ^8} = \frac{1}{2}\frac{(2\,\mathrm{TeV})^4\ \sigma^{(6)}_{g,\, \Lambda=2\,\mathrm{TeV}}}{\Lambda_{\min}^4},
   \end{equation}
   
   where $\sigma^{(8)}_{i,\, \Lambda=2\,\mathrm{TeV}}$ is the contribution
   to the given bin arising from the dimension-8 squared amplitude and has
   $\sigma^{(8)}_{i}\propto |\mathcal{M}^{(8)}_i|^2$. This gives a minimum value
   of $\Lambda$ for this bin:\footnote{Note that the value of
     $\Lambda_{\min}$ found via this method is independent of the mass
     scale chosen to evaluate the cross sections. However, finding the
     cross sections implicitly involves choosing some mass scale for
     the EFT (we choose $\Lambda=2\,$TeV.)}

 \begin{equation}
    \label{eq:empirical_lambdamin}
    \Lambda_{\min} = (2\,\mathrm{TeV})\left(2\times\frac{\sigma^{(8)}_{3,\, \Lambda=2\,\mathrm{TeV}}}{\sigma^{(6)}_{g,\, \Lambda=2\,\mathrm{TeV}}}\right)^\frac{1}{4}.
   \end{equation}

\begin{figure}[htbp]
  \includegraphics[width=.63\textwidth]{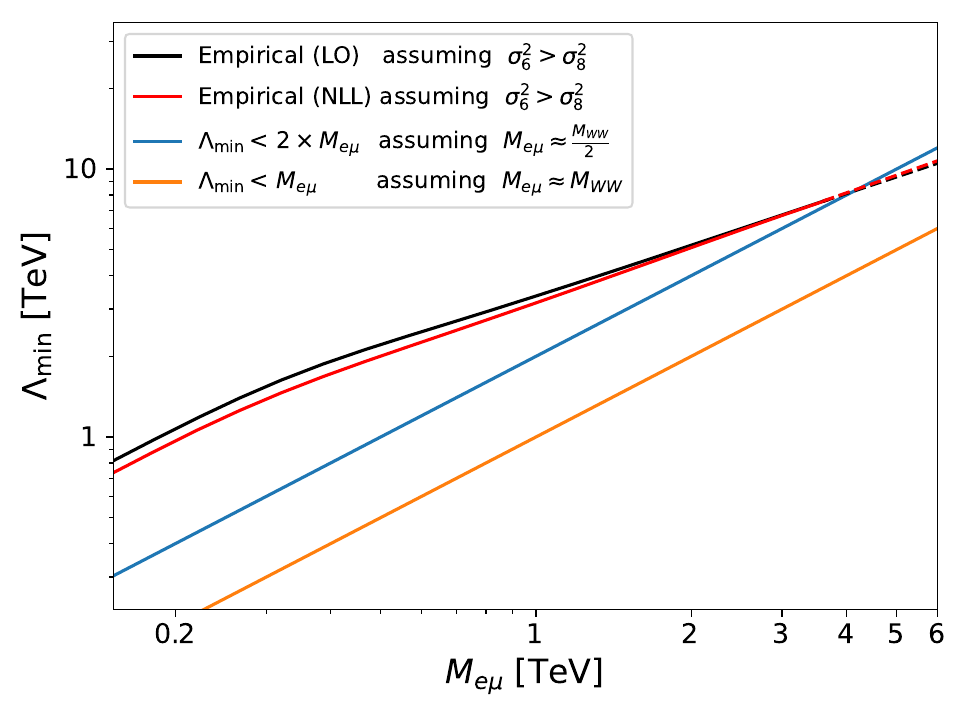}
  \centering
  \caption{Minimum value of $\Lambda$ that should be used for each bin in
    $M_{e\mu}$ under various assumptions. The blue and orange lines assume
    a linear relation between $M_{e\mu}$ and $M_{WW}$, namely 
    $M_{e\mu}= M_{WW}/2$ and $M_{e\mu}= M_{WW}$
    respectively. The other curves correspond to the empirical approach of eq.~\eqref{eq:empirical_lambdamin}, which compares the size of
    dimension-6 and the largest dimension-8 operator directly, ensuring
    $\Lambda$ is big enough to keep the hierarchy in the EFT
    expansion. The plotted values of $\Lambda_{\min}$ are determined using the resummed (red) and fixed order
    (black) predictions.}
  \label{fig:min_lambda_with_mll}

\end{figure}

\begin{figure}[htbp]
  \includegraphics[width=.63\textwidth]{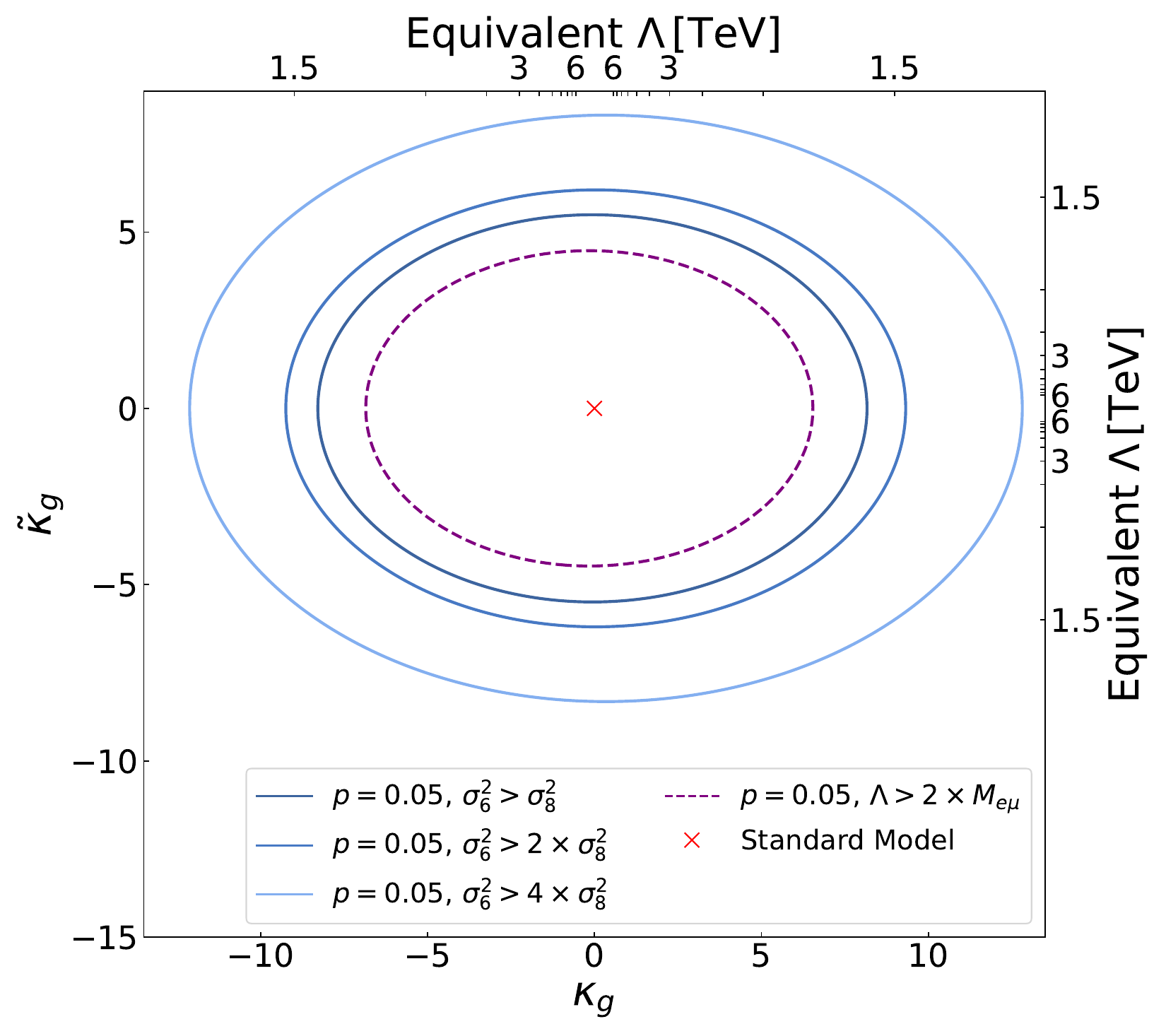}
  \centering
  \caption{Comparison of the contour plots resulting from the various
    assumptions on EFT validity (demonstrated in
    figure~\ref{fig:min_lambda_with_mll}) using ATLAS data.}
  \label{fig:contour_comp}
\end{figure}

   We then compared the minimum value of $\Lambda$ found with
  eq.~\eqref{eq:empirical_lambdamin} to the value obtained using the
  method of eq.~\eqref{eq:maxbin} by assuming $M_{e\mu}= M_{WW}/2$ and also under the assumption
  $M_{e\mu} = M_{WW}$. This is shown in
  figure~\ref{fig:min_lambda_with_mll}. We found that, at lower
  energies, the assumption $M_{e\mu}= M_{WW}/2$ does not hold. This is
  due to the fact the cross section grows with energy, leading to
  higher energy $M_{WW}$ bins having an outsized effect on lower
  energy $M_{e\mu}$ bins. This means that, at low energies, one cannot
  assume a simple relation between the two. Furthermore, close to the
  kinematical boundary $M_{WW}\lesssim 14\,$TeV, events with
  high values of $M_{e\mu}$ take larger and larger fractions of the
  di-boson energy. For this reason, in the following we adopted the
  value $\Lambda_{\min}$ derived from
  eq.~\eqref{eq:empirical_lambdamin_derivation}, which captures the
  best of both behaviours. We show in figure~\ref{fig:contour_comp} that, depending on which
  assumption one takes, a variety of different constraints can be found,
  in turn depending on how conservative you would like to be with the
  empirical approach. To create this demonstrative contour plot, current ATLAS data is used to fit the CP-even and CP-odd version of the dimension-6 operator
$\mathcal{O}_{GH}$. It can be seen that the contour plot shows large dependence on the assumption taken. The naive assumption that $M_{e\mu}= M_{WW}/2$ results in a very strong constraint. This motivates better profiling of the size of
  EFT errors which we leave to future work. For the rest of the plots
  in this paper we adopt the constraint
  $\sigma^{(6)}_{g} > 2\times \sigma^{(8)}_{3}$.
\newpage


\newpage
\section{Numerical Predictions}
\label{sec:Numerics}

In this section we study the SM and BSM predictions for $WW$
production at the LHC. We first develop the best SM prediction in
order to demonstrate how EW corrections and jet-veto resummation
affect the SM prediction, which has obvious consequences for how large
new physics effects needs to be in order to be visible in this
channel. We then present results for the dimension-8 EFT operators
previously considered and compare them both to the dimension-6
operators and to the best SM prediction. We also demonstrate the
effect of jet-veto resummation on the BSM contributions which have
large effects the size of new physics contributions.

We present results with a centre-of-mass energy $\sqrt s=14\,$TeV,
with jets reconstructed according to the anti-$k_t$
algorithm~\cite{Cacciari:2008gp} with a jet radius $R = 0.4$. In order
to eliminate contamination from $ZZ$ production, we consider only
events with an electron and a muon. Also, we do not consider decays
into $\tau$ leptons. We adopt the fiducial cuts on leptons and jets
detailed in table~\ref{tab:fiducial-cuts}. These are the cuts of the
experimental analysis performed by the ATLAS collaboration
in~\cite{ATLAS:2019rob}, which we assume to also be similar for
studies of this channel at the HL-LHC.
\begin{table}[!htbp]
\begin{center}
\begin{tabular}{c|c}
  \hline\hline
  Fiducial selection requirement  & Cut value \\
  \hline\hline
  $p_T^{\ell}$ & $>27\,\mathrm{GeV}$ \\
  $|y_{\ell}|$ & $<2.5$ \\
  $M_{e\mu}$ & $>55\,\mathrm{GeV}$ \\
  $|\vec{p}_T^{\ e}+\vec{p}_T^{\ \mu}|$ & $>30\,\mathrm{GeV}$ \\
  Number of jets with $p_T> 35\,\mathrm{GeV}$ & 0 \\
  $\ETmiss$ & $>20\,\mathrm{GeV}$ \\
  \hline\hline
\end{tabular}
 \end{center}
 \caption{
   Definition of the $WW\rightarrow e\mu$ fiducial phase space, where
   $\vec{p}_T^{\ \ell},y_\ell$ are the transverse momentum and rapidity of either an
   electron or a muon, $M_{e\mu}$ is the invariant mass of the electron-muon
   pair, and $\ETmiss$ is the missing transverse energy.
 }
   \label{tab:fiducial-cuts}
\end{table}

For the following results we set electroweak constants using the $G_\mu$ scheme. We use input parameters as given in table~\ref{tab:input-params}.

\begin{table}[!htbp]
\begin{center}
\begin{tabular}{c|c}
  \hline\hline
   Input Parameter & Value \\
  \hline\hline
  $G_\mu$ & $1.16637\times 10^{-5}\,\mathrm{GeV}^{-2}$ \\
  $M_W$ & $80.385\,\mathrm{GeV}$ \\
  $M_Z$ & $91.1876\,\mathrm{GeV}$ \\
  $m_t$ & $173\,\mathrm{GeV}$ \\
  $m_b$ & $4.66\,\mathrm{GeV}$ \\
  $M_H$ & $125\,\mathrm{GeV}$ \\
  $\Gamma_W$ & $2.093\,\mathrm{GeV}$ \\
  $\Gamma_Z$ & $2.4952\,\mathrm{GeV}$ \\
  $\Gamma_t$ & $1.4777\,\mathrm{GeV}$ \\
  $\Gamma_H$ & $4.07\times 10^{-3}\,\mathrm{GeV}$ \\
  \hline\hline
\end{tabular}
 \end{center}
 \caption{Input parameters used for the numerical results presented below.}
   \label{tab:input-params}
\end{table}

\subsection{SM $q\bar q$ + EW Predictions}
\label{sec:SM}
Fixed order precision predictions for $WW$ production have existed for some
time. The current QCD state-of-the-art is NNLO accuracy for the
$q\bar q$-initiated
contribution~\cite{Gehrmann:2014fva,Grazzini:2016ctr} and approximate
NLO for the $gg$-initiated
contribution~\cite{Grazzini:2020stb}. Electroweak (EW) corrections
have also been computed at NLO accuracy~\cite{Grazzini:2019jkl}. Such
accuracy might however not be enough to accurately describe the cross
sections we are interested in. In fact, since the cuts in
table~\ref{tab:fiducial-cuts} involve a tight veto on accompanying
jets, we expect large logarithms of the ratio of veto threshold
$p_{T, \mathrm{veto}}$ (in our case $35\,$GeV) and the invariant mass of
the $WW$ pair $M_{WW}$ to appear at all orders in perturbation
theory. These logarithms give rise to a double-logarithmic Sudakov
form factor $\sim\exp[-\alpha_s \ln^2(p_{T, \mathrm{veto}}/M_{WW})]$ which
suppresses the $WW$ cross section as $M_{WW}$ increases. Such effects
generally spoil the convergence of fixed-order calculations, and are
best taken into account through resummed calculations that account for
large logarithms at all orders in QCD perturbation theory. The state
of the art of logarithmic resummations for jet-processes is the
so-called next-to-next-to-leading logarithmic (NNLL) accuracy,
accounting for all terms up to
$\alpha_s^n \ln^{n-1}(p_{T, \mathrm{veto}}/M_{WW})$ in the \emph{logarithm}
of $d\sigma/dM_{WW}$. This accuracy can be upgraded to NNLL$^\prime$
by including exactly all constant terms at relative order
$\alpha_s^2$, which are formally N$^3$LL if one performs a strict
logarithmic counting. For the $q\bar q$ contribution, NNLL resummation
is implemented in the program
MCFM-RE~\cite{Arpino:2019fmo}. NNLL$^\prime$ accuracy can be achieved
automatically when performing the matching with exact NNLO using a
multiplicative matching scheme. In this work, we choose to use the
multiplicative scheme presented in~\cite{Kallweit:2020gva}, as
implemented in the program MATRIX+RadISH. Last, NNLL$^\prime$+NNLO
(which is equivalent to NNLL+NNLO) accuracy is embedded in existing
SCET resummations as implemented in MCFM 10~\cite{Campbell:2023cha}
and in GENEVA~\cite{Gavardi:2023aco}. We also cross-checked matched
NNLL+NNLO results to those obtained with GENEVA. Resummation for the
$gg$ contribution is only implemented at NLL accuracy in MCFM-RE,
because the NLO corrections are only approximately known. We also
consider EW corrections at NLO, as obtained from
MATRIX+OpenLoops~\cite{Grazzini:2019jkl}. This also gives the NLO
photon induced contribution arising from $\gamma\gamma\to WW$. To
augment NNLL+NNLO QCD predictions with the NLO EW corrections
we adopt the prescription given in~\cite{Grazzini:2019jkl} where the
NNLO QCD correction is replaced by the resummed and matched QCD
correction, as follows
\begin{equation}
	\label{eq:sm_best_res}
	\mathrm{d}\sigma_{\mathrm{NNLL+NNLO\,\,QCD}\times \mathrm{EW}_{q\bar{q}}}=\mathrm{d}\sigma^{q\bar{q}}_{\mathrm{NNLL+NNLO\,\,QCD}}\big(1+\delta^{q\bar{q}}_{\mathrm{EW}}\big)+\mathrm{d}\sigma^{\gamma\gamma}_{\mathrm{NLO}}+\mathrm{d}\sigma^{gg}_{\mathrm{NLL}},
\end{equation}
where $\delta^{q\bar{q}}_{\mathrm{EW}}$ are the NLO EW corrections to
the LO quark induced process. This combination scheme is one such scheme that
could be employed to augment the QCD predictions. One method to estimate
the size of the missing QCD-EW $(\alpha_s\alpha)$ terms is to take the difference between the 
additive and multiplicative schemes presented in~\cite{Grazzini:2019jkl}. This difference
gives an estimate for the size of the cross-terms which can then be added as an additional 
source of theoretical uncertainty. This comparison has been performed for exactly this process
in~\cite{Banerjee:2024eyo}, where an alternative exponentiated scheme was also implemented. The effect of the
scheme change was found to be small (within QCD scale uncertainties) up to $M_{e\mu}\sim 1\,\mathrm{TeV}$.
Given that there is currently no consensus on the best way to estimate EW missing higher order
uncertainties, and that the `best' prediction would also include the resummation of EW Sudakov logarithms~\cite{Denner:2024yut},
we consider further investigation of this uncertainty to lie beyond the scope of this work.

In all predictions, care must be exercised in handling the interference with top production. We neglect
it in the present study by utilising a four-flavour scheme for parton
distribution functions, the
NNPDF31$\_$nnlo$\_$as$\_$0118$\_$luxqed$\_$nf$\_$4 PDF
set~\cite{Bertone:2017bme}.
\newpage
\begin{figure}[htbp]
  \includegraphics[width=.5\textwidth]{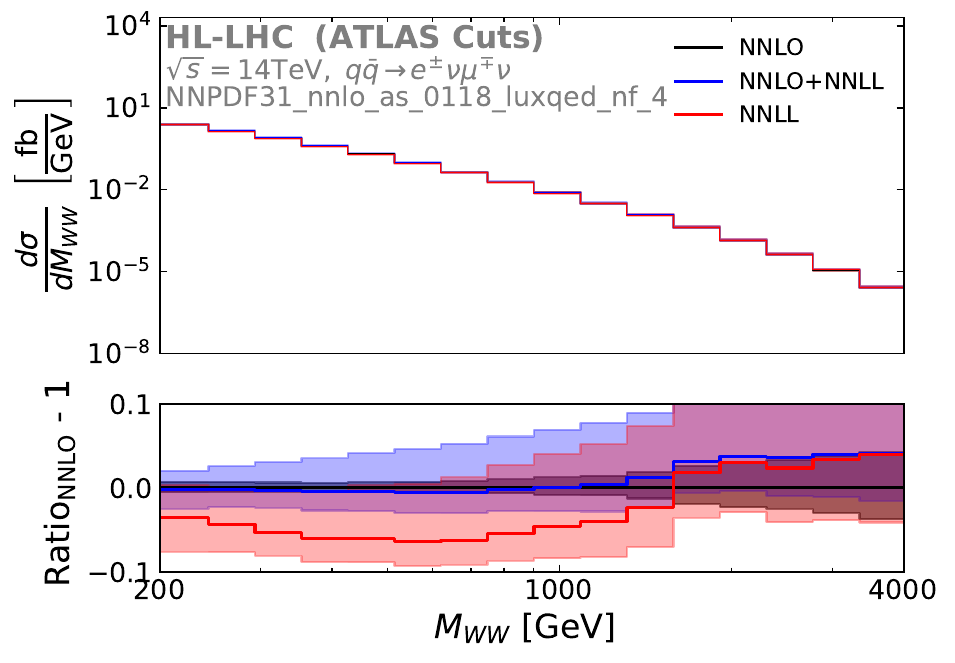}
  \includegraphics[width=.5\textwidth]{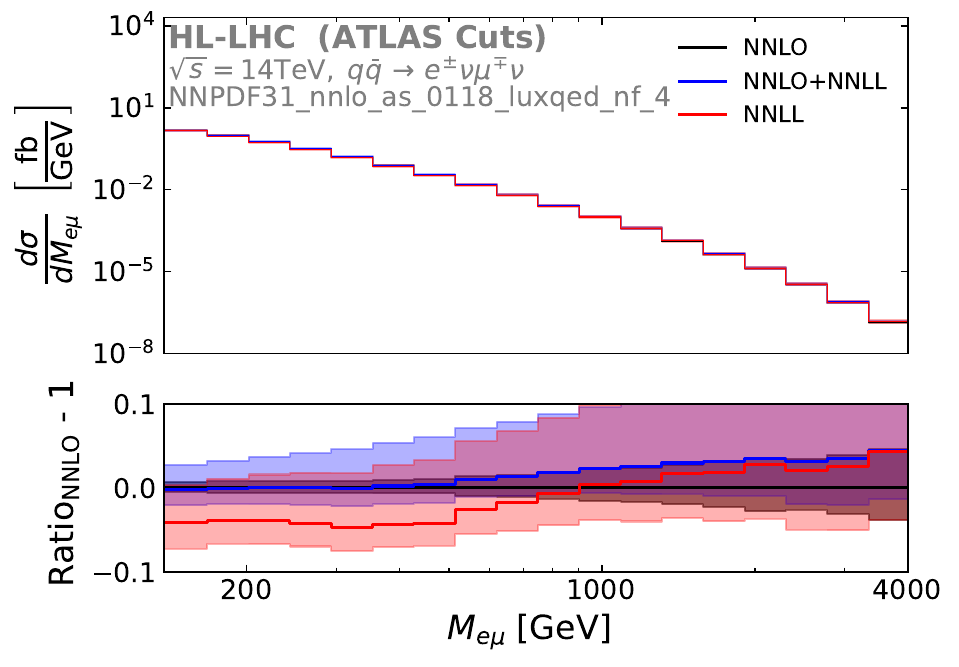}  
  \caption{The distribution in the invariant mass of the $WW$ (left) and lepton pair (right), in various approximations. See text for details.}
  \label{fig:Ressumation_Effects}
\end{figure}

In figure~\ref{fig:Ressumation_Effects} we compare predictions for $d\sigma/dM_{WW}$
(left) and $d\sigma/dM_{e\mu}$ (right) for the fiducial cuts in
table~\ref{tab:fiducial-cuts}, in different approximations, namely
NNLO, pure NNLL, and matched NNLL+NNLO. These predictions do not
include any $gg$ initiated contributions, for which only a NLL
resummation is available, and is implemented only in MCFM-RE. In all
cases, we choose $M_{WW}/2$ as renormalisation scale $\mu_R$ and
factorisation scale $\mu_F$ for the ``central'' predictions for each
approximation. We then estimate theoretical uncertainties for NNLO by
performing 7-point scale variations, i.e.\
$M_{WW}/4\le \mu_{R,F}\le M_{WW}$ with $1/2 \le \mu_R/\mu_F\le 2$. For resummed predictions, we also include variation of the resummation scale $Q$ in the range $[M_{WW}/4,M_{WW}]$ for $\mu_R=\mu_F=M_{WW}/2$.

We observe that NNLL resummed predictions for $M_{WW}$ are, within
errors, compatible with NNLL+NNLO ones.  Pure NNLL predictions miss a
constant term at order $\alpha_s^2$. We observe that the impact of
this missing term is of the order 5\% throughout the whole $M_{WW}$
distribution. This term could be obtained by augmenting the NNLL
resummation to NNLL$^\prime$ accuracy. The situation is similar for
$M_{e\mu}$. Note that, for the distribution in $M_{e\mu}$, the
difference between the central values of NNLL and NNLL+NNLO are below
5\%, so within each other's theoretical uncertainties. We also notice
that NNLO predictions follow NNLL+NNLO, but with smaller
uncertainties, since they correspond to scale variations
only.\footnote{It is known that, in the presence of a jet-veto, scale
  variations tend to underestimate NNLO uncertainties~\cite{Stewart:2011cf}, so they are overly optimistic.}
 \newpage
\begin{figure}[htbp]
  \includegraphics[width=.5\textwidth]{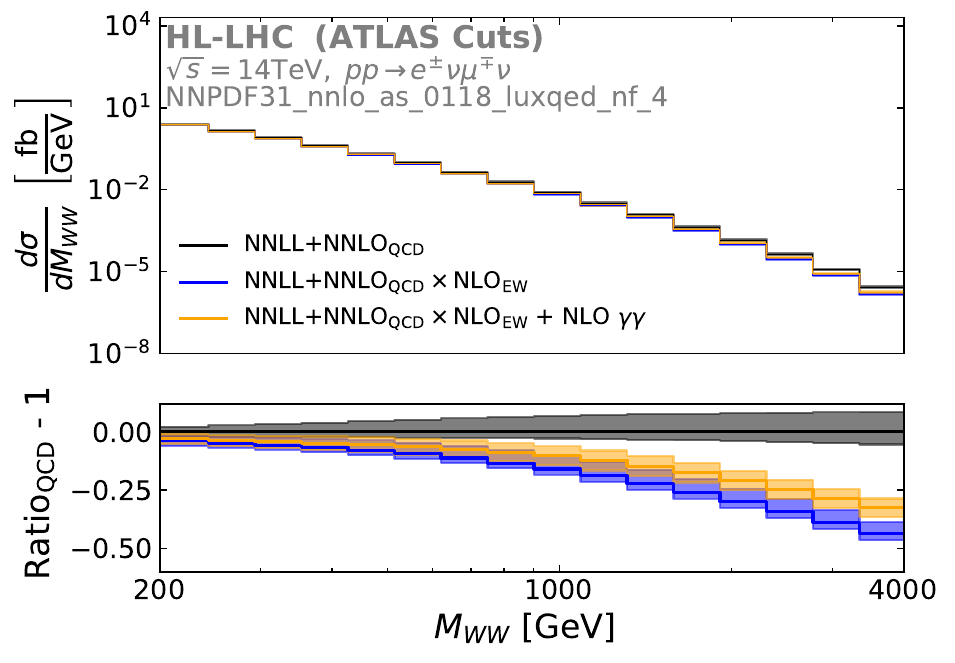}
  \includegraphics[width=.5\textwidth]{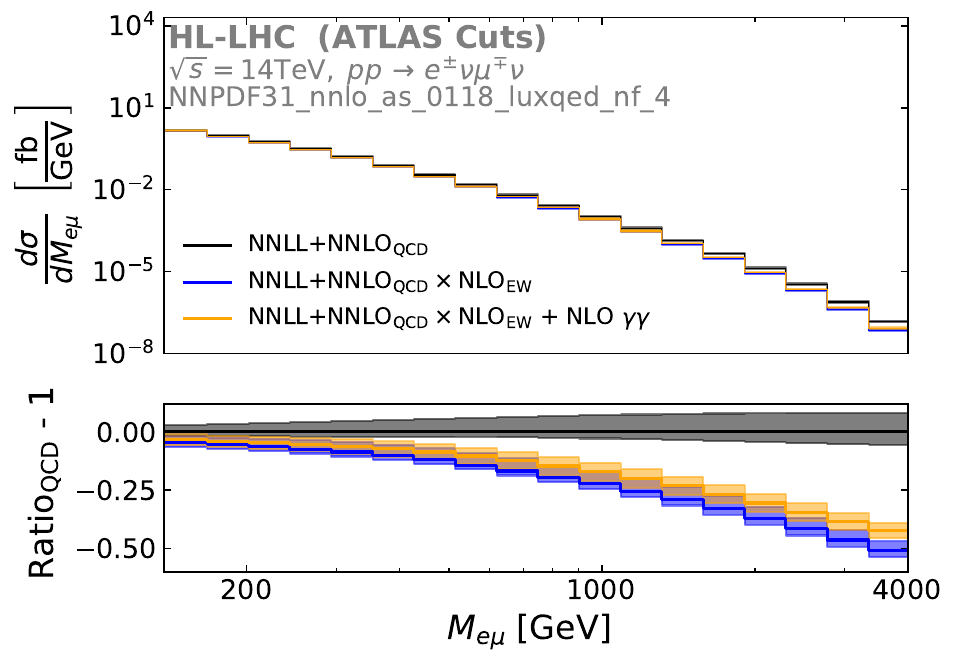}  
  \caption{The distribution in the invariant mass of the $WW$ (left)
    and lepton pair (right), with and without electroweak
    corrections. See text for details.}
  \label{fig:EW_Effects}
\end{figure}

In figure~\ref{fig:EW_Effects}, we demonstrate the impact of the EW
corrections on both the $M_{WW}$ and $M_{e\mu}$ distributions. As
expected, EW corrections result in a reduction of the cross sections,
the size of this reduction growing with increasing $M_{WW}$ and
$M_{e\mu}$. This is due to the presence of Sudakov logarithms arising
from EW virtual corrections. The addition of the $\gamma\gamma$
contribution has a non-negligible effect, and gives an enhancement of
the cross sections up to about 15\%. Note that the Sudakov suppression
does not occur in the $gg$ channel at the considered order. This might contribute to enhancing
the BSM signal we consider over the $q\bar q$ dominated background.
\subsection{SM $gg$ Predictions}
\label{sec:GG}
\begin{figure}[htbp]
\centering
  \includegraphics[width=.49\textwidth]{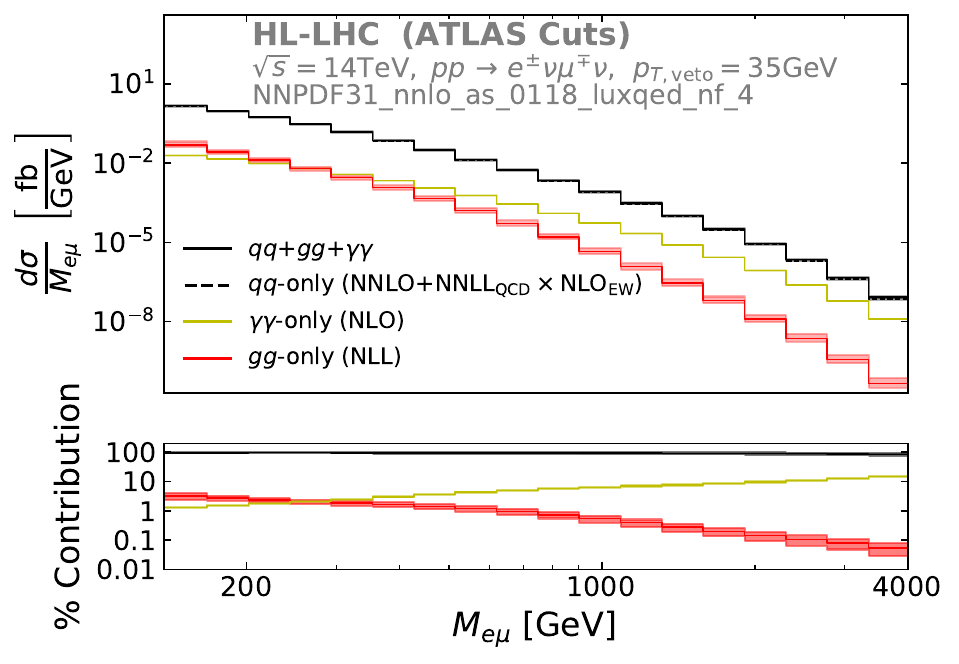}
  \includegraphics[width=.49\textwidth]{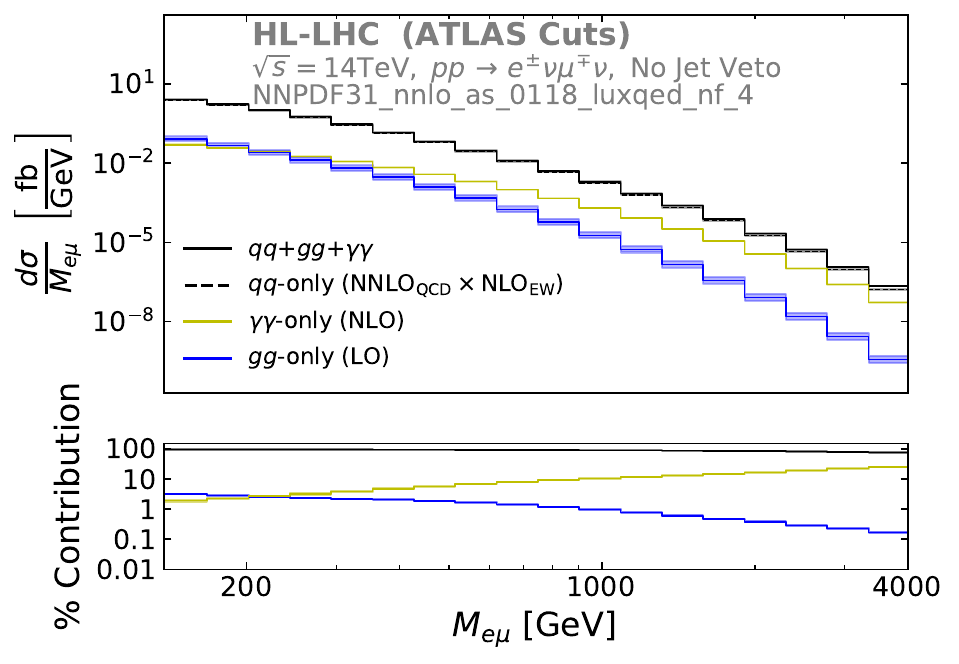}
  \caption{The distribution in the invariant mass of the lepton pair,
    in different approximations. See text for details.}
  \label{fig:gg-SM}
\end{figure}

Here we assess the impact of the SM $gg$ channel in
figure~\ref{fig:gg-SM}, and compare the size of the $gg$ channel both
with and without the presence of the jet-veto given in
table~\ref{tab:fiducial-cuts}. Although this channel is not the
largest contribution to the SM cross section, this is the contribution
that SMEFT operators will interfere with and so its size must be
accurately gauged. In the presence of a strong jet-veto the fully
resummed (NNLL+NNLO / NLL) predictions should be included due to QCD
effects. However, when lifting the jet-veto condition, the fixed order
(NNLO / LO) predictions can be used.
The solid black line corresponds to our best
prediction, which is NNLL+NNLO$_{\rm
  QCD}$+NLO$_{\rm EW}$+NLO$_{\gamma\gamma}$ for the $q\bar q$ channel
and NLL for the $gg$ channel. We see that the $gg$ channel, both LO
and NLL, gives a contribution that is at least two orders of magnitude
smaller than the $q\bar q$ channel. The main reason for this is the fact that it is loop-induced, so not only does it start at order $\alpha_s^2$ but also decreases with energy. Furthermore, the $gg$
luminosity is smaller than the $q\bar q$ one at the considered energy
scales. We note that, since the $gg$ contribution to the SM is so small, it can be considered negligible in the high energy limit.
This implies that the dimension-8 interference term will likely be
undetectable by itself in the EFT regime. This is due to the fact that a large interference term would imply that the squared
term is also detectable, and therefore needs to be included.

We also show how the jet-veto affects the lepton-pair invariant mass distribution. As
expected, the presence of a jet-veto has a bigger impact on the $gg$
channel, due to the fact that gluons have a larger colour factor than
quarks. Notably, in the high-energy tail, the cross section for the
$q\bar q$ channel is reduced by a factor of three, as opposed to an
order of magnitude for the $gg$ channel.

\begin{figure}[htbp]
\centering
  \includegraphics[width=.8\textwidth]{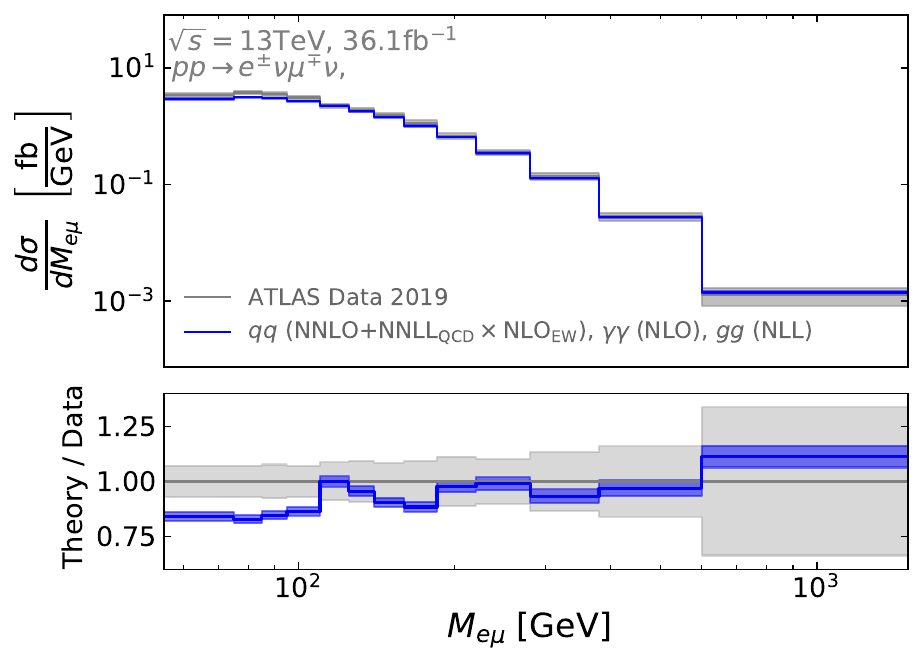}
  \caption{Comparison of our best prediction for the dilepton invariant mass distribution with ATLAS data~\cite{ATLAS:2017bbg}. The band around the experimental data gives the combined statistical and systematic uncertainties quoted by the ATLAS collaboration.}
  \label{fig:AtlasComparison}
\end{figure}

Last, in figure~\ref{fig:AtlasComparison} we show a comparison of our
best prediction with $\sqrt s=13\,$TeV ATLAS data~\cite{ATLAS:2017bbg}. We
observe with $M_{e\mu}>110\,$GeV agreement within experimental uncertainties, slightly worse in
the low-energy bins, a feature already seen in~\cite{ATLAS:2017bbg, Gavardi:2023aco}.
Having established the SM contribution to $WW$ production, we now turn
to the effect of dimension-8 operators in the $gg$ channel.

\subsection{BSM Predictions}
\label{sec:bsm_preds}

\begin{figure}[htbp]
  \includegraphics[width=.5\textwidth]{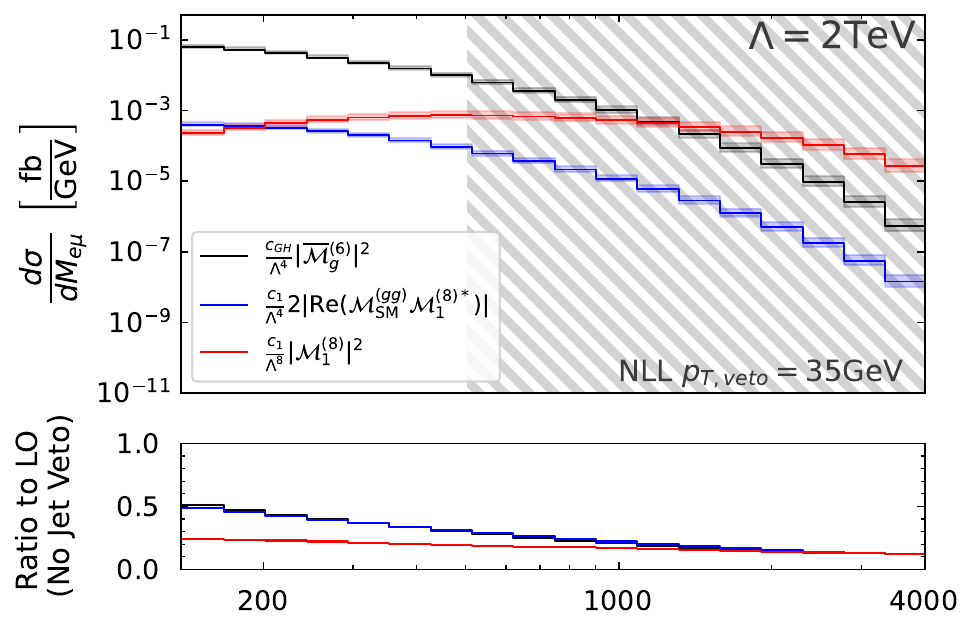}
  \includegraphics[width=.5\textwidth]{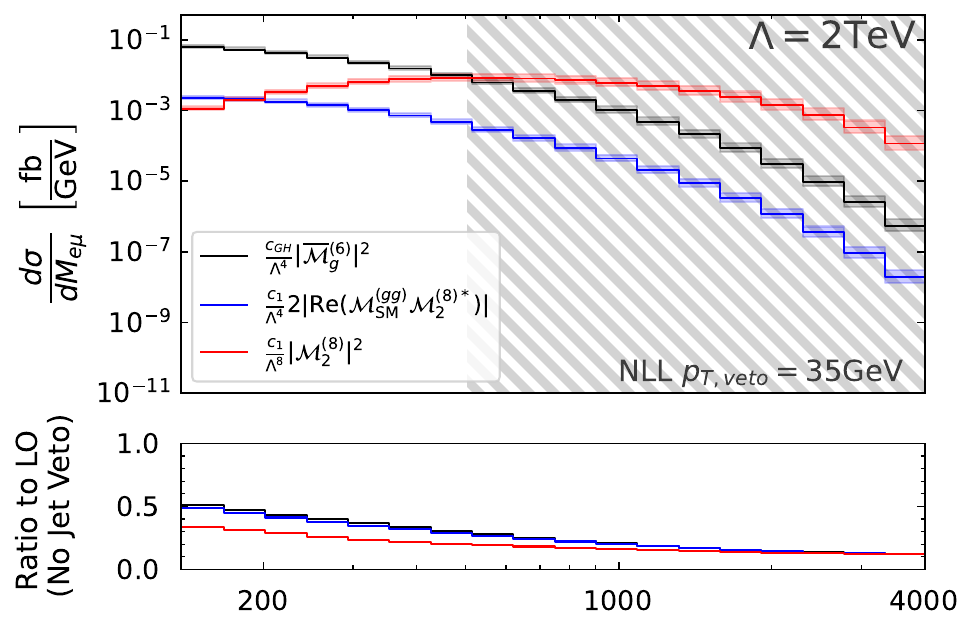}  
  \includegraphics[width=.5\textwidth]{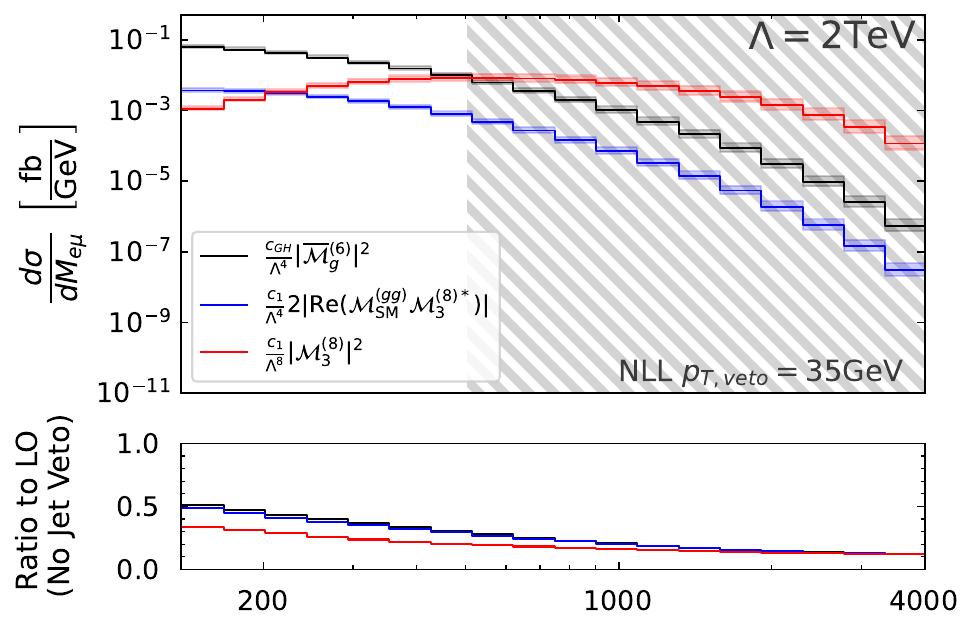}  
  \includegraphics[width=.5\textwidth]{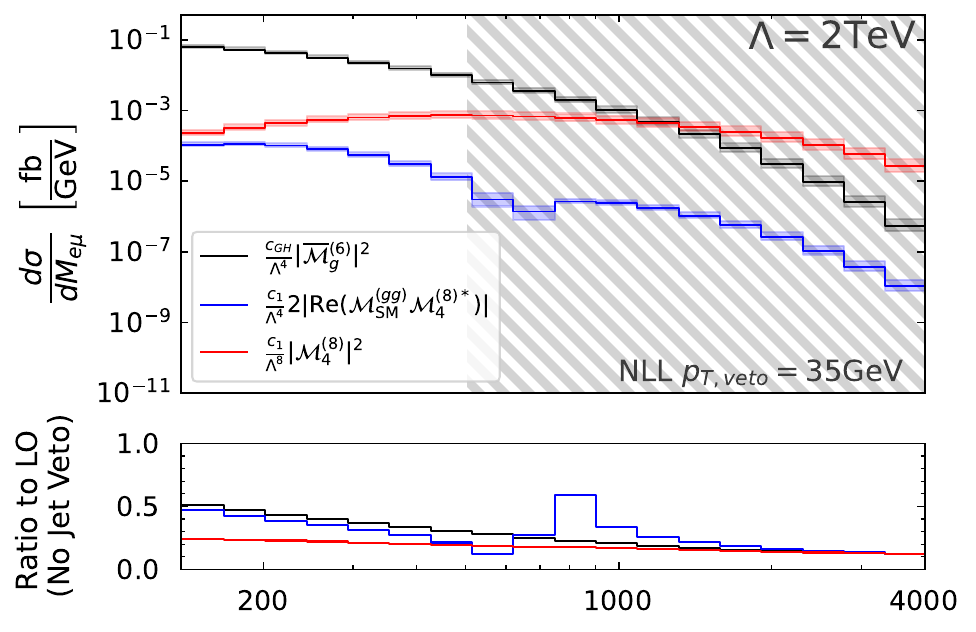}  
  \includegraphics[width=.5\textwidth]{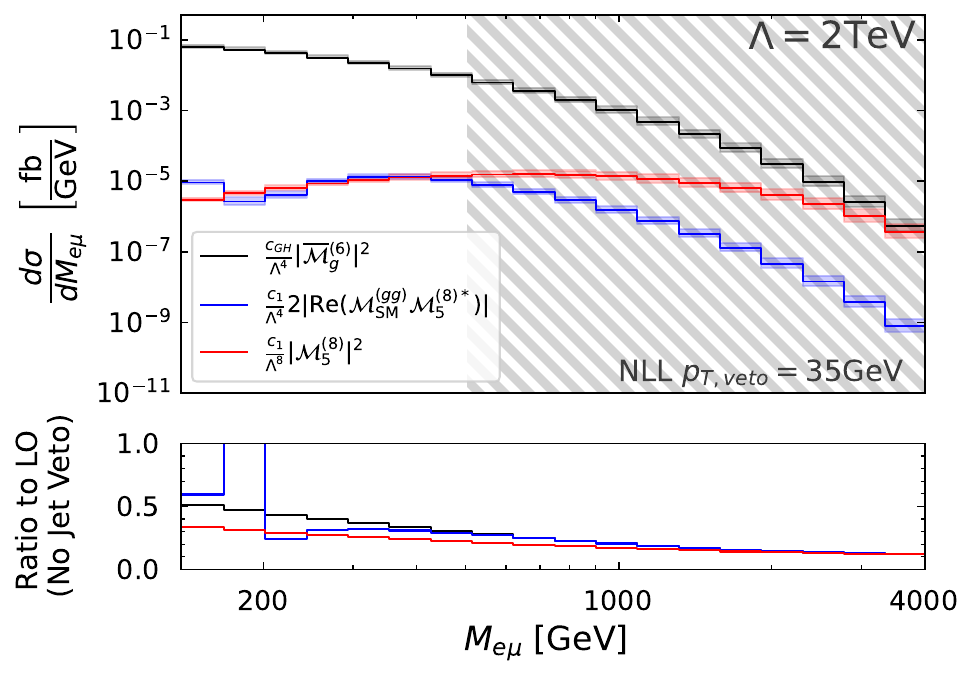}  
  \includegraphics[width=.5\textwidth]{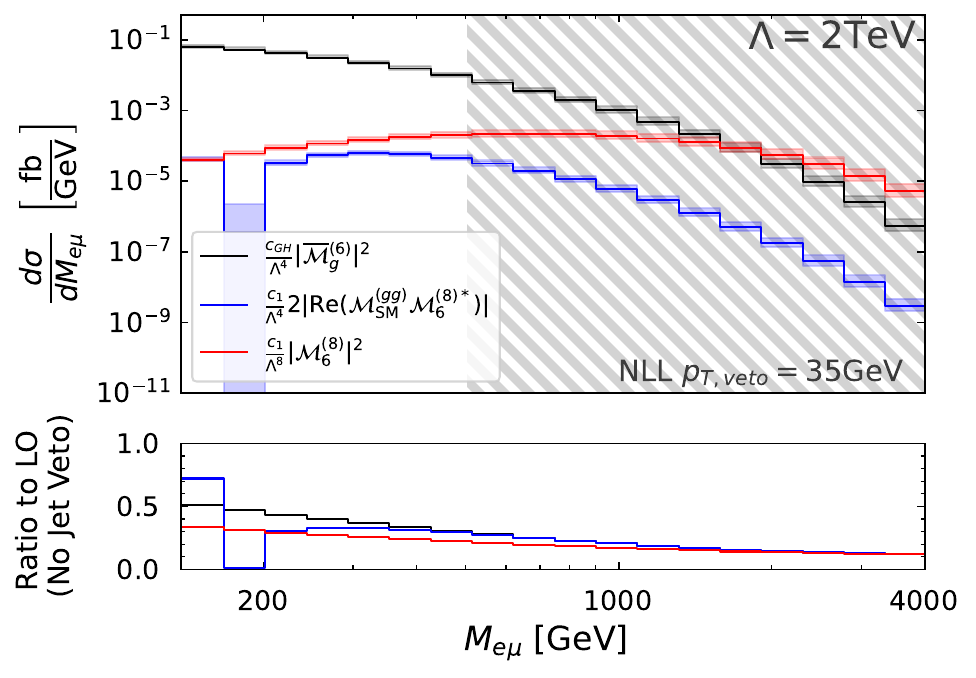}  
  \caption{Comparison of dimension-8 interference (blue) and
    dimension-8 squared (red) operators with the dimension-6 (black)
    operator at EFT mass scale $\Lambda=2\,$TeV. These contributions
    are shown at (NLL) accuracy with a jet-veto resummation, the ratio
    with the leading order contribution is shown in the lower panel of
    each plot. It can be seen that most bins have a NLL contribution
    at least half as big as the fixed order contribution, with
    reductions below $10\%$ in the high energy bins which are relevant
    to constraints.}
  \label{fig:BSMOperatorPrediction}
\end{figure}

  In figure~\ref{fig:BSMOperatorPrediction} we present predictions for
  $M_{e\mu}$ obtained from the helicity amplitudes calculated in
  section~\ref{sec:dim8ops}. We show the leading contribution from the
  EFT expansion which is the interference with the SM as
  well as the corresponding squared dimension-8 contributions. We
  show predictions at a reference value of $\Lambda=2\,$TeV, for the
  ATLAS cuts from table~\ref{tab:fiducial-cuts}, and at an energy
  of $\sqrt s=14\,$TeV.
  
We also include the contribution of the CP-even dimension-6 operator
$\mathcal{O}_{GH}$ for reference. More
precisely, we consider Feynman rules stemming from the SMEFT
Lagrangian in eq.~(\ref{eq:effective-L}), setting individual
coefficients to one and all others to zero. It can be seen that, in
general, at $\Lambda=2\,$TeV the dimension-8 interference
term (labelled $2|\mathrm{Re}(\mathcal{M}^{(gg)}_{\rm SM}\mathcal{M}^{(8)\,*}_i)|$ in the figure, with
$i=1,2, \dots, 6$) is almost always smaller than its dimension-8
squared counterpart ($|\mathcal{M}^{(8)}_i|^2$).\footnote{Note that interference contributions can become negative. Since we want to plot them in logarithmic scale, we have decided to plot their absolute value. These leads to apparent discontinuities in figure~\ref{fig:BSMOperatorPrediction}, see e.g.\ the contribution of operator 4.} As mentioned earlier, this is due to
the SM gluon-fusion amplitude being very small. It can
also be seen that, for each operator, at some value of $M_{e\mu}$, the
contribution of a squared dimension-8 operator becomes non-negligible
relative to the corresponding contribution at dimension-6
($|\mathcal{M}^{(6)}_{g}|^2$). The values of $M_{e\mu}$ at which
this transition happens differ between the six dimension-8
operators. For instance, for operator 3, this occurs at around
$M_{e\mu}\simeq 0.4\,$TeV, whereas for operator 5 this does not occur until after
$M_{e\mu}\simeq  3\,$TeV. This is consistent with figure~3 which shows that using bins up to $M_{e\mu}\simeq 0.4\,$TeV requires $\Lambda_{\min} \simeq 2\,$TeV.

\begin{figure}[htbp]
  \includegraphics[width=.8\textwidth]{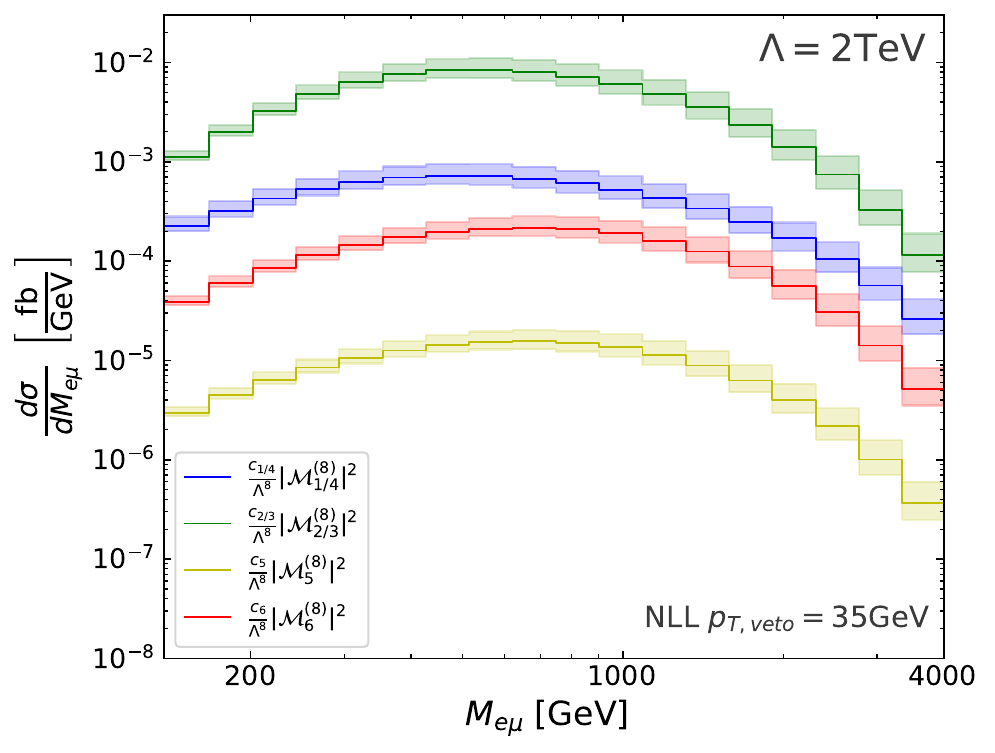}
  \centering
  \caption{Comparison of the size of the contribution to the cross section of the squared amplitude ($|\mathcal{M}^{(8)}_i|^2$) generated by each operator. It can be seen that operators $2$ and $3$ have the same size as operators $1$ and $4$. Operators $6$ is somewhat smaller than operators $1$ and $4$ and operator $5$ is substantially smaller than the other operators.}
  \label{fig:Squaredopscomparison}
\end{figure}

We also wish to stress the effect of the jet-veto condition on
$gg$-mediated contributions especially for the BSM signal. Using the
LO prediction without at least a parton shower, or better a full NLL
resummation, in effect ignores the jet-veto which gives predictions
for the signal up to a factor of $10$ larger. In general, this effect
does not depend on which amplitude we are considering as it is an
effect generated by the initial-state gluons. It does however depend
on the energy scale being considered, the jet-veto suppression being
stronger at larger values of $M_{e\mu}$. We also note that the
operators have very different sizes. At $\Lambda=2\,$TeV, operators
$2$ and $3$ are the largest with operators $1$ and $4$ being a factor
of $10$ smaller. Operator $6$ is a factor of about $50$ smaller than
operators $2$ and $3$ and operator $5$ is a factor of $1000$ smaller
than operators $2$ and $3$. This is shown in
figure~\ref{fig:Squaredopscomparison}. We note that the large
differences in size between these operators mean that some will be
much better constrained than others.

\begin{figure}[htbp]
  \includegraphics[width=.5\textwidth]{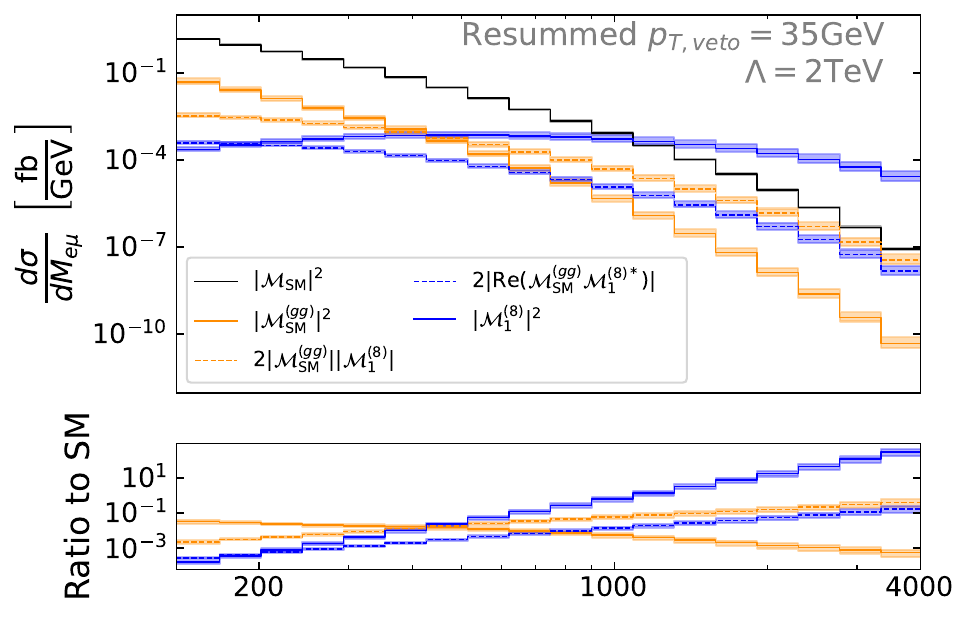}
  \includegraphics[width=.5\textwidth]{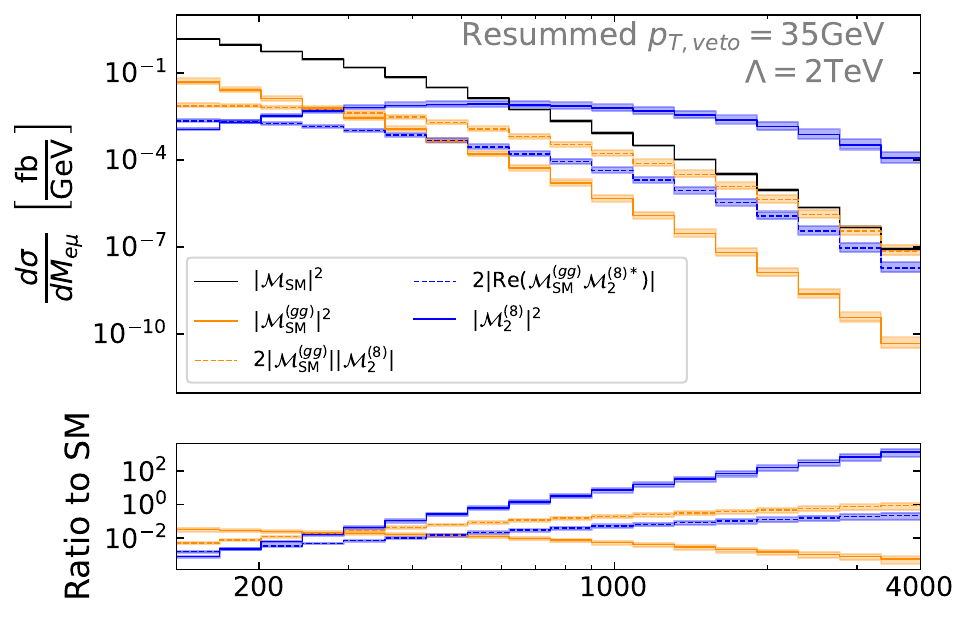}  
  \includegraphics[width=.5\textwidth]{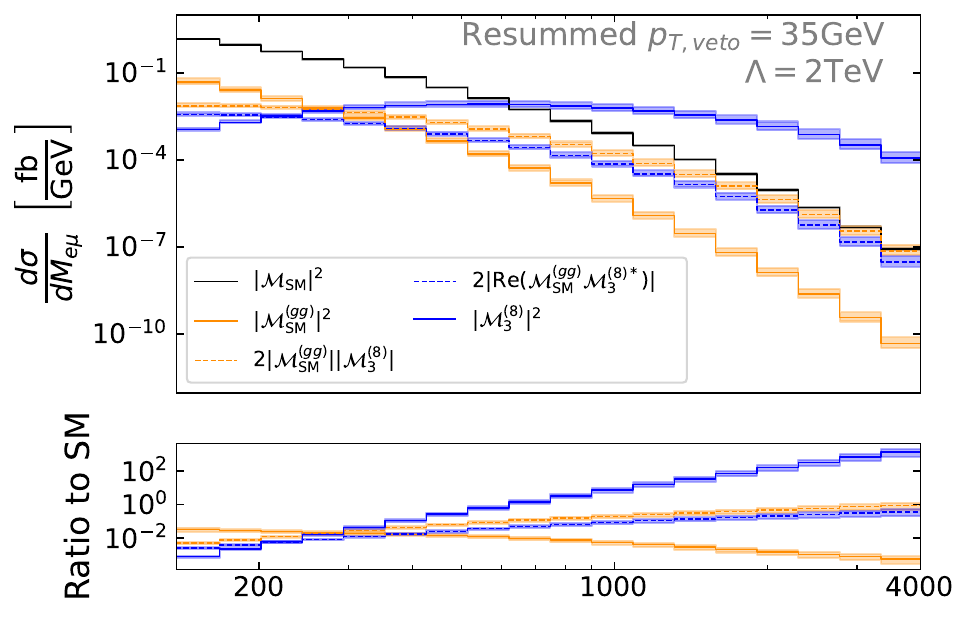}  
  \includegraphics[width=.5\textwidth]{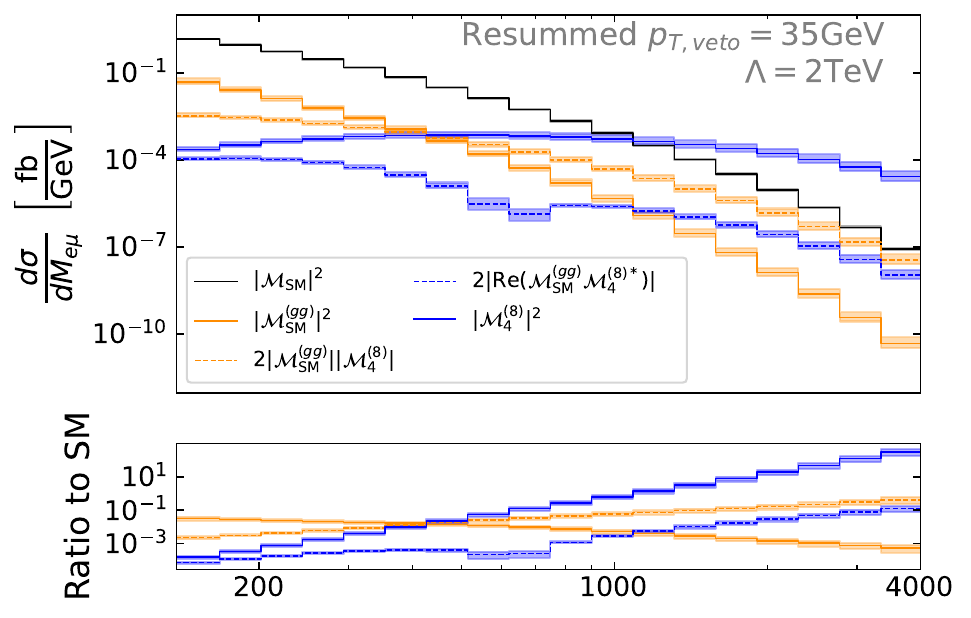}  
  \includegraphics[width=.5\textwidth]{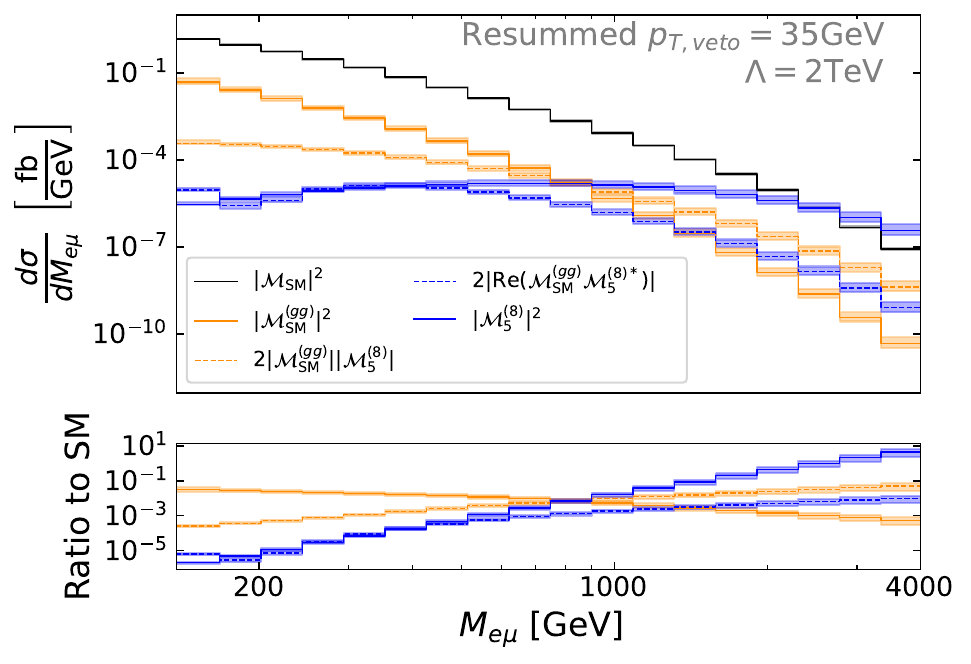}  
  \includegraphics[width=.5\textwidth]{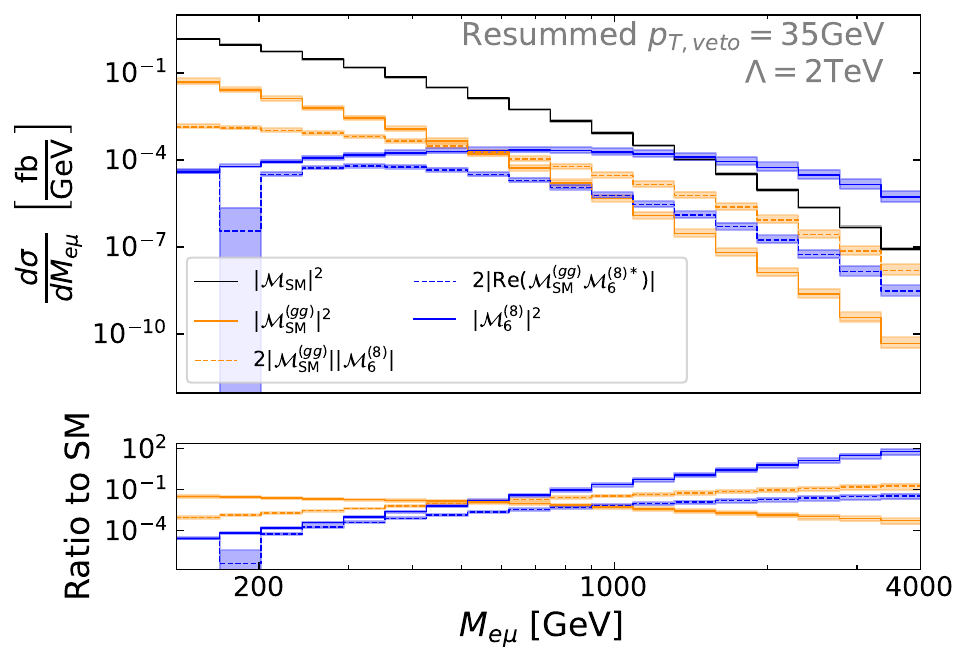}  
  \caption{Comparison of dimension-8 interference (blue-dashed) and
    dimension-8 squared (blue) contributions with the SM (black)
    operator at EFT mass scale $\Lambda=2\,$TeV. The SM $gg$
    contribution (orange) is also shown for comparison. For the SM we use the resummed prediction given by~\eqref{eq:sm_best_res} with jet-veto $p_{T, \mathrm{veto}}=35\,$GeV for all $gg$ predictions we use NLL accuracy with this same jet-veto.}
  \label{fig:SMBSMOperatorPrediction}
\end{figure}

We are now in a position to look at the prospects of constraining
dimension-8 operators from interference by comparing the BSM signal to
the SM background in Figure~\ref{fig:SMBSMOperatorPrediction}. For
both the signal and background we use the best resummed
predictions. Each interference term is bounded from above by the
purple dashed line, corresponding to perfect overlap of the BSM and SM
amplitudes (labelled $2|\mathcal{M}^{(gg)}_{\rm SM}||\mathcal{M}^{(8)}_i|$). The
closer $2|\mathrm{Re}(\mathcal{M}^{(gg)}_{\rm SM} \mathcal{M}^{(8)\,*}_i)|$ is to this
upper bound, the better the interference of the corresponding BSM
amplitude with the SM $gg$ channel. It can be seen
    that due to the small $gg$ contribution, the interference terms
    are suppressed in this channel and even in the case of perfect
    interference between SM $gg$ and dimension-8
    (orange-dashed). Their contribution is too small to be used for
    constraints with current luminosity and theoretical
    uncertainties. We also observe that only operator $4$
shows a poor overlap with the SM. In all other cases, the interference
terms, even with sizeable overlap with the SM, are small because
$|\mathcal{M}^{(gg)}_{\mathrm{SM}}|$ is itself small. If the dimension-8 squared term becomes non-negligible then it would also make the interference term visible. However, this corresponds to the regime in which the EFT approximation breaks down.

\newpage

In the next section we will not constrain dimension-8 operators using
their squared amplitudes due to the fact we would need to account for
dimension-10 operators in order to consistently study their effect
within the EFT framework. Instead we will turn our attention to the
CP-even and CP-odd dimension-6 operators, which we have just
demonstrated can be constrained from their squared contributions
safely without including the dimension-8 interference terms. Then, in
section~\ref{sec:constd8}, we will assume a hypothetical (though
motivated) scenario in which the contribution of dimension-6 operators
is negligible, and obtain some constraints on dimension-8 operators
using their square amplitudes.

We remark that, in both section~\ref{sec:sensitivity}
and~\ref{sec:constd8}, we will use only the $M_{e\mu}$ distribution to
find constraints, leaving the exploration of other observables to
future work.

\section{Constraining Dimension-6 Operators}
\label{sec:sensitivity}

In this section we present constraints on dimension-6 operators using
current and future data. As we have seen in
section~\ref{sec:bsm_preds}, due to the small SM $gg$ contribution,
when considering operators which contribute via gluon fusion, we can
consider the dimension-6 squared contribution whilst assuming that the
dimension-8 terms will be negligible (as long as we are in the EFT
regime). We start by describing the statistical methods we used to
both constrain operators with current data and produce sensitivity
studies for the HL-LHC. We then compare the current constraints from
this channel to results generated by Higgs studies. We then present
sensitivity studies of the dimension-6 operators and discuss how
removing the jet-veto and hypothetical reduction of the uncertainties
can improve sensitivity. We also compare these to projections of
constraints from future Higgs studies.

Using eq.~\eqref{eq:Mgg-squared} and the best SM
prediction found in section~\ref{sec:SM} we can define, for a set of
$\kappa_i$ (which we also take to include values for $c_i$ and
$\Lambda$), a prediction at either the LHC or HL-LHC which we call
$\{m_j\left(\kappa_i\right)\}$. We can then compare this to data
points $\{n_j\}$. For the LHC, we take this data
from ATLAS~\cite{ATLAS:2019rob}. However for the HL-LHC sensitivity studies
$\{n_j\}$ are obtained from the best current SM
predictions. As mentioned, we only take $\{n_j\}$ bins up to the
largest bin $N$ which satisfies eq.~\eqref{eq:maxbin} for the given
$\Lambda$ or $\kappa_i$ ($\kappa_i$ as converted with
eq.~\eqref{eq:kappaconversion}).

For the generation of exclusion plots and sensitivity studies we then use a delta chi-squared test statistic defined as:
\begin{equation}
  \label{eq:deltachisq}
    \Delta \chi^2\left(\kappa_i\right) \equiv \chi^2\left(\kappa_i\right) - \chi^2\left(\hat\kappa_i\right)\,,
\end{equation}
where $\chi^2\left(\kappa_i\right)$ is defined as:
\begin{equation}
  \label{eq:chisq}
    \chi^2\left(\kappa_i\right) \equiv \sum_{j=1}^N \frac{\left(n_j - m_j\left(\kappa_i\right)\right)^2}{(\Delta m_j)^2}\,,
\end{equation}
and $\hat\kappa_i$ are values of the considered $\kappa_i$ which
minimise $\chi^2\left(\kappa_i\right)$. For each value of $N$, the
$\hat\kappa_i$ must be found separately.
In order to account for
theoretical and systematic errors, following~\cite{Arpino:2019fmo}, we use
\begin{equation}
  \label{eq:Dni-exp}
  (\Delta m_j)^2 = m_j\left(\kappa_i\right) + (\Delta^{(\rm th)}_j/2)^2  + (\Delta^{(\rm sys)}_j/2)^2 \,.
\end{equation}
In the above equation, $\Delta^{(\rm th)}_j$ is the theoretical
uncertainty associated with the SM prediction for $n_j$,
namely the difference between the maximum and minimum value of $n_j$.
The quantity $\Delta^{(\rm sys)}_j$ gives the experimental systematic
error. For real data, this is the one quoted by the ATLAS
collaboration. For projected data, this is computed by extrapolating
current systematic errors to higher energies. How this is done in
practice will be explained when the constraints on the BSM parameters
are presented in section~\ref{sec:projections}. 

While for actual data we can obtain constraints assuming
$\Delta \chi^2\left(\{\kappa_i\}\right)$ is distributed according to a
$\chi^2$ distribution, for the HL-LHC sensitivity studies we use the
method of median significance. This is done by generating many sets of
$\{n_j\}$ using the expected $\{\bar n_j\}$ given by the Standard
Model best prediction and a Poisson distribution for each bin
independently. For these simulated data sets we obtain the probability
distribution for $\Delta\chi^2\left(\{\kappa_i\}\right)$, whose median
makes it possible to calculate the p-value associated with the
considered $\{\kappa_i\}$. We then exclude all values of
$\{\kappa_i\}$ whose p-value is less than 0.05.

\subsection{Constraints from Current Data}
\label{sec:constraints}

The values of $\delta\kappa_t$ and $\kappa_g$ are already well
constrained by Higgs production~\cite{ATLAS:2019nkf}. The
best fit parameters were $\delta\kappa_t=0.09 $ and $\kappa_g =
-0.1$, and within  $2\sigma$ we have
$-0.19 \textless \delta\kappa_t \textless 0.39$ and
$-0.21 \textless \delta\kappa_t + \kappa_g \textless 0.21$. Therefore, we simplify
the region of allowed phase space for $\delta\kappa_t$ and $\kappa_g$
as a parallelogram enclosed by the four points:
\begin{equation}
	\label{eq:kappat-kappag}
(\delta\kappa_t, \kappa_g)= (0.39, -0.60),(0.39, -0.18),(-0.19, -0.02),(-0.19, 0.40),
\end{equation}
This constraint can be converted in a corresponding lower bound for
$\Lambda$ using eq.~\eqref{eq:kappaconversion} and taking
$\alpha_s(M_{H})=0.113$ (which we also take for all future
conversions), giving $\Lambda\gtrsim 5\,$TeV. The parameters
$\tilde \kappa_g$ and $\tilde \kappa_t$ have been previously
constrained in~\cite{ATLAS:2019nkf}, resulting in
$-1 < \tilde\kappa_t + \tilde\kappa_g < 1$ when
$ \delta\kappa_t + \kappa_g = 0$. The parameter $\tilde \kappa_t$ has
also been previously constrained~\cite{ATLAS:2020ior,ATLAS:2023cbt,
  CMS:2022dbt, Bahl:2020wee}, giving a constraint of
$-1 < \tilde\kappa_t < 1$. Combining these constraints gives
$-2 < \tilde\kappa_g < 2$. Unlike in Higgs studies, we will be able to
access $\kappa_g$ and $\tilde\kappa_g$ independently of
$\delta\kappa_t$ and $\tilde\kappa_t$. This is due to the fact that,
at high energies, the contribution of top loops will be suppressed,
hence enhancing the sensitivity to contact interactions.

\begin{figure}[htbp]
\centering
  \includegraphics[width=.8\textwidth]{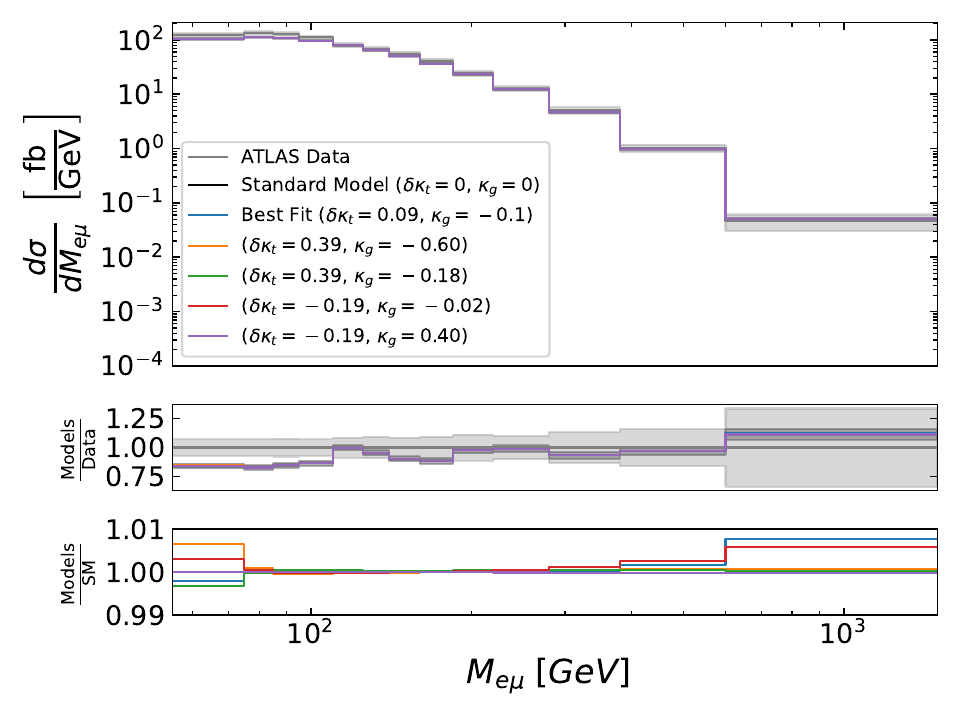}
  \caption{Comparison to ATLAS data of the extremal models not already ruled out by previous studies as in eq.~\eqref{eq:kappat-kappag}. The largest dimension-8 contribution is also included at a mass scale consistent with the size of $\kappa_g$.}
  \label{fig:currentmodel_comparisons}
\end{figure}

First, we are able to verify that values of $\delta\kappa_t, \kappa_g$
within current constraints are all compatible with the most recent
ATLAS data for $WW$ production~\cite{ATLAS:2019rob} (see
figure~\ref{fig:currentmodel_comparisons}). Rephrasing these bounds in terms of a
scale for the EFT results in $\Lambda>5\,$TeV. We also checked separately the size of the largest dimension-8 squared contribution
(operator 3) corresponding to $\Lambda = 5\,$TeV (which is well into the EFT
regime) and we observed compatibility with data within two standard
deviations, similar to the SM.

\begin{figure}[htbp]
	\centering
  \includegraphics[width=.8\textwidth]{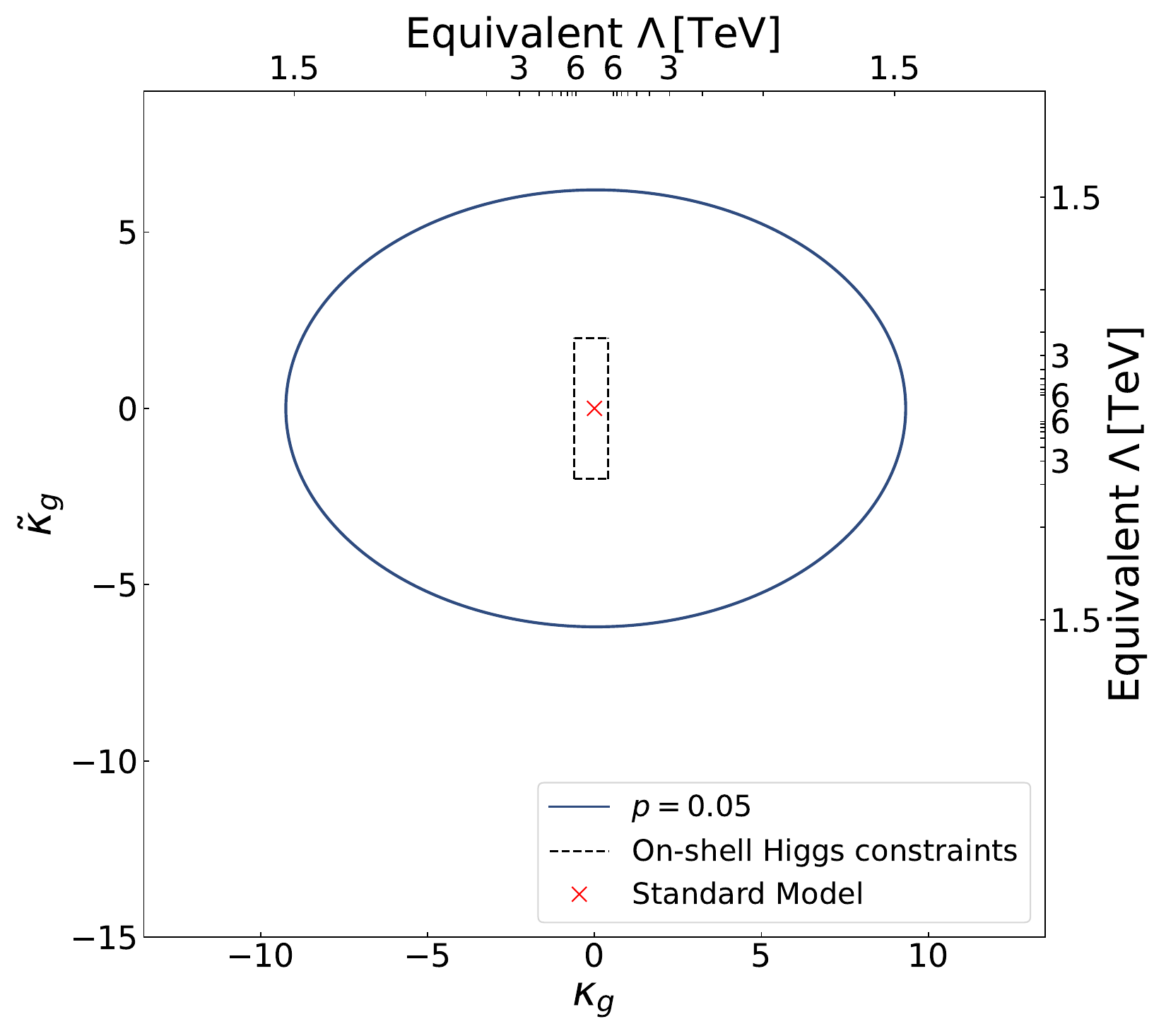}
  \caption{Constraints on $\kappa_g$ and $\tilde\kappa_g$ arising from current ATLAS data with $WW$ production~\cite{ATLAS:2019rob}. The exclusion contour is placed at a p-value of 0.05 which corresponds to $\sim 2\sigma$. Anything outside of this contour is excluded. The current constraints are also taken at $2\sigma$.}
  \label{fig:ATLASContourplot}
\end{figure}

Given the fact that the gluon channel interference between SM and
dimension-8 amplitudes is very small, we can treat the
dimension-8 operators as unconstrained in the EFT regime. We could
then try to see if we can use current ATLAS $WW$ data to constrain
$\tilde\kappa_g$. Since low-$M_{e\mu}$ bins are not expected to
be sensitive to higher-dimensional interactions, we have neglected the first three bins (which did not agree perfectly with data) to concentrate on
the high-$M_{e\mu}$ bins. We then performed a simultaneous fit of $\kappa_g$ and
$\tilde\kappa_g$, and obtain the contour plots in
figure~\ref{fig:ATLASContourplot}.

We find unfortunately that the constraints we obtain are not
competitive with those already found in earlier works, even when
taking into account the fact that $\kappa_g$ and $\tilde\kappa_g$ are
not measured independently of $\kappa_t$ and $\tilde\kappa_t$
respectively. It should be noted that the constraints on
$\tilde\kappa_t$ are not strong enough to be
interpreted within the SMEFT framework unless
$|\kappa_t|, |\tilde\kappa_t| \ll 1$, which leads to an EFT scale
$\Lambda \gg v$ as per eq.~\eqref{eq:kappaconversion}. For this reason,
we have chosen not to include the $\kappa_t$ and $\tilde\kappa_t$
constraints which, even if ignoring EFT regime considerations, are not
competitive with current constraints.

\subsection{Projections at HL-LHC}
\label{sec:projections}

\begin{figure}[htbp]
\centering
  \includegraphics[width=.9\textwidth]{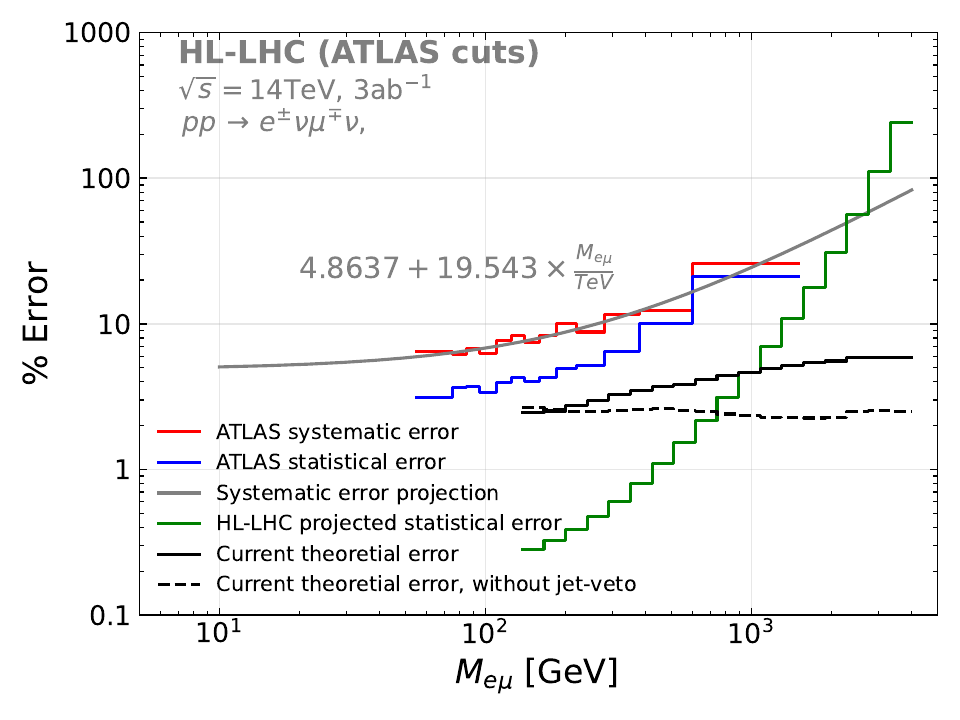}
  \caption{Projections for the sources of error for the dilepton invariant mass distribution at the HL-LHC ($14\,$TeV, $3\,$ab$^{-1}$). The statistical errors assume ATLAS cuts both with and without a jet-veto.}
  \label{fig:HLLHC_error_predictions}
\end{figure}

We expect the constraints we have obtained in the previous section to
be improved when considering the High Luminosity LHC as the EFT
effects will mostly appear in the tail of distributions which will
receive better statistics in future runs. We first show in
figure~\ref{fig:HLLHC_error_predictions} how the dilepton invariant mass
distribution is affected by statistical, theoretical, and systematic
errors. Using the current ATLAS systematic errors we can extrapolate a
linear expression for how these may grow with energy assuming no
improvement in their handling between now and HL-LHC's first runs. We also show the expected
statistical and theoretical errors. We can see that the systematic errors
will dominate due to the large growth with energy and that the
SM will stop producing any events after an energy of
$4\,$TeV. For this reason, and due to the growing systematic errors we
choose this to be the approximate cut-off for our analysis. Whilst speculative at this point, it is possible that the current
systematic errors can be brought in line with the maximum between theoretical and statistical errors.
If this were achieved, then there would be high motivation to get below $1\%$
agreement between theory and data at low energies. This will aid constraining power at $M_{e\mu} \lesssim
  2\,$TeV. From figure~\ref{fig:min_lambda_with_mll}, we can understand that, if an
  operator has already been constrained to be over $\sim4\,$TeV, then
  it was probably using bins with $M_{e\mu}$ between $1\,$TeV and $2\,$TeV. Therefore,
  reductions in the theory uncertainty to $1\%$ will give limited
  improvements. However, for any operators that could not be previously
  constrained or are constrained under $4\,$TeV, the sensitivity will
  be improved substantially as theoretical errors are reduced. Note
  that this applies assuming the presence of the jet-veto.

\begin{figure}[htbp]
	\centering
  \includegraphics[width=.8\textwidth]{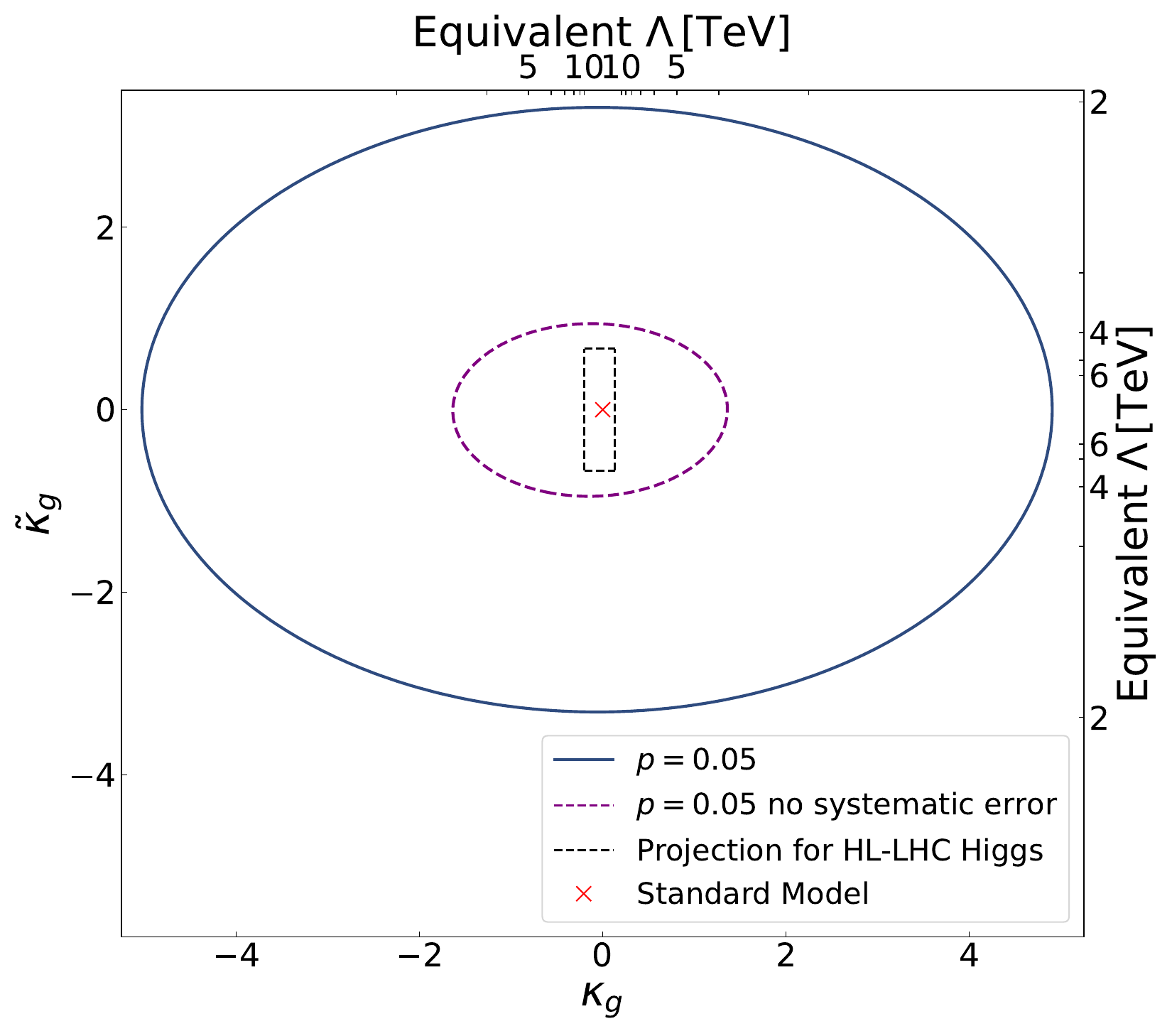}
  \caption{Sensitivity plots for $\kappa_g$ and $\tilde\kappa_g$ at
    HL-LHC with ATLAS cuts ($14\,$TeV, $3\,$ab$^{-1}$) using $WW$
    production. The exclusion contour is placed at a p-value of 0.05 which corresponds to $\sim 2\sigma$. The current constraints are taken at $ 2\sigma$. }
  \label{fig:HLLHCContourplot_resum}
\end{figure}

\newpage
In figure~\ref{fig:HLLHCContourplot_resum}, we present the contour
plots corresponding to projections from the HL-LHC. In order to ensure
that the plots remain within the EFT regime the bins used in the
statistical analysis are cut off once the EFT regime breaks down
in accordance with equation eq.~\eqref{eq:empirical_lambdamin}. This leads to
discontinuities in the contour plots which could be reduced by using a
finer binning or in the ideal case a variable binning. We describe how we have dealt with these discontinuities in appendix~\ref{sec:AppendixDiscont}. We also include
a contour plot without systematic errors to show the ideal case for
this channel at the HL-LHC considering we do not know how the
systematic errors will be improved upon between now and the first runs
of HL-LHC.

In order to compare our constraints with those of the Higgs channel we
use the projections given by \cite{Cepeda:2019klc,
  Englert:2014uua}. Together they suggest a conservative factor of $3$
improvement in the constraints for $\kappa_g$, which we also take to
apply for $\tilde \kappa_g$. Although we see improvement in the
constraints at HL-LHC for the $WW$ channel, they are not competitive
with the predicted constraints from Higgs studies. However by removing
systematic errors we see that the $WW$ channel could provide
complementary constraints on $\tilde\kappa_g$. We find that
$|\tilde\kappa_g| < 0.9$ would give
$\Lambda_{\tilde\kappa_g} > 3.9\,$TeV, up from the current
value of $\Lambda_{\tilde\kappa_g} > 2\,$TeV. Improvement in theoretical uncertainties down to $1\%$ could further improve this to $\Lambda_{\tilde\kappa_g} > 4.7\,$TeV. Once again, it is found that
$\tilde\kappa_t$ cannot be constrained within EFT considerations. This
can be explained by the fact that the SMEFT operator which generates
the $\tilde\kappa_t$ whilst being dimension-$6$, appears as loop
induced in the SM and is therefore not a leading order SMEFT
contribution to this channel.

\subsection{Effect of the Jet-veto}
\label{sec:noveto}

One way to improve the constraints on the $gg$ operators would be to
remove the jet-veto. The jet-veto further suppresses the $gg$ channel
relative to the $q\bar{q}$ channel as seen in
figures~\ref{fig:gg-SM} and~\ref{fig:BSMOperatorPrediction} and so removing it could give
increased sensitivity to gluon induced operators. This could be done
by tagging $b$-jets and setting the veto to only remove those
jets~\cite{CMS:2020mxy}. This would probably not be perfectly efficient however by
considering the fixed order predictions without a jet-veto we can imagine a scenario in
which such a perfect background removal process could be
designed. This allows us to highlight the effect of the jet-veto on
the gluon operator sensitivity.

\begin{figure}[htbp]
  \includegraphics[width=.5\textwidth]{figures/contour_14TeVkgkgtilde.pdf}
    \includegraphics[width=.5\textwidth]{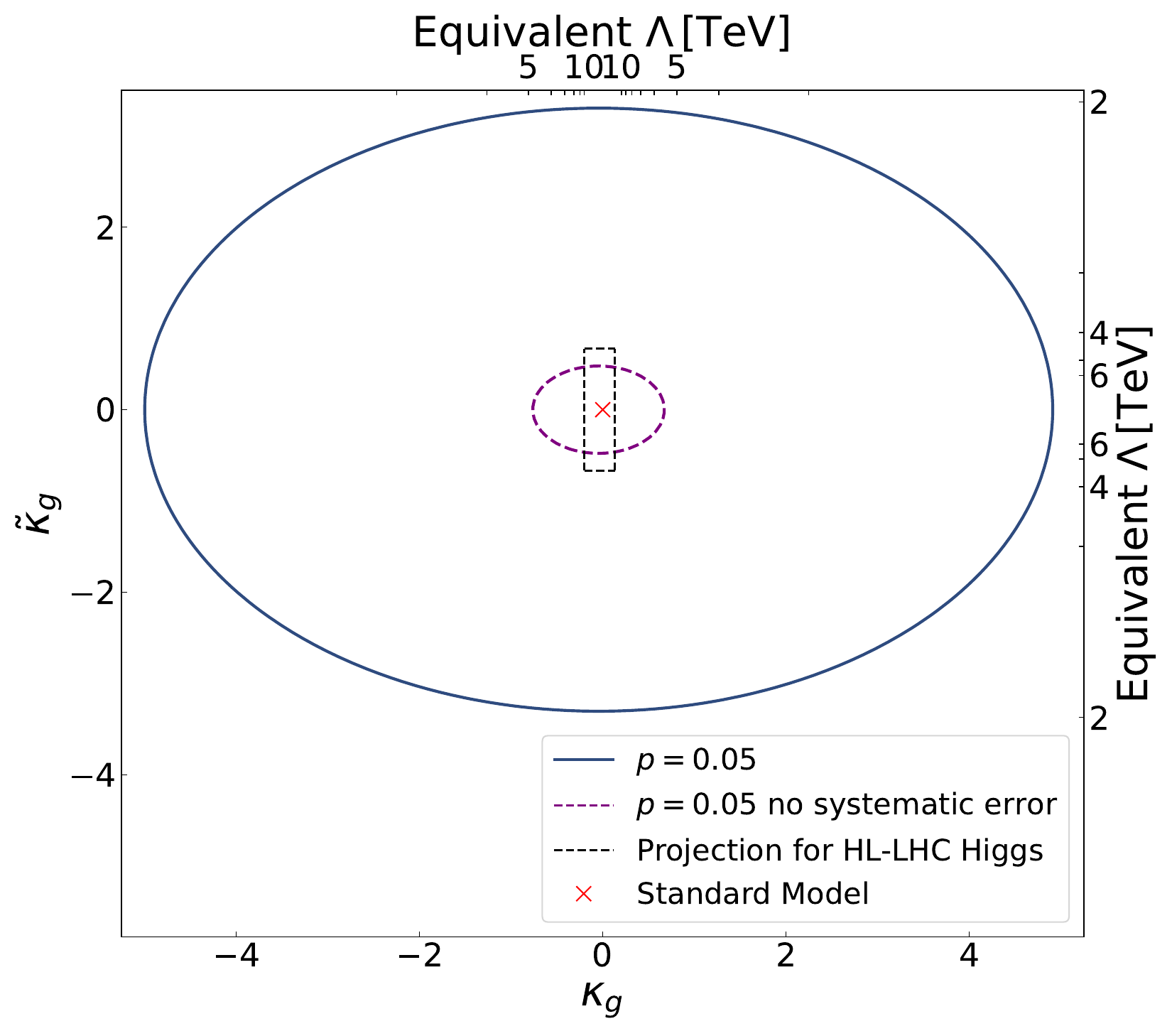}
  \caption{Sensitivity plots for $\kappa_g$ and $\tilde\kappa_g$ at HL-LHC with ATLAS cuts ($14\,$TeV, $3\,$ab$^{-1}$) using $WW$ production - however with the jet-veto condition lifted. The exclusion contour is placed at a p-value of 0.05 which corresponds to $\sim 2\sigma$. The current constraints are taken at $2\sigma$. We show the plots with (left) and without (right) systematic errors for comparison.}
  \label{fig:HLLHCContourplot_fo}
\end{figure}

It can be seen from figure~\ref{fig:HLLHCContourplot_fo} that removing
the jet-veto can improve the sensitivity of this channel to gluon
induced operators subject to an improvement in the systematic error
predictions. Without a reduction in the systematic errors, removing the jet-veto does not improve constraining value. By
removing the systematic errors, the value can be further constrained to
$|\tilde\kappa_g| < 0.5$. This is equivalent to
$\Lambda_{\tilde\kappa_g} > 5.2\,$TeV. In this case, the constraints on
$\kappa_g$ become competitive with the projected constraints from
Higgs production. This constraint cannot be substantially improved by
reducing theoretical uncertainties for the reasons discussed in section~\ref{sec:projections}.

\section{Constraining Dimension-8 Operators}
\label{sec:constd8}

In section~\ref{sec:constraints} we saw that the
constraining power for $O_{GH}$ and the anomalous $t\bar th$ coupling from the $WW$ channel with
current LHC data is not competitive with that of on-shell Higgs
studies. Although future projections - particularly in the case of the
CP-odd dimension-6 operator - are more optimistic, projections for
improvements in the single Higgs channel at HL-LHC give the ability to reduce
the uncertainty in constraining $\kappa_g$ by a factor of three
\cite{Cepeda:2019klc, Englert:2014uua} (as mentioned earlier). This implies
$|\kappa_g| \lesssim 0.2$, which corresponds to
$\Lambda \gtrsim 10\,$TeV. For $\tilde\kappa_g$ the constraint is
weaker at $|\tilde\kappa_g| \lesssim 0.7$, which corresponds to
$\Lambda \gtrsim 4.4\,$TeV.  

With the dimension-6 operators already well constrained by Higgs production, we can posit a scenario in which
the dimension-6 and dimension-8 terms are decoupled and live at
completely different mass scales, or $c_6 \ll c_8 = O(1)$ with the same EFT scale $\Lambda$, or even that dimension-6 operators
are not generated at all by the UV theory. In this scenario, the strong dimension-6
constraints from Higgs production do not rule out that the dimension-8
$ggWW$ operators have negligible contribution. We can therefore put
further constraints on the dimension-8 operators within this
assumption. Starting from eq.~\eqref{eq:Mgg}, we separate the various
contributions to the amplitude $\mathcal{M}^{(gg)}$ as follows:
\begin{equation}
  \label{eq:Mgg_higherorder}
  \mathcal{M}^{(gg)}  = \mathcal{M}^{(gg)}_{\rm SM} + \frac{c_6}{\Lambda^2}  \mathcal{M}^{(gg)}_6 + \frac{c_8}{\Lambda^4} \mathcal{M}^{(gg)}_8 + \frac{c_{10}}{\Lambda^6} \mathcal{M}^{(gg)}_{10} + \frac{c_{12}}{\Lambda^8} \mathcal{M}^{(gg)}_{12}\,+\dots.
\end{equation} 
Once again squaring we obtain
\begin{equation}
  \label{eq:Mgg_higherordersquared}
  \begin{split}
  |\mathcal{M}^{(gg)}|^2 & = |\mathcal{M}^{(gg)}_{\rm SM}|^2 + \frac{2 c_6}{\Lambda^2}\mathrm{Re}\left(\mathcal{M}^{(gg)}_{\rm SM}\mathcal{M}^{(gg)}_6\right) + \frac{1}{\Lambda^4}\left[c_6^2|\mathcal{M}^{(gg)}_6|^2 + 2 c_8 \mathrm{Re}\left(\mathcal{M}^{(gg)}_{\rm SM}\mathcal{M}^{(gg)}_8\right)\right] \\
     & + \frac{2}{\Lambda^6}\left[c_6 c_8 \mathrm{Re}\left(\mathcal{M}^{(gg)}_{6}\mathcal{M}^{(gg)}_8\right) +c_{10}\mathrm{Re}\left(\mathcal{M}^{(gg)}_{\rm SM}\mathcal{M}^{(gg)}_{10}\right)\right] \\
     & + \frac{1}{\Lambda^8}\left[c_8^2|\mathcal{M}^{(gg)}_8|^2+2c_6 c_{10}\mathrm{Re}\left(\mathcal{M}^{(gg)}_{6}\mathcal{M}^{(gg)}_{10}\right) + 2c_{12}\mathrm{Re}\left(\mathcal{M}^{(gg)}_{\rm SM}\mathcal{M}^{(gg)}_{12}\right) \right] \\
     & +\mathcal{O}\left(\frac{1}{\Lambda^{10}}\right)\,.
  \end{split}
\end{equation} 
If now, motivated by the constraints arising from Higgs production, we
assume that our BSM model has $c_6 \to 0$, we can first remove all terms with $\mathcal{M}^{(gg)}_{\rm SM}$ in
eq.~\eqref{eq:Mgg_higherordersquared}, because its interference with
all higher-dimensional operators is either zero (with dimension-6) or
very small (with dimension-8 and higher). The assumption $c_6 \to 0$ allows us to remove all other remaining terms
except $|\mathcal{M}^{(gg)}_8|^2/\Lambda^8$, which can be
used to constrain the dimension-8 operators.\footnote{Note that
  the CP-odd dimension-6 operator does not interfere with CP-even
  higher-order operators.} We still need the $c_{10}$ and $c_{12}$ terms to be smaller than the $c_8$ terms and we can do this by staying in the EFT regime such that each of the amplitudes in eq.~\eqref{eq:Mgg_higherorder} get smaller sequentially (due to increasing negative powers of $\Lambda$). To achieve this we keep the constraint from~\eqref{eq:empirical_lambdamin} inputting the mass scale of the dimension-8 operator. This ensures the hierarchy of EFT operators greater than dimension-8 and justifies the exclusion of terms such proportional to $c_6 c_{10}$ (which is always smaller than $c_6 c_8$) and $c_{10}, c_{12}$ which are smaller than $c_8$.\footnote{Note that
  although we use the dimension-6 amplitude to calculate if we are in the EFT regime, we still subsequently set $c_6 \to 0$.}

To ensure that this assumption is not in contradiction with current
data and future projections, we first constrain dimension-8 operators, and a
posteriori we check that the largest dimension-6-dimension-8 ($c_6 c_8$ piece)
interference term is $1/4$ the size of the dimension-8 squared
operator for each of the bins used to constrain the dimension-8
operator (taking the coefficient of the dimension-6 amplitude to be the maximum previously constrained by on-shell Higgs data~\cite{ATLAS:2019nkf}, or in the case of HL-LHC the expected improvement~\cite{Cepeda:2019klc, Englert:2014uua}.). This condition gives us an intrinsic limit on how well
dimension-8 operators could be constrained. For completeness, we have
also considered the CP-odd dimension-8 interference with a CP-odd dimension-6
operator. The largest contribution of the CP-odd dimension-6-dimension-8 term to the $WW$
cross-section is the CP-odd $\mathcal{\tilde O}_{GH}$'s interference with a CP-odd version of operator
6. This has the same contribution as its CP-even counterpart but
$\tilde c_6 / \Lambda^2$ has not been constrained as well as $c_6 / \Lambda^2$. In the
following, we assume $\tilde c_6=0$, leaving a more complete analysis
of the CP-odd dimension-8 operators to future work.

\subsection{Constraints from Current Data}
\label{sec:constraints_d8}
\begin{figure}[htbp]
  \centering
  \includegraphics[width=.8\textwidth]{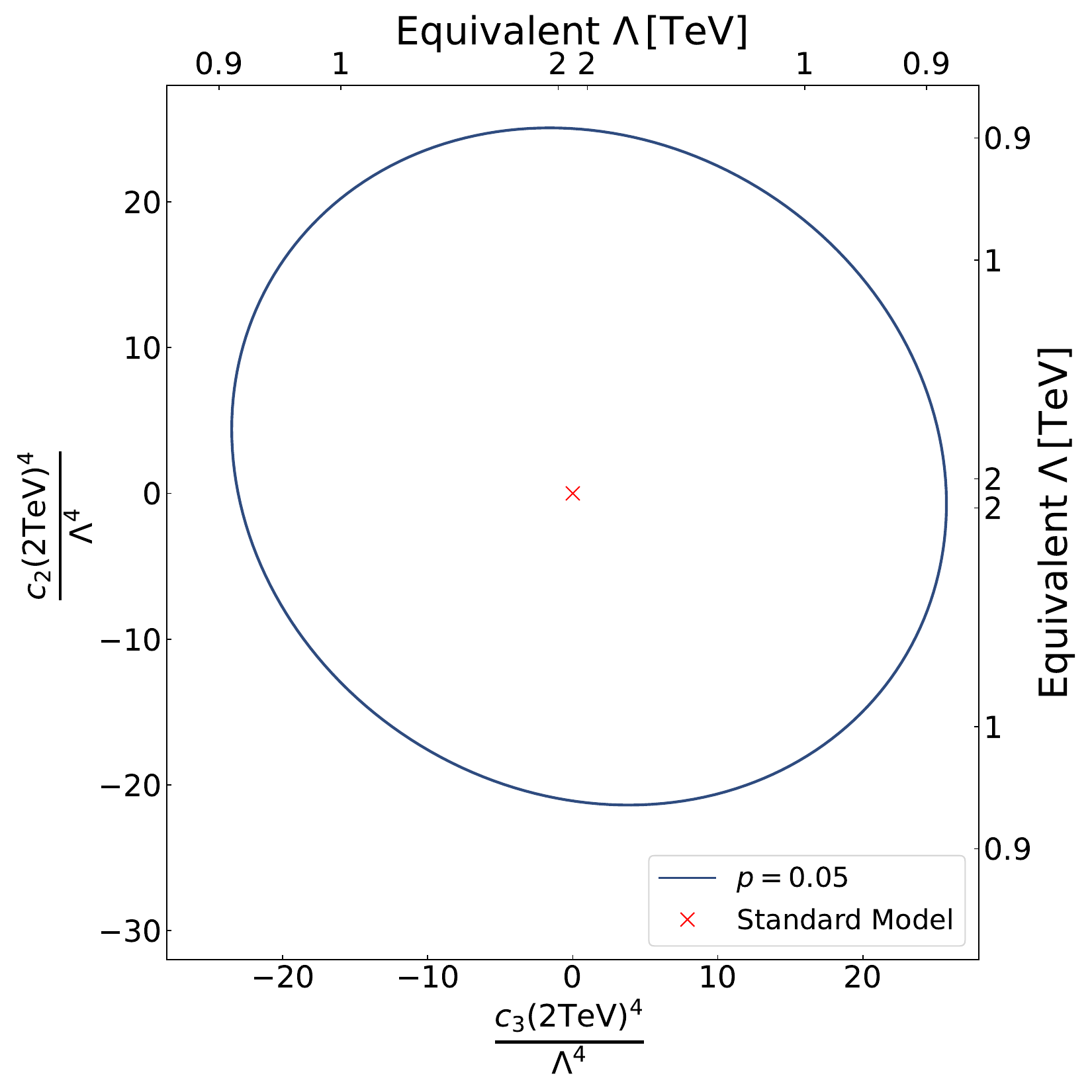}
  \caption{Constraints for operators 2 and 3 obtained using current
    ATLAS data. The contour is placed at a p-value of 0.05 which corresponds to $\sim 2\sigma$. Both operators can be constrained to have
    $\Lambda \gtrsim 900\,$GeV. The contour is approximately circular
    because the amplitudes corresponding to the two operators have the
    same magnitude and small interference (either with each other or
    with the SM).}
  \label{fig:ATLAS-O23}
\end{figure}

We start with operators 2 and 3, the ones with the largest
contribution to the $WW$ cross section. These are the only operators
that can be constrained using current ATLAS data, and we find
 $\Lambda \gtrsim 900\,$GeV, see figure~\ref{fig:ATLAS-O23}. This is already a new result.

It can be seen in figure~\ref{fig:ATLAS-O23} that the contour is approximately circular. This can be explained by noting that the squared contributions to the $M_{e\mu}$ distribution in this channel are identical as seen in figure~\ref{fig:Squaredopscomparison}. If we study the forms of equations~\eqref{eq:M2subamps} and~\eqref{eq:M3subamps}, then it can be noted that $\mathcal{M}_8^{(2)}$ and $\mathcal{M}_8^{(3)}$ can be written as:

\begin{subequations}
  \begin{align}	
   \mathcal{M}^{(2)}_{8\ ++/--} & \propto \mathcal{M}^{(a)}_8 + \mathcal{M}^{(b)}_8 \,, \\
 	\mathcal{M}^{(3)}_{8\ ++/--} & \propto -\mathcal{M}^{(a)}_8 + \mathcal{M}^{(b)}_8 \,.
  \end{align}
\end{subequations}
Where $\mathcal{M}^{(a)}_8 = \langle34\rangle\langle56\rangle[46]^2$ and  $\mathcal{M}^{(b)}_8 = [34][56]\langle35\rangle^2$. Since, from figure~\ref{fig:Squaredopscomparison}, $|\mathcal{M}^{(2)}_8|^2 = |\mathcal{M}^{(3)}_8|^2$, we can infer that $2\mathrm{Re}\left(\mathcal{M}_8^{(a)}\left(\mathcal{M}_8^{(b)}\right)^*\right) = 2\mathrm{Re}\left(\langle34\rangle^2\langle56\rangle^2[46]^2[35]^2\right)$ gives zero contribution to the $M_{e\mu}$ distribution. From this we can also deduce that 

\begin{equation}	
   \mathcal{M}_8^{(2)}\left(\mathcal{M}_8^{(3)}\right)^*\propto |\mathcal{M}_8^{(b)}|^2 -|\mathcal{M}_8^{(a)}|^2  + 2\mathrm{Re}\left(\mathcal{M}_8^{(a)}\left(\mathcal{M}_8^{(b)}\right)^*\right) \,.
\end{equation}
Since we have $|\mathcal{M}_8^{(a)}|^2 = s_{34}s_{56}s_{46}$ and
$|\mathcal{M}_8^{(b)}|^2 = s_{34}s_{56}s_{35}$, then
$|\mathcal{M}_8^{(b)}|^2 -|\mathcal{M}_8^{(a)}|^2 = 0$ as
$s_{46}=s_{35}$. The interference between operators 2 and 3
($\mathcal{M}_8^{(2)}\left(\mathcal{M}_8^{(3)}\right)^*$) is therefore
only proportional to
$2\mathrm{Re}\left(\mathcal{M}_8^{(a)}\left(\mathcal{M}_8^{(b)}\right)^*\right)$
and therefore gives no contribution to the $M_{e\mu}$ distribution of
this channel.

Since the SM $gg$-contribution is also small, these operators cannot be
readily distinguished using their interference with the SM
background. In practice, for the $M_{e\mu}$ distribution or the $WW$
channel, these two operators are indistinguishable and therefore a
constraint can only be placed on their combined contribution. Whether
this degeneracy between the operators can be lifted either by studying
their contributions to other channels (i.e. $ZZ$ production) or by
looking at other distributions, is a question we leave to future work.

\subsection{Projections at HL-LHC}
\label{sec:projections_d8}
\begin{figure}[htbp]
  \includegraphics[width=.5\textwidth]{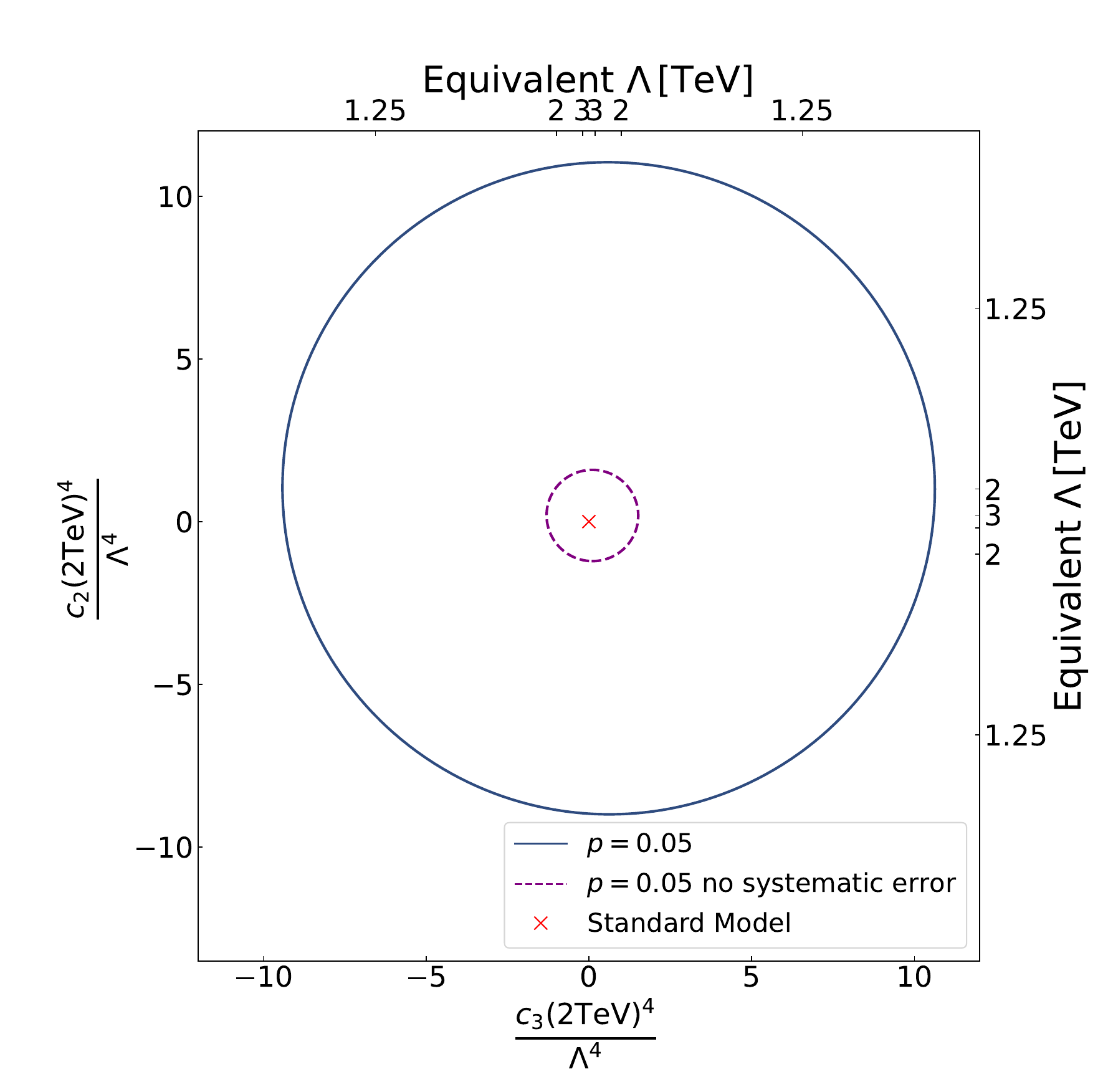}
  \includegraphics[width=.5\textwidth]{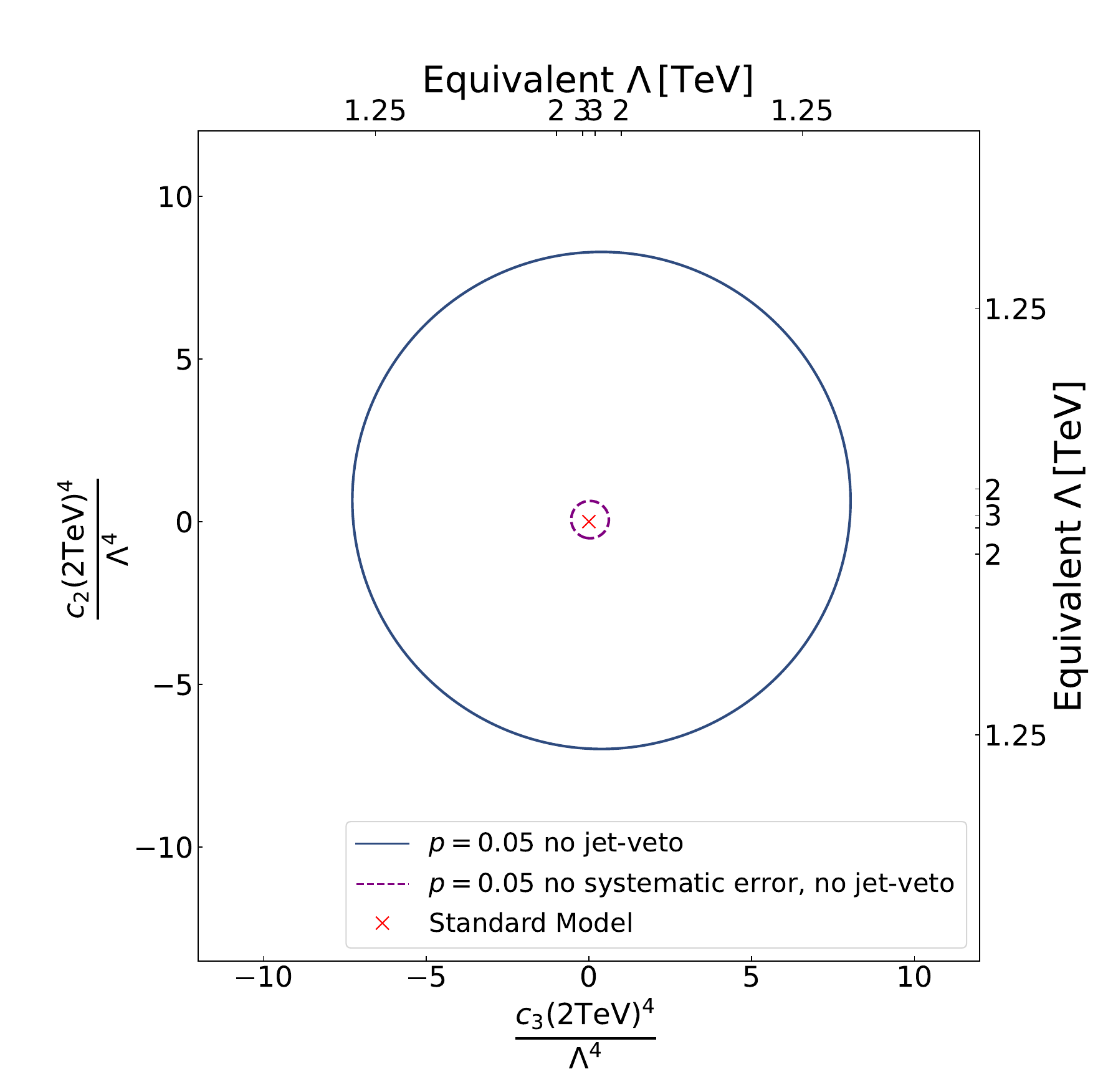}
  \caption{Sensitivity plots for operators 2 and 3 at the
    HL-LHC with ATLAS cuts($14\,$TeV, $3\,$ab$^{-1}$) using $WW$ production. The contours are placed at p=0.05 or $\sim 2\sigma$. As in the previous section, contours on the
    right panel correspond to the situation in which a jet-veto
    condition is applied, whereas those on the right are obtained
    without a jet-veto.  In both panels, we show contours
    corresponding to no systematic errors. Again the circular plots correspond to two operators with the same squared amplitude and negligible interference with each other and the SM.}
  \label{fig:HLLHCContourplot_k2k3}
\end{figure}

We now see how operators 2 and 3 can be further constrained at the
HL-LHC. The result is shown in
figure~\ref{fig:HLLHCContourplot_k2k3}. As expected, removing the
jet-veto condition improves the sensitivity to these
operators. Furthermore, in the assumption that systematic
uncertainties could be reduced to be much less than statistical and
theoretical uncertainties, we obtain the ultimate constraint
  $\Lambda \gtrsim 3\,$TeV. A reduction of the theoretical
   uncertainties to $1\%$ could push this ultimate constraint up to
   $\Lambda \gtrsim 4\,$TeV in the no jet-veto case. In the jet-veto
   case, the constraint of $\Lambda \gtrsim 2\,$TeV could rise to
   $\Lambda \gtrsim 3\,$TeV if theoretical uncertainties are reduced
   to $1\%$. Note that the contours in
figure~\ref{fig:HLLHCContourplot_k2k3} are still almost circular. This
shows that increasing sensitivity in this channel does not lift the
degeneracy between these two operators. For this not be the case we
need to have a strong interference either with another operator or we
need this channel to have errors reduced such that it becomes
sensitive enough for the SM interference of operators to no longer be
negligible.

In a scenario in which operators $2$ and $3$ are zero, we can try
  to constrain operators $1$ and $4$.  Unfortunately, it is not
  possible to constrain these operators, or operators $5$ and $6$ with
  the uncertainties we have quoted so far. It might be possible to
  constrain operators $1$ and $4$ if an overall $1\%$ accuracy is
  reached at the HL-LHC. However, in that case, one needs a prescription to
  profile the uncertainties which arise from the exclusion of other EFT
  contribution (for example dimension-6-dimension-10 interference), which we
  leave to future work. In general, a better strategy to constrain operators with different dimensions could be to keep all the bins, and attach a futher ``EFT uncertainty'' to each bin.

\section{Conclusions}
\label{sec:the-end}

In this paper, we addressed the question of the importance of
dimension-8 operators in constraining EFT parameters in $WW$ production. This process is difficult to model
with current automated tools because of the presence of a
jet-veto. Here, we study operators arising in gluon fusion, which have
been primarily considered at the level of dimension-6 operators. These are typically
constrained by using not only their interference with the SM, but also
their amplitude squared. The latter is formally of the same order as
the interference of dimension-8 amplitudes with the SM.

We considered all six CP-even dimension-8 operators contributing to
this process, we computed the corresponding amplitudes and
implemented them in the program MCFM-RE, which provides predictions for
$WW$ production with a jet-veto at state-of-the-art accuracy.  

We found that, due to the fact that the gluon induced
SM amplitudes become small at high energies, so do their interference
with the dimension-8 amplitudes. Therefore, unless we break the EFT
hierarchy the interference of the dimension-8 amplitudes with the SM
is much smaller than the square of the dimension-6 amplitudes and can
be safely neglected when performing EFT fits. We further
found that the jet-veto suppression affects the BSM signal more than
the SM background. This is due to the fact that the background occurs
mainly via quark-antiquark annihilation, and quarks radiate less than
gluons.

With this view, we investigated what constraints could be placed on the
coefficients of dimension-6 operators using current and future data
from the LHC. We found that, if we keep the jet-veto condition, these
bounds are not competitive from those which could be inferred from
Higgs cross-sections. However, relaxing the jet-veto condition and
with the optimistic assumption of systematic uncertainties below
theoretical uncertainties, it might be possible to have competitive
constraints on the CP-odd dimension-6 operator.

Before placing constraints on the EFT operators we also ensured that
we obtained the best possible prediction for the SM background. The
best current QCD prediction is given by matching NNLL to NNLO. This prediction gives a larger and more realistic
QCD scale variation error than using NNLO alone would
provide. Furthermore, we found that these predictions should be
augmented to include EW corrections at NLO which have large effects in
the high energy tails of these distributions, which is where we are placing
constraints on new physics.

Finally, inspired by the strong existing constraints on dimension-6 operators,
we postulated a scenario in which they are negligible, and
investigated what bounds could be placed on dimension-8 operators. We
found that two out of the six CP-even operators can be constrained
with current data, corresponding to a scale of new physics
$\Lambda \gtrsim 900\,$GeV. With future data, this constraint can be
improved, and we obtain $\Lambda \gtrsim 2\,$TeV with a jet-veto and $\Lambda \gtrsim 3\,$TeV in the best case scenario. Even with future data, it
is not possible to constrain operators $1$, $4$, $5$, and $6$. It may be possible to constrain
operators $1$ and $4$ if the combined theoretical and statistical uncertainties are
brought under the $1\%$ level.

We comment
  on prospects of constraints at the FCC-hh. While the increased
  luminosity and energy will improve statistical uncertainties, these
  are conditional on improvements in systematics and also in the
  theoretical uncertainties for lower energy bins. At higher energy,
  EW corrections grow to such an extent that logarithms
  $\ln\left(M_W/M_{WW}\right)$ will need to be
  resummed before meaningful constraints can be derived. This requires
  dedicated theoretical studies along the lines of~\cite{Denner:2024yut}, which we
  leave for future work.

After this study there are two natural steps. One is to perform a
comprehensive analysis of dimension-8 operators for all diboson
channels, for instance $ZZ$ and $Z\gamma$. These are comparatively
straightforward to study as fixed order predictions can be used for
both the signal and the background. It is also interesting to complete
the analysis of dimension-8 operators in $WW$ production by including
those occurring quark-antiquark annihilation, of which there are
many. Their interference with the SM could potentially be sizeable due
to the fact that the corresponding SM amplitude is not
loop-induced. In all these studies, it will be important to find good
proxies for the invariant mass of a $WW$ pair, especially if these
could disentangle the effects of degenerate operators. It would also
be very useful if alternative jet-veto conditions could be developed
such that a much larger fraction of the available signal events could
be kept.

\section*{Acknowledgements}
We thank Ennio Salvioni and Jonas Lindert for both useful discussions and for comments on the manuscript. AB is supported by the UK STFC under the Consolidated Grant ST/X000796/1, and thanks Royal Holloway, University of London for hospitality while this work was completed. The work of AM is partially supported by the National Science Foundation under Grant Numbers PHY-2112540 and PHY-2412701. MAL is supported by the UKRI guarantee scheme for the Marie Sk\l{}odowska-Curie postdoctoral fellowship, grant ref. EP/X021416/1. We acknowledge the use of computing resources made available by the Cambridge Service for Data Driven Discovery (CSD3), part of which is operated by the University of Cambridge Research Computing on behalf of the STFC DiRAC HPC Facility (www.dirac.ac.uk). The DiRAC component of CSD3 was funded by BEIS capital funding via STFC capital grants ST/P002307/1 and ST/R002452/1 and STFC operations grant ST/R00689X/1. DiRAC is part of the UK National e-Infrastructure.

\section*{Data Availability}
The data and code used to produce the plots for this paper as well as Mathematica files to derive the dimension-8 amplitudes are made available online~\cite{results_repo}. The implementation of dimension-8 amplitudes is available in a new branch of MCFM-RE~\cite{mcfm_re_d8}. If either are used please cite this paper.

\appendix

\section{Diagrams for the Dimension 6 Operators}

Here we include the diagrams which mediate the dimension-6 operators
that were considered in this paper. Figure~\ref{fig:Feynman-dim6kg}
shows the diagram generated from the operators which couple gluons to
the Higgs boson. Figure~\ref{fig:Feynman-dim6kt} shows how the operators which modify the top-Higgs coupling appear in the loop of the $gg$ fusion channel.
\begin{figure}[htbp]
  \begin{center}
    \begin{tikzpicture}[scale=1.5, transform shape]
      \begin{feynman}
        \vertex (a); 
        \vertex [right = 1cm  of a, dot] (v0) {};
        \vertex [right = 1cm  of v0] (v1);
        \vertex [right = 1cm  of v1] (c);
        \vertex [right = 0.5cm  of v1] (aux);
        \vertex [node font=\tiny, above = 0.55cm  of aux] (label1){\(W^+\)};
        \vertex [node font=\tiny, below = 0.55cm  of aux] (label2){\(W^-\)};
        \vertex [node font=\tiny, above = 0.9cm of a] (a1) {\(g\)};
        \vertex [node font=\tiny, below = 0.9cm of a] (a2) {\(g\)};
        \vertex [above = 0.9cm of c] (v2);
        \vertex [below = 0.9cm of c] (v3);
        \vertex [right = 0.8cm  of v2] (d);
         \vertex [right = 0.8cm  of v3] (e);
         \vertex [node font=\tiny, above = 0.6cm of d] (l1) {};
         \vertex [node font=\tiny, right = 0.2cm of l1] (l12) {};
         \vertex [node font=\tiny, below = 0.1cm of l12] (l13){\(\nu_{e}\)};
        \vertex [node font=\tiny, below = 0.6cm of d] (l2);
        \vertex [node font=\tiny, right = 0.2cm of l2] (l22) {};
        \vertex [node font=\tiny, above = 0.1cm of l22] (l23){\(e^+\)};
        \vertex [node font=\tiny, above = 0.6cm of e] (l3);
        \vertex [node font=\tiny, right = 0.2cm of l3] (l32) {};
        \vertex [node font=\tiny, above = 0cm of l32] (l33){\(\mu^{-}\)};
        \vertex [node font=\tiny, below = 0.6cm of e] (l4);
        \vertex [node font=\tiny, right = 0.2cm of l4] (l42) {};
        \vertex [node font=\tiny, below = 0cm of l42] (l43){\(\bar\nu_{\mu}\)};
        \diagram* {
          (a1) -- [gluon, momentum={\(p_1\)}] (v0), 
          (a2) -- [gluon, momentum'={\(p_2\)}] (v0),
          (v0) -- [scalar, momentum'={\(p_1 + p_2\)}] (v1), 
          (v1) -- [boson] (v2), 
          (v1) -- [boson] (v3), 
          (v2) -- [fermion, momentum={\(p_3\)}] (l1), 
          (v2) -- [anti fermion, momentum={\(p_4\)}] (l2), 
          (v3) -- [fermion, momentum={\(p_5\)}] (l3), 
          (v3) -- [anti fermion, momentum={\(p_6\)}] (l4),
        };
      \end{feynman}
       \end{tikzpicture}
  \end{center}
  \caption{Feynman diagram corresponding to the amplitude for the process $g(p_1)\>g(p_2)\! \to \! W^+(\to \nu(p_3)\>e^+(p_4))\>W^-(\to \mu^-(p_5)\> \bar \nu(p_6))$ occurring through the dimension-6 $ggh$ coupling.
  \label{fig:Feynman-dim6kg}
  }
\end{figure}

\begin{figure}[htbp]
  \begin{center}
    \begin{tikzpicture}[scale=1.5, transform shape]
      \begin{feynman}
        \vertex (a); 
        \vertex [right = 1cm  of a, dot] (v0) {};
        \vertex [right = 0.6cm  of a] (q0);
        \vertex [right = 1cm  of v0] (v1);
        \vertex [right = 1cm  of v1] (c);
        \vertex [right = 0.5cm  of v1] (aux);
        \vertex [node font=\tiny, above = 0.55cm  of aux] (label1){\(W^+\)};
        \vertex [node font=\tiny, below = 0.55cm  of aux] (label2){\(W^-\)};
        \vertex [node font=\tiny, above = 0.9cm of a] (a1) {\(g\)};
        \vertex [node font=\tiny, below = 0.9cm of a] (a2) {\(g\)};
        \vertex [node font=\tiny, above = 0.5cm of q0] (q1);
        \vertex [node font=\tiny, below = 0.5cm of q0] (q2);
        \vertex [above = 0.9cm of c] (v2);
        \vertex [below = 0.9cm of c] (v3);
        \vertex [right = 0.8cm  of v2] (d);
         \vertex [right = 0.8cm  of v3] (e);
         \vertex [node font=\tiny, above = 0.6cm of d] (l1) {};
         \vertex [node font=\tiny, right = 0.2cm of l1] (l12) {};
         \vertex [node font=\tiny, below = 0.1cm of l12] (l13){\(\nu_{e}\)};
        \vertex [node font=\tiny, below = 0.6cm of d] (l2);
        \vertex [node font=\tiny, right = 0.2cm of l2] (l22) {};
        \vertex [node font=\tiny, above = 0.1cm of l22] (l23){\(e^+\)};
        \vertex [node font=\tiny, above = 0.6cm of e] (l3);
        \vertex [node font=\tiny, right = 0.2cm of l3] (l32) {};
        \vertex [node font=\tiny, above = 0cm of l32] (l33){\(\mu^{-}\)};
        \vertex [node font=\tiny, below = 0.6cm of e] (l4);
        \vertex [node font=\tiny, right = 0.2cm of l4] (l42) {};
        \vertex [node font=\tiny, below = 0cm of l42] (l43){\(\bar\nu_{\mu}\)};
        \diagram* {
          (a1) -- [gluon, momentum={\(p_1\)}] (q1), 
          (a2) -- [gluon, momentum'={\(p_2\)}] (q2),
          (q1) -- [fermion] (v0), 
          (q2) -- [anti fermion] (v0),
          (q2) -- [fermion] (q1),
          (v0) -- [scalar, momentum'={\(p_1 + p_2\)}] (v1), 
          (v1) -- [boson] (v2), 
          (v1) -- [boson] (v3), 
          (v2) -- [fermion, momentum={\(p_3\)}] (l1), 
          (v2) -- [anti fermion, momentum={\(p_4\)}] (l2), 
          (v3) -- [fermion, momentum={\(p_5\)}] (l3), 
          (v3) -- [anti fermion, momentum={\(p_6\)}] (l4),
        };
      \end{feynman}
       \end{tikzpicture}
  \end{center}
  \caption{Feynman diagram corresponding to the amplitude for the process $g(p_1)\>g(p_2)\! \to \! W^+(\to \nu(p_3)\>e^+(p_4))\>W^-(\to \mu^-(p_5)\> \bar \nu(p_6))$ occurring through the dimension-6 modified $t\bar{t}h$ coupling.}
  \label{fig:Feynman-dim6kt}
\end{figure}

\section{Helicity amplitudes for operators 3 and 4}

\label{sec:AppendixO3O4}
Due to the $\mathrm{CP}$-odd fields in operators 3 and 4 the
evaluation of the helicity amplitudes becomes more involved. In
particular, we made use of the identity
\begin{equation}
	\label{eq:spinor_epsiilon_id}
	i\epsilon_{\mu\nu\rho\sigma}[a|\gamma^\mu|b\rangle[c|\gamma^\nu|d\rangle[e|\gamma^\rho|f\rangle[g|\gamma^\sigma|h\rangle = 4\big([ac]\langle df\rangle[eg]\langle hb\rangle - \langle bd\rangle[ce]\langle fh\rangle [ga] \big)\,.
\end{equation}
This can be proven as follows using the formalism of \cite{Dreiner:2008tw}:
\begin{equation}
	i\epsilon^{\mu\nu\rho\kappa} = \eta^{\mu\nu}\eta^{\rho\kappa} - \eta^{\mu\rho}\eta^{\nu\kappa} + \eta^{\mu\kappa}\eta^{\nu\rho} -
		\frac{1}{2}\mathrm{Tr}[\bar{\sigma}^{\mu}\sigma^{\nu}\bar{\sigma}^{\rho}\sigma^{\kappa}]\,.
\end{equation}
Therefore considering
\begin{equation}
  \label{eq:epsilon}
	i\epsilon^{\mu\nu\rho\kappa}
		[a|\bar{\sigma}^\mu|b\rangle
		[c|\bar{\sigma}^\nu|d\rangle
		[e|\bar{\sigma}^\rho|f\rangle
		[g|\bar{\sigma}^\kappa|h\rangle\,,
\end{equation}
we have a term proportional to
\begin{equation}
	\frac{1}{2}\mathrm{Tr}[\bar{\sigma}^{\mu}\sigma^{\nu}\bar{\sigma}^{\rho}\sigma^{\kappa}]
		[a|\bar{\sigma}_\mu|b\rangle
		[c|\bar{\sigma}_\nu|d\rangle
		[e|\bar{\sigma}_\rho|f\rangle
		[g|\bar{\sigma}_\kappa|h\rangle\,,
\end{equation}
Using the identities
\begin{subequations}
  \begin{align}
	{\sigma}^{\mu}_{\alpha\dot{\alpha}}\bar{\sigma}_{\mu}^{\dot{\beta}\beta}&=2\delta_{\alpha}^{\beta}\delta^{\dot{\beta}}_{\dot{\alpha}}\,,\\
	{\sigma}^{\mu}_{\alpha\dot{\alpha}}\sigma_{\mu}^{\dot{\beta}\beta}&=2\epsilon_{\alpha\beta}\epsilon_{\dot{\alpha}\dot{\beta}}\,,\\
	\bar{\sigma}^{\mu,\dot{\alpha}\alpha}\bar{\sigma}_{\mu}^{\dot{\beta}\beta}&=2\epsilon^{\alpha\beta}\epsilon^{\dot{\alpha}\dot{\beta}}\,,
  \end{align}
\end{subequations}
we can write the product as
\begin{equation}
	\frac{1}{2}\bar{\sigma}^{\mu,\dot\alpha^\prime\alpha^\prime}\sigma^{\nu}_{\alpha^\prime\dot\beta^\prime}\bar{\sigma}^{\rho, \dot\beta^\prime\beta^\prime}\sigma^{\kappa}_{\beta^\prime\dot\alpha^\prime}
		a^\dagger_{\dot\alpha}\bar\sigma_{\mu}^{\dot\alpha\alpha}b_\alpha c^\dagger_{\dot\beta}\bar\sigma_{\nu}^{\dot\beta\beta}d_\beta e^\dagger_{\dot\gamma}\bar\sigma_{\rho}^{\dot\gamma\gamma}f_\gamma g^\dagger_{\dot\lambda}\bar\sigma_{\kappa}^{\dot\lambda\lambda}h_\lambda\,.
\end{equation}
Evaluating all contractions of $\sigma$ matrices, this becomes:
\begin{equation}
  \begin{split}
	8\epsilon^{\alpha^\prime\alpha}\epsilon^{\dot{\alpha^\prime}\dot{\alpha}}\delta_{\alpha^\prime}^\beta\delta_{\dot\beta^\prime}^{\dot\beta}\epsilon^{\beta^\prime\gamma}\epsilon^{\dot{\beta^\prime}\dot{\gamma}}\delta_{\beta^\prime}^\lambda\delta_{\dot\alpha^\prime}^{\dot\lambda}
		a^\dagger_{\dot\alpha}b_\alpha c^\dagger_{\dot\beta}d_\beta e^\dagger_{\dot\gamma}f_\gamma g^\dagger_{\dot\lambda}h_\lambda& =    8\epsilon^{\beta\alpha}\epsilon^{\dot{\lambda}\dot{\alpha}}\epsilon^{\lambda\gamma}\epsilon^{\dot{\beta}\dot{\gamma}}
		 a^\dagger_{\dot\alpha}b_\alpha c^\dagger_{\dot\beta}d_\beta e^\dagger_{\dot\gamma}f_\gamma g^\dagger_{\dot\lambda}h_\lambda\\
                & = 8a^{\dagger\dot\lambda}b^\beta c^{\dagger{\dot\gamma}}d_\beta e^\dagger_{\dot\gamma}f^\lambda g^\dagger_{\dot\lambda}h_\lambda\,.
  \end{split}
\end{equation}
Removing explicit indexes, we obtain
\begin{equation}	
  \label{eq:Trace_relation}
  \frac{1}{2}\mathrm{Tr}[\bar{\sigma}^{\mu}\sigma^{\nu}\bar{\sigma}^{\rho}\sigma^{\kappa}]
  [a|\bar{\sigma}_\mu|b\rangle
  [c|\bar{\sigma}_\nu|d\rangle
  [e|\bar{\sigma}_\rho|f\rangle
  [g|\bar{\sigma}_\kappa|h\rangle = 8(a^\dagger g^\dagger)(bd) (c^\dagger e^\dagger) (fh) = 8[ag]\langle bd\rangle [ce] \langle fh\rangle\,.
\end{equation}
Inserting this result into eq.~\eqref{eq:epsilon}, we obtain
\begin{equation}
\begin{aligned}
	i\epsilon_{\mu\nu\rho\sigma}[a|\gamma^\mu|b\rangle[c|\gamma^\nu|d\rangle[e|\gamma^\rho|f\rangle[g|\gamma^\sigma|h\rangle = 4\big([ac] \langle db\rangle [eg] \langle hf\rangle - [ae]\langle fb\rangle [cg] \langle hd\rangle + \\
	[ag] \langle hb\rangle [ce]\langle fd\rangle + 2[ag]\langle bd\rangle[ce]\langle fh\rangle\big)\,.
\end{aligned}
\end{equation}
Through repeated application of the Schouten identity and the
anti-symmetry of the spinor product one can
obtain equation~\eqref{eq:spinor_epsiilon_id} as:
\begin{subequations}
  \begin{align}
    i\epsilon_{\mu\nu\rho\sigma}[a|\gamma^\mu|b\rangle[c|\gamma^\nu|d\rangle[e&|\gamma^\rho|f\rangle[g|\gamma^\sigma|h\rangle
    = 4\big([ac] \langle db\rangle [eg] \langle hf\rangle - [ae]\langle fb\rangle [cg] \langle hd\rangle \nonumber  \\
    & +[ag] \langle hb\rangle [ce]\langle fd\rangle + [ag]\langle bd\rangle[ce]\langle fh\rangle
    + [ag]\langle bd\rangle[ce]\langle fh\rangle\big)\,.\\
    i\epsilon_{\mu\nu\rho\sigma}[a|\gamma^\mu|b\rangle[c|\gamma^\nu|d\rangle[e&|\gamma^\rho|f\rangle[g|\gamma^\sigma|h\rangle
    = 4\big([ac] \langle db\rangle [eg] \langle hf\rangle - [ae]\langle fb\rangle [cg] \langle hd\rangle \nonumber \\
   & + [ag] [ce] \left(\langle hb\rangle \langle fd\rangle + \langle bd\rangle \langle fh\rangle\right)
    + [ag]\langle bd\rangle[ce]\langle fh\rangle\big)\,.\\
    i\epsilon_{\mu\nu\rho\sigma}[a|\gamma^\mu|b\rangle[c|\gamma^\nu|d\rangle[e&|\gamma^\rho|f\rangle[g|\gamma^\sigma|h\rangle
   = 4\big([ac] \langle db\rangle [eg] \langle hf\rangle - [ae]\langle fb\rangle [cg] \langle hd\rangle \nonumber \\
    & + [ag] [ce] \left(\langle dh\rangle \langle bf\rangle\right)  + [ag]\langle bd\rangle[ce]\langle fh\rangle\big)\,. \\
      i\epsilon_{\mu\nu\rho\sigma}[a|\gamma^\mu|b\rangle[c|\gamma^\nu|d\rangle[e&|\gamma^\rho|f\rangle[g|\gamma^\sigma|h\rangle
 = 4\big([ac] \langle db\rangle [eg] \langle hf\rangle - \langle fb\rangle \langle hd\rangle \left([ae] [cg] + [ga] [ce]  \right) \nonumber \\         & + [ag]\langle bd\rangle[ce]\langle fh\rangle\big)\,.\\
            i\epsilon_{\mu\nu\rho\sigma}[a|\gamma^\mu|b\rangle[c|\gamma^\nu|d\rangle[e&|\gamma^\rho|f\rangle[g|\gamma^\sigma|h\rangle = 4\big([ac] \langle db\rangle [eg] \langle hf\rangle - \langle fb\rangle \langle hd\rangle \left([ac] [eg] \right) \nonumber \\
           & + [ag]\langle bd\rangle[ce]\langle fh\rangle\big)\,.\\
              i\epsilon_{\mu\nu\rho\sigma}[a|\gamma^\mu|b\rangle[c|\gamma^\nu|d\rangle[e&|\gamma^\rho|f\rangle[g|\gamma^\sigma|h\rangle = 4\big([ac] [eg] \left( \langle db\rangle \langle hf\rangle + \langle bf\rangle \langle hd\rangle \right) \nonumber \\
&              + [ag]\langle bd\rangle[ce]\langle fh\rangle\big)\,.\\
                i\epsilon_{\mu\nu\rho\sigma}[a|\gamma^\mu|b\rangle[c|\gamma^\nu|d\rangle[e&|\gamma^\rho|f\rangle[g|\gamma^\sigma|h\rangle = 4\big([ac] [eg] \langle fd\rangle \langle bh\rangle + [ag]\langle bd\rangle[ce]\langle fh\rangle\big)\,.\\
        i\epsilon_{\mu\nu\rho\sigma}[a|\gamma^\mu|b\rangle[c|\gamma^\nu|d\rangle[e&|\gamma^\rho|f\rangle[g|\gamma^\sigma|h\rangle = 4\big([ac]\langle df\rangle [eg] \langle hb\rangle - \langle bd\rangle[ce]\langle fh\rangle[ga]\big)\,.
      \end{align}
    \end{subequations}
There are also three other cases which can be proven in a similar way.
\begin{subequations}
\begin{align}
&  \frac{1}{2}\mathrm{Tr}[\bar{\sigma}^{\mu}\sigma^{\nu}\bar{\sigma}^{\rho}\sigma^{\kappa}]
  \langle b|\sigma_\mu|a]
  \langle d|\sigma_\nu|c]
  [e|\bar{\sigma}_\rho|f\rangle
  [g|\bar{\sigma}_\kappa|h\rangle\,,\\
&  \frac{1}{2}\mathrm{Tr}[\bar{\sigma}^{\mu}\sigma^{\nu}\bar{\sigma}^{\rho}\sigma^{\kappa}]
  \langle b|\sigma_\mu|c]
  [c|\bar{\sigma}_\nu|d\rangle
  [e|\bar{\sigma}_\rho|f\rangle
  [g|\bar{\sigma}_\kappa|h\rangle\,,\\
&  \frac{1}{2}\mathrm{Tr}[\bar{\sigma}^{\mu}\sigma^{\nu}\bar{\sigma}^{\rho}\sigma^{\kappa}]
  \langle b|\sigma_\mu|c]
  [c|\bar{\sigma}_\nu|d\rangle
  [e|\bar{\sigma}_\rho|f\rangle
  [g|\bar{\sigma}_\kappa|h\rangle\,,
\end{align}
\end{subequations}
which are all equal to $8[ag]\langle bd\rangle [ce] \langle fh\rangle$
as
in~\eqref{eq:Trace_relation}. Relation~\eqref{eq:spinor_epsiilon_id}
is then used to evaluate the Levi-Civitas in the helicity amplitudes
for operators $3$ and $4$. The simpler applications of the above
identity is the amplitude for operator 3, since the incoming
polarisation vectors contract with the incoming momenta as:
\begin{equation}
	i\epsilon_{\mu\nu\rho\sigma}[a|\gamma^\mu|b\rangle[c|\gamma^\nu|d\rangle[1|\gamma^\rho|1\rangle[2|\gamma^\sigma|2\rangle = 4\big([ac]\langle d1\rangle[12]\langle 2b\rangle - \langle bd\rangle[c1]\langle 12\rangle [2a] \big)\,.
\end{equation}
From here we can see that the $[ac]$ and $\langle bd\rangle$ will only
be non zero in the cases that the two incoming helicities are the
same, i.e.\ only for $++$ and $--$ configurations. 

In case of operator 4, we can rewrite the Feynman rules for the vertex as:
\begin{equation}
	\mathcal{O}_4  : 8 i \,\frac{c^{(GW)}_4}{\Lambda^4}\delta_{a_1 a_2}  \left[\left(\left( \epsilon^{\mu_1\mu_{(34)}}_{\quad\>\alpha\beta}p_1^\alpha p_{(34)}^\beta \right)\times \left( 1\to2, 3\to5, 4\to6\right)\right)\, + \, \left(1\to2\right) \right]\,.
\end{equation}
In this way the calculation of its amplitude can be remarkably simplified.

\section{Smoothing Contour Plots}
\label{sec:AppendixDiscont}
\begin{figure}[htbp]
\centering
  \includegraphics[width=.6\textwidth]{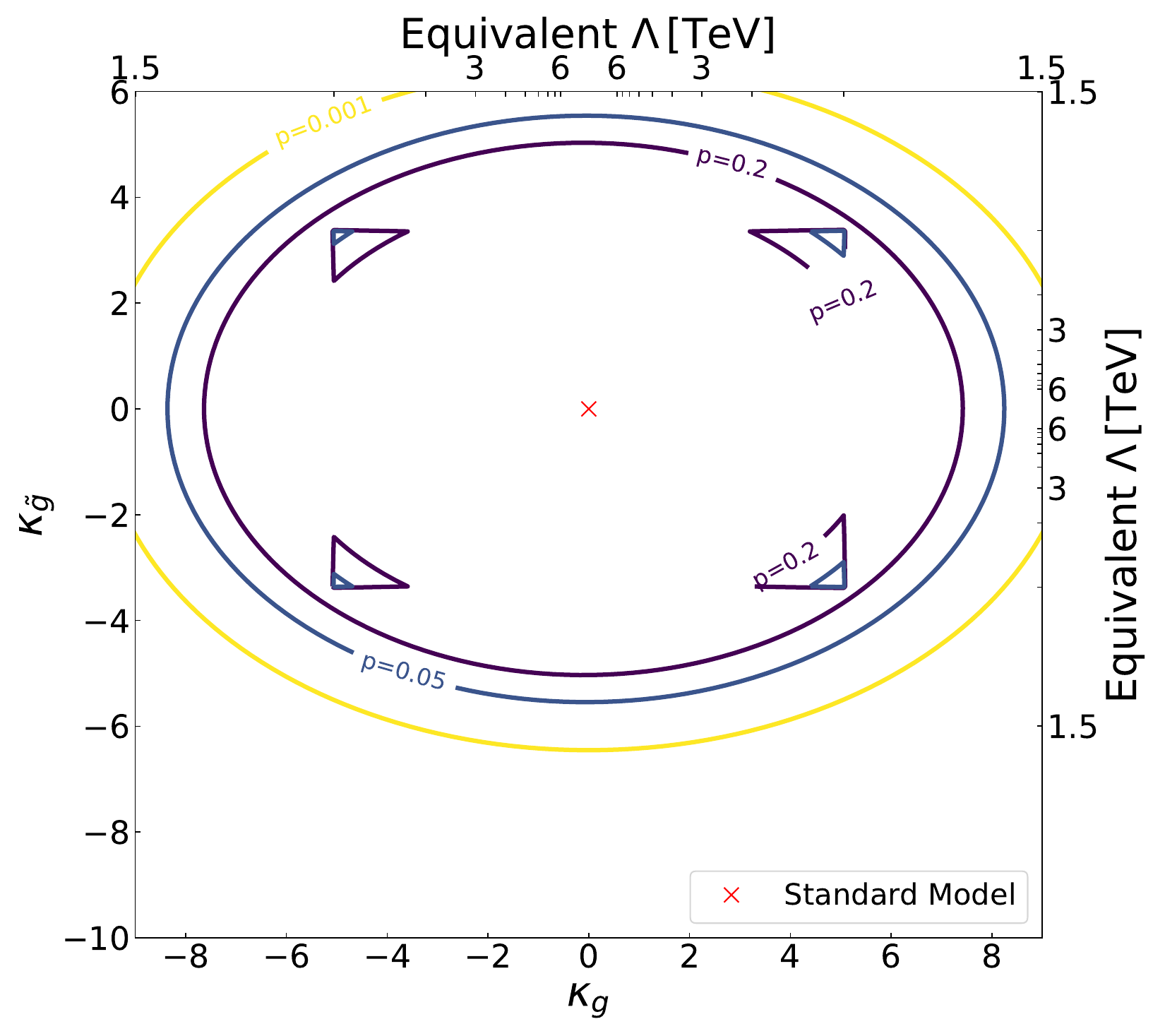}
  \caption{Unprocessed contour plot for the dimension-6 CP-even and CP-odd operators at ATLAS. The contours are placed at values of $p=0.2$, $p=0.05$, and $p=0.001$. It can be seen that by turning on both operators at the same time, a constraint could be made at around $\kappa=3$, $\tilde\kappa_g=5$. However this constraint does not encompass the cases where either operator is small and so we choose the lower constraint given by the complete ellipse at $p=0.02$.}
  \label{fig:Contourplot_k2k3_all}
\end{figure}

\begin{figure}[htbp]
  \includegraphics[width=.5\textwidth]{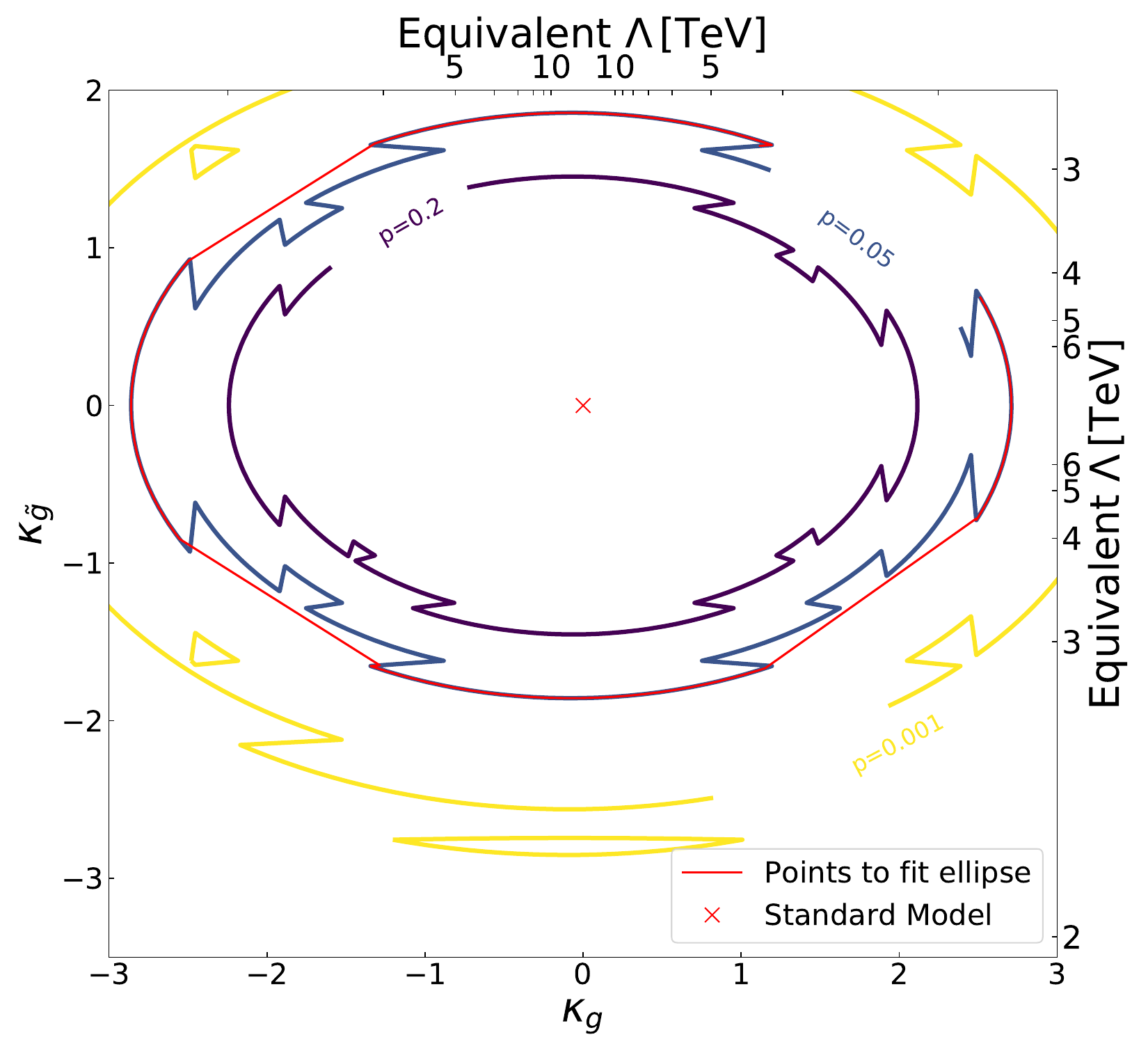}
  \includegraphics[width=.5\textwidth]{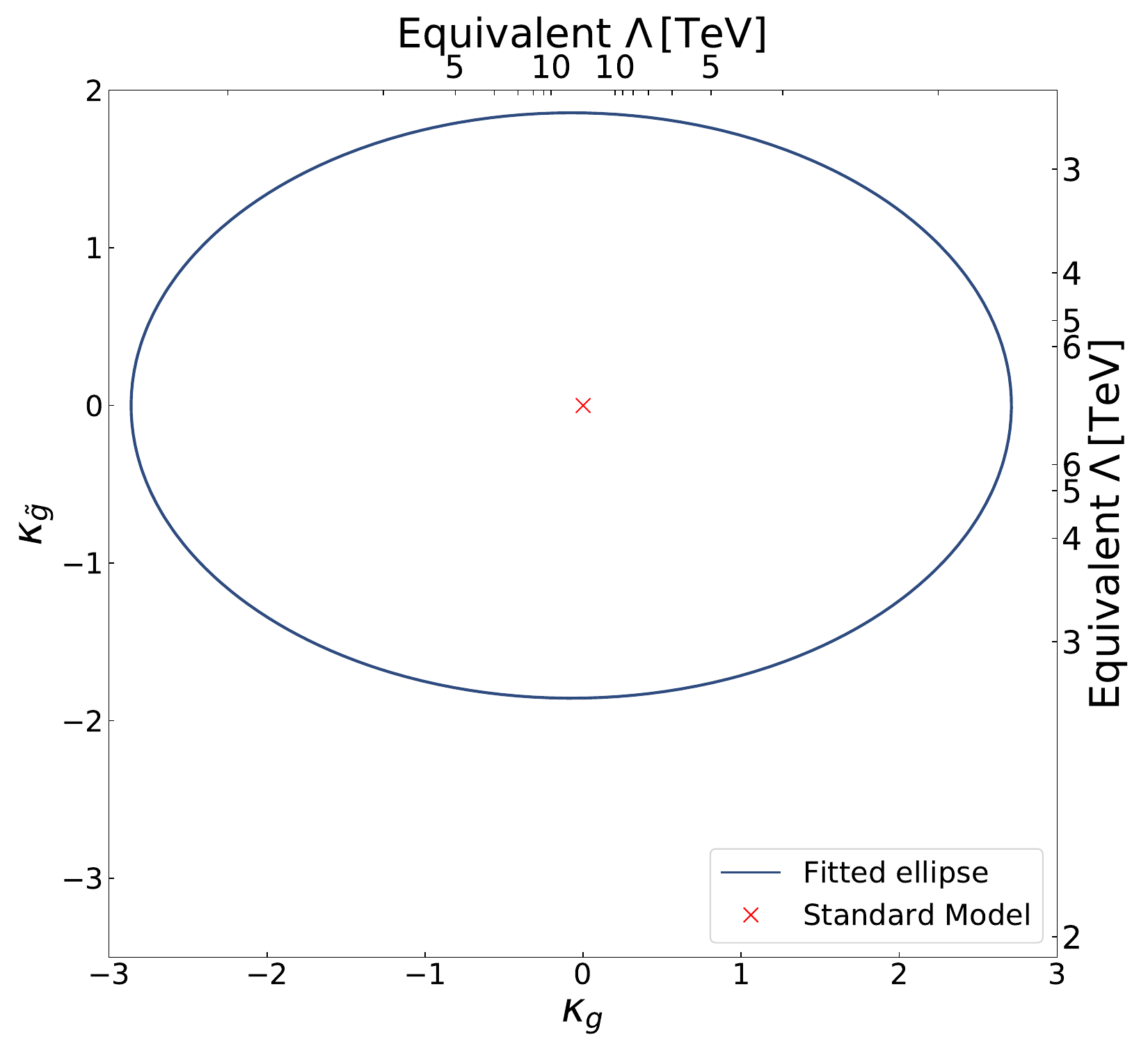}
  \caption{Unprocessed (left) and processed (right) contour plots for the dimension-6 CP-even and CP-odd operators at HL-LHC. The steps at each bin can be much more clearly seen in this plot. We select the outer most points of the completed boundary at $p=0.02$ and fit an ellipse to them in order to extract our constraint.}
  \label{fig:HLLHCContourplot_k2k3_findelipse}
\end{figure}

Due to the EFT validity constraints discussed in
section~\ref{sec:EFTValidity}, the number of bins which can be used in
a constraint depends on the energy being constrained. As the value of
$\Lambda$ increases, its value can be constrained using higher values
of $M_{e\mu}$.  However since we have chosen a set of fixed with bins
for the HL-LHC predictions and the ATLAS data is also given by a set
of fixed bins, we decide only to use a bin based on $\Lambda$ being
large enough such that the larger edge of the bin is within the EFT
regime. This divides the space of possible $\Lambda$ values into a
series of concentric squares which in turn leads to discontinuities in
the contour plots. To overcome these discontinuities we take a
conservative approach by choosing the outermost contour which
constrains the operators in all directions at a given accuracy, see
for example figure~\ref{fig:Contourplot_k2k3_all}. In the case of
figure~\ref{fig:HLLHCContourplot_k2k3_findelipse}, the contour does
not form a complete ellipse. In this case we take the parts of the
ellipse at $p=0.05$ and fit an ellipse to the points in order to give
a conservative constraint. This process is shown in
figure~\ref{fig:HLLHCContourplot_k2k3_findelipse}.

\clearpage
\bibliographystyle{JHEP}

\bibliography{dim8WW.bib}

\end{document}